# Photocatalysis with TiO$_2$ Nanotubes: "Colorful" Reactivity and Designing Site-Specific Photocatalytic Centers into TiO$_2$ Nanotubes

Xuemei Zhou,[†] Ning Liu,[†] and Patrik Schmuki*,[†,‡]

[†] Department of Materials Science WW4, LKO, University of Erlangen-Nuremberg, Martensstrasse 7, 91058 Erlangen, Germany

[‡] Department of Chemistry, Faculty of Science, King Abdulaziz University, P.O. Box 80203, Jeddah 21569, Saudi Arabia

Corresponding Author: *E-mail: schmuki@ww.uni-erlangen.de. Tel.: +49 91318517575. Fax: +49 9131 852 7582.

**ABSTRACT**: Photocatalytic reactions on $TiO_2$ have recently gained an enormous resurgence because of various new strategies and findings that promise to drastically increase efficiency and specificity of such reactions by modifications of the titania scaffold and chemistry. In view of geometry, in particular, anodic $TiO_2$ nanotubes have attracted wide interest, as they allow a high degree of control over the separation of photogenerated charge carriers not only in photocatalytic reactions but also in photoelectrochemical reactions. A key advantage of ordered nanotube arrays is that nanotube modifications can be embedded site specifically into the tube wall; that is, cocatalysts, doping species, or junctions can be placed at highly defined desired locations (or with a desired regular geometry or pattern) along the tube wall. This allows an unprecedented level of engineering of energetics of reaction sites for catalytic and photocatalytic reactions, which target not only higher efficiencies but also the selectivity of reactions. Many recent tube alterations are of a morphologic nature (mesoporous structures, designed faceted crystallites, hybrids, or 1D structures), but a number of color-coded (namely, black, blue, red, green, gray) modifications have attracted wide interest because of the extension of the light absorption spectrum of titania in the visible range and because unique catalytic activity can be induced. The present Perspective gives an overview of $TiO_2$ nanotubes in photocatalysis with an emphasis on the most recent advances in the use of nanotube arrays and discusses the underlying concepts in tailoring their photocatalytic reactivity.

## 1. INTRODUCTION

Over the last 40 years, photocatalytic reactions on $TiO_2$ have attracted tremendous scientific and technological interest.[1−4] The term photocatalysis, in general, is used for any semi- conductor where light with $h\nu > E_g$ is employed to generate charge-carrier (electron− hole) pairs that are then ejected from the semiconductor and react with suitable red/ox-couples in the environment.[1,5,6] The origin for focusing on $TiO_2$ as a semiconductor is an almost unique set of material properties that allows for an effective direct conversion of light into species which can be either a highly reactive intermediate or a desired final chemical product. Photocatalysis on $TiO_2$ addresses several contemporary global challenges such as pollutant degradation, $H_2$ generation from aqueous solutions, or nonfossil fuel production from $CO_2$. Moreover, it is used to establish functionality on surfaces such as controllable wetting and self- cleaning properties.[6−14] Although traditionally efforts have concentrated on energy and environmental uses, a multitude of other applications have been explored including biomedical devices (with antibacterial, biocompatibility, and drug release features) or the use of photocatalysis in photo-organic synthesis.[15−19]

$TiO_2$ is in general considered economically viable and nontoxic, and it provides an outstanding photocorrosion resistance (the latter being a key reason for the success of $TiO_2$ in photocatalysis). In spite of these desired features of titania, inherent to using $TiO_2$ in photocatalysis are two fundamental drawbacks: First, there is a wide band gap for $TiO_2$ ($E_g \approx 3$ eV for rutile and 3.2 eV for anatase), which means that in solar-light-driven processes, only UV light can be exploited; that is, only ≈4−7% of the entire solar spectrum is efficiently absorbed. Therefore, many research efforts have tried to reduce the $TiO_2$ bandgap by doping or band gap engineering. These efforts have resulted in various $TiO_2$ modifications, which may be categorized by their colored appearance such as yellow, green, red, blue, black, and numerous shades of gray.[20−31] Second, many desired charge-transfer reactions (or pathways to a desired reaction product) are kinetically hindered, and a wide range of strategies have been developed to "co-catalyze" the photocatalytic reactions in a favorable direction.[4,23,32]

Common strategies to enhance the photocatalytic perform- ance of $TiO_2$ include intrinsic measures such as optimizing crystal structures (polymorphs and faceting)[33−35] or modified surfaces, for example, by decorating $TiO_2$ with charge-transfer- mediators,[4] visible light absorber/charge injection moieties,[36−39] or secondary semiconductors[40−42] or metals to create desired electronic heterojunctions.[7,43]

Most photocatalytic reactions are carried out either under open-circuit conditions (electron and hole transfer occur from the same electrode) using $TiO_2$ nanoparticle suspensions or $TiO_2$ layers fixed on a support, or alternatively in a photo- electrochemical setting (under applied electrochemical bias) where $TiO_2$ is generally used as a photoanode together with an inert or catalytic cathode such as Pt, C, among others.

In both settings (open-circuit conditions and photoelectro- chemical approaches), it is important to consider the nature of the electronic junction formed at a semiconductor/electrolyte interface, as it determines not only the energetics of phase boundaries but also to a large extent the kinetics of the reaction.

In the past decade, a plethora of novel morphologies of $TiO_2$ such as nanowires, nanosheets, and nanotubes have become increasingly synthetically controllable and can be designed to an unprecedented degree.[44−46] These geometries have greatly affected research in various photocatalytic fields (such as hydrogen generation, pollution degradation or "self-cleaning" surfaces).[4,47−52]

Numerous excellent reviews have described general synthesis strategies and key advantages of 1D nanostruc- tures.[3,5,8,11,12,47,53−59] Therefore, in the present work, we will only give a short outline of some key fundamentals and concepts with an emphasis on aspects that are highly relevant to the case of $TiO_2$ nanotubes.

In the present Perspective, we will focus on the most widely investigated morphology over recent years: *self-ordered nanotubes* grown by self-organizing electrochemical anodization (oxida- tion) (SOA) of a metallic Ti substrate as illustrated in Figure 1a. For detailed overviews on the growth of such tubes, see refs 60−66. Meanwhile, it is possible to grow not only hexagonally close packed tube arrangements but also tubes with a defined tube-to-tube interspace (Figure 1b). For various applications such as to construct flow-through membranes (Figure 1c), the tube layers can be lifted-off the substrate and can be used in free- standing approaches or transferred to a foreign substrate. We will look at key aspects of photocatalysis on TiO$_2$ nanotube arrays and namely *modifications* that allow *site-specific* property alterations that integrate a functional feature into the nano- tubular geometry in a highly defined form and at a desired location.

Owing to the control over the nanoscale tube geometry (diameter, length) and tube crystal structure (amorphous, anatase, rutile), electrochemical anodization is in many cases the most straightforward nanotube synthesis path, leading to assemblies for photocatalysis and electrodes for photoelectro- chemical applications.

Common and less-common approaches that target a modification of the reactivity (selectivity or yield of a photoelectrochemical reaction) of these TiO$_2$ nanotubes are illustrated in Figure 1d−i. Such approaches include surface alterations of the chemistry and of physical properties (e.g., the attachment of active species for light harvesting or of bioactive molecules, Figure 1h) to induce electronic effects (such as doping or band gap engineering), to establish electronic heterojunctions (e.g., secondary semiconductor particle deco- ration or approaches to increase the surface area, Figure 1h), to alter the crystallinity (Figure 1i), or to create core−shell type of tubes (Figure 1g). Moreover, several techniques achieve a direct embedding of features into the tube wall (ion implantation or in- growth, Figure 1f). Some techniques are unique to anodic nanotubes such as intrinsic doping with a species X by anodizing a Ti-X alloy or embedding metallic Au or Pt into nanotube walls by anodizing noble metal-containing titanium alloys (Figure 1d). Finally, some approaches are very promising in view of selective catalysis such as the filling of the tubes with specific capturing agents (e.g., zeolite in Figure 1e).

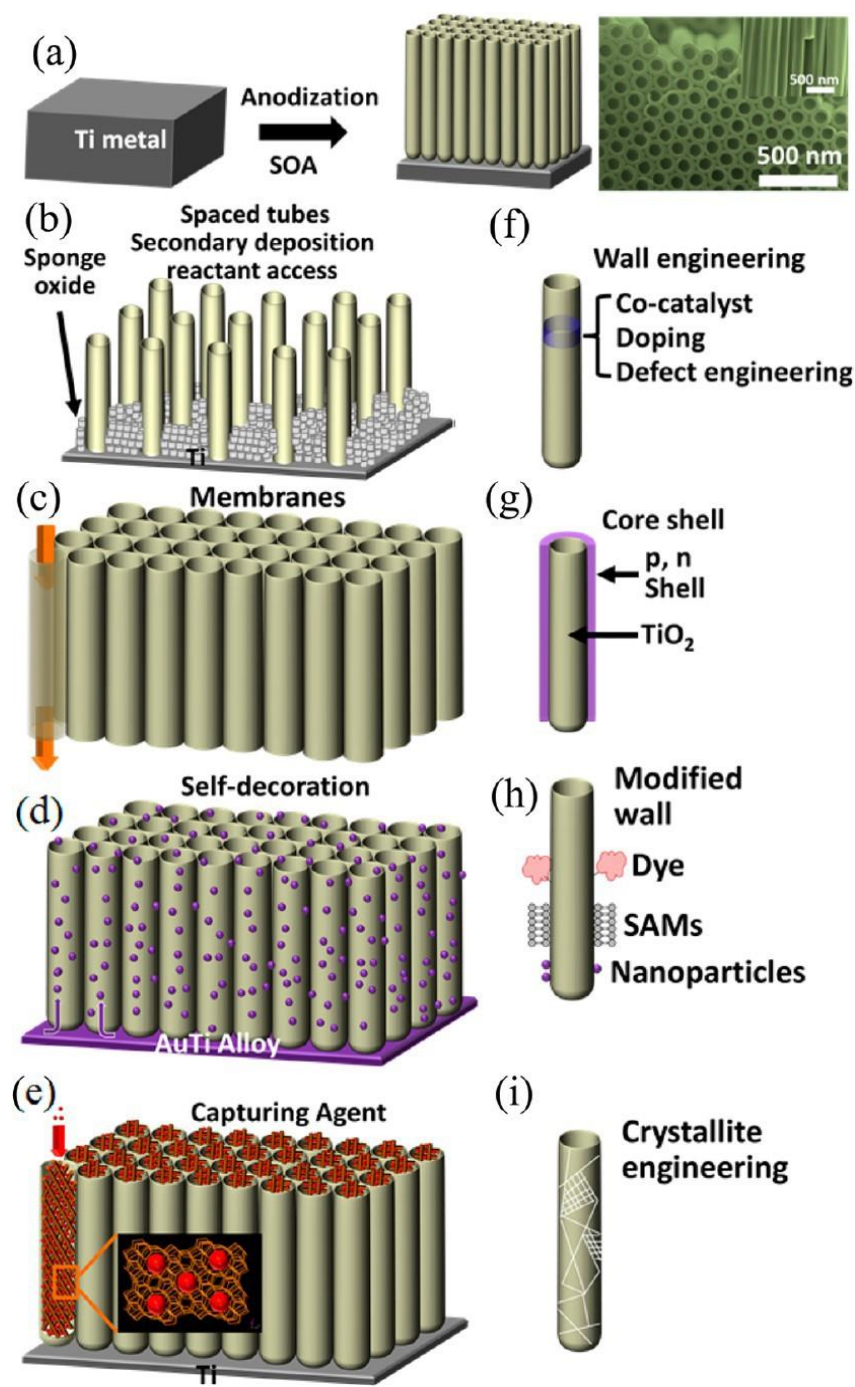

Figure 1. Schematic drawing of (a−b) formation and (c−i) modification of anodic nanotube arrays (as discussed in the text).

In the present work, we will discuss these modification approaches and underlying goals and concepts. We will give examples of applications of such nanotubes and specifically modified tubes toward enhanced photocatalytic $H_2$ generation and pollution degradation; additionally, we include some discussion on more exotic uses.

In section 2, we give a brief overview on some background to photocatalysis; hence, we do not aim at a comprehensive treatment of the fundamentals of photocatalysis but intend to highlight some key points and principles essential or crucial for the discussion of anodic tube properties and modification discussed in section 3.

## 2. SOME BASIC CONSIDERATIONS

### 2.1 General.

In any $TiO_2$ structure, a photocatalytic reaction consist of the steps outlined in Figure 2a. If a semiconductor such as $TiO_2$ is illuminated with light of an energy higher than its band gap, electrons are promoted from the valence band to the conduction band, leaving holes ($h^+$) in the valence band.[7,47,56]

Holes and electrons diffuse or migrate then on their respective band to the semiconductor surface and react with a suitable redox species in the environment.[7] Whether or not an electron transfer across the phase boundary can take place, is determined by the energetic position of the valence and the conduction band at the semiconductor surface relative to the level of the redox potentials available in the surrounding (e.g., $H_2O$, $CO_2$, $O_2$, and other examples provided in Figure 2b).[8–10,77] For an electron transfer to the solution to occur, the oxidized state in the solution has to lie energetically lower than the conduction band edge; for a hole- transfer, the reduced state in the solution has to lie energetically higher than the valence band edge of the semiconductor.[8–10,77]

In the absence of organics, for the exit of electrons to an aqueous electrolyte, the most-relevant capture agents are $H^+$ and $O_2$. In the $H^+$ case, this can be exploited to form $H_2$, and in the $O_2$ case, superoxide radicals can be formed (that can contribute to pollution degradation as well as a wide range of other reaction products including the formation of $H_2O_2$.[13,43,65,78–80]) For the exit of holes, the main pathway is the reaction of $OH^-$ or water to $O_2$ or $H_2O_2$. Please note that this $H_2O_2$ is generated by a hole transfer from the valence band that forms $OH^•$ and may continue to react to $H_2O_2$ (i.e., $H_2O_2$ production on $TiO_2$ may be caused by a conduction or a valence band mechanism).

Importantly, the "exit energy" of the hole allows, in an aqueous environment, the reaction of water to directly form $OH^•$ radicals (by $h^+$ transfer to $HO^-$ or water).[82] The $OH^•$ radicals are able to oxidize a wide range of inorganic and organic compounds.[65,80] In fact, most organic compounds can be oxidized by $OH^•$ radicals fully to $CO_2$ and $H_2O$; in other words, such reactions can be exploited, for example, for the oxidative destruction of virtually any organic pollutant or organic monolayers.

In many processes, undesired competition can take place; for example, for the conduction band reaction, the formation of $H_2$ from $H^+$ competes with a reduction of $O_2$ in the surrounding. Moreover, various multiple reaction processes can occur such as hole capture from the valence band followed by a second oxidation of the product at the conduction band (by injecting an electron to the conduction band). This phenomenon, "current doubling", is often observed for multiple oxidation-state reactants, and a prominent example is methanol-containing solutions.[83,84] Such competing situations can be overcome by suitable counter measures (e.g., degassing, changing the solvent, or band gap engineering).

More recently, conduction band electrons are explored for the photocatalytic reduction of $CO_2$ to useful fuels.[8,11,48,85] Reduction reactions of carbon dioxide to CO or $CH_x$ (in a wet gas phase) or $CH_3OH$ (in liquid $H_2O$) are thermodynamically possible on $TiO_2$ (see Figure 2b).[10,12,77,86,87] Nevertheless, these reactions are usually kinetically hindered due to the energetically disadvantageous nature of the two electron transfer steps involved.[10] It is important to note that reduction reactions involving a kinetically preferred one-electron transfer such as $CO_2 + e^- \rightarrow CO^{-\bullet}$ are not expected to be thermodynamically feasible on neat $TiO_2$ ($E^0 = -1.9$ V vs NHE).[10]

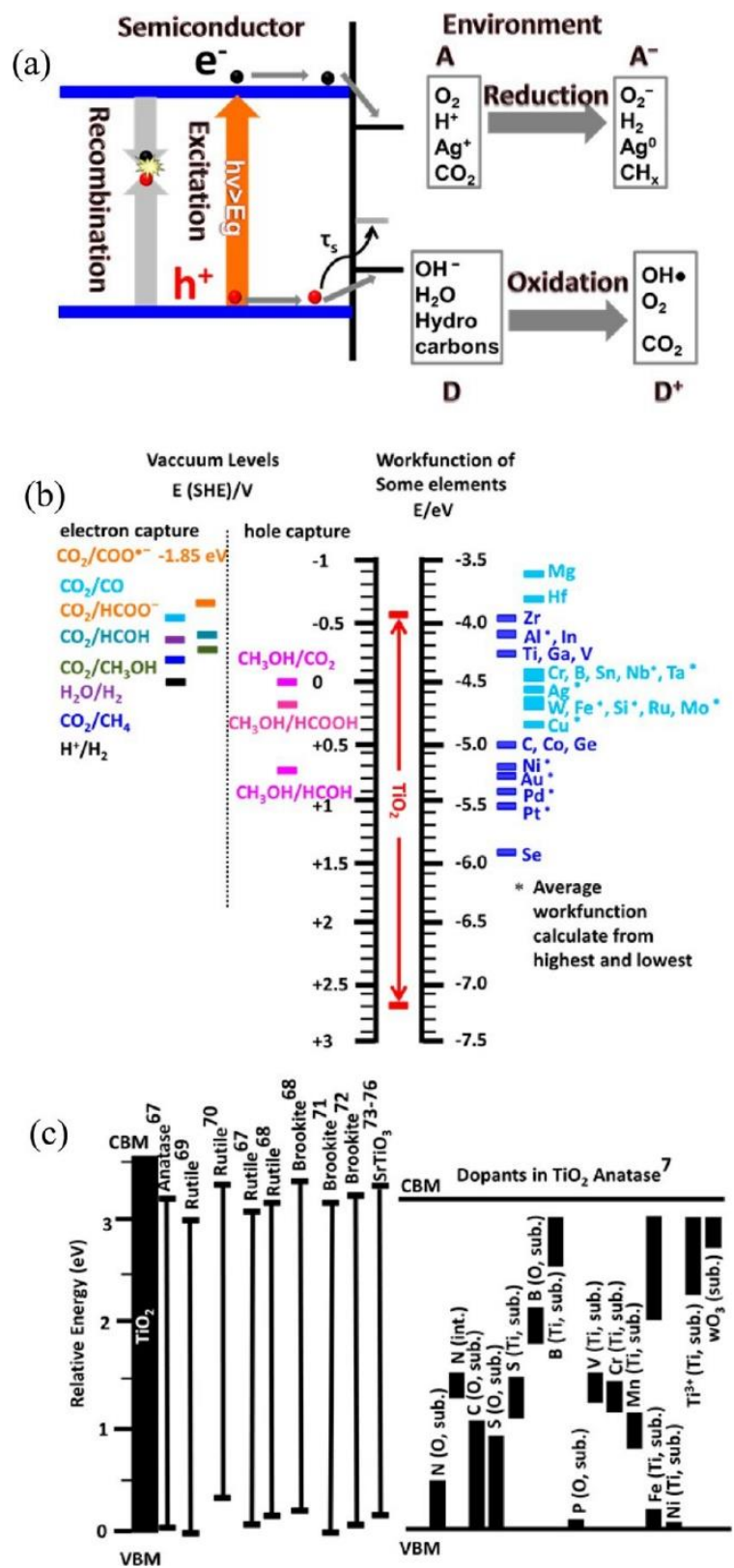

Figure 2. (a) Scheme of photocatalytic processes at a $TiO_2$ semiconductor/electrolyte interface. Light ($h\nu$) excites valence band electron to conduction band. Electron and hole react with environment. Acceptor species (A) are reduced and donor species (D) are oxidized (= photocatalytic reactions). Competing with the desired reactions is trapping and recombination of electrons and holes (= reducing the photocatalytic efficiency). (b) Relative energetic positions of various redox couples and metals relative to the band-edges of $TiO_2$. (c) Schematic illustration of energy level positions for various polymorphs and typical dopants in $TiO_2$ anatase. Values are obtained from refs 67−76 and ref 7.

In comparison with the wide application of photocatalysis in $H_2$ generation, for $CO_2$ conversion and pollutant degradation the use of the photogenerated electron−hole pairs in organic synthetic reactions are much less exploited. The main reason for this is that a multitude of redox or radical-based reaction pathways become accessible after a photoinduced electron or hole transfer from $TiO_2$ to an organic species has taken place.

Regarding thermodynamics, an important (and sometimes overlooked) factor for $TiO_2$ photocatalysis is the polymorph of $TiO_2$ used. $TiO_2$ is known in three main crystalline forms: anatase, rutile, and brookite. In addition, a synthetic layered phase, so-called $TiO_2$ (B),[88] and some high pressure polymorphs exist.[89,90] These polymorphs have a different relative position of valence and conduction band as well as different bandgap energies ($E_g$) as illustrated in Figure 2c.[67–72] The different gap (rutile $E_g \approx 3.0$ eV compared to anatase $E_g \approx$ 3.2 eV) has a significant consequence on the amount of light absorbed if solar (or solar simulator) light is used; this, in turn, has a considerable effect on achievable conversion efficiencies.

In this context, it is also important to note that energetic considerations based on Figure 2 are made and hold for the case of a nonelectrochemically biased situation and are significantly modified when junctions are formed (see section 2.2).

Moreover, the positions of the band edges as in Figure 2 determine the thermodynamics of a photocatalytic reaction but not its kinetics.[5] The kinetics is determined by the time scales of electron transfer, trapping, and recombination of the photo- generated charge carriers.[43,47] Corresponding fundamentals are discussed in numerous excellent reviews.[55,58,91–94] Key is the lifetime of excited carriers versus their reaction time with the surrounding. A most effective means to affect carrier lifetime is the formation of electronic junctions as discussed below.

### 2.2 Junctions.

One of the essential motive for modifying reactivity and selectivity of $TiO_2$ photocatalysts and, in particular, nanotubes is junction engineering.

*Formation and Characterization.* Electronic junctions are formed when a semiconductor is in contact with a phase of a different Fermi level $E_F$ as illustrated in Figure 3a. Such junctions give electron and hole transport an opposite direction, and thus drastically enhance carrier lifetime. In photocatalysis, most relevant junctions to the semiconductor are formed: (i) by the contact with an electrolyte,[12] (ii) by the contact with a metal,[23] (iii) by a contact with another semiconductor[95–97] (including another polymorph or crystal facet[98,99]), or (iv) by local variations in the doping level within the $TiO_2$.[4,100]

Junctions are commonly described by an induced surface barrier $U_s$ and a corresponding depletion width $W$ in the semiconductor. For a junction with a metal, $U_s$ is determined by the work function of the metal with $U_s = U_{fb} − E_F$ (Figure 3a). For a contact of $TiO_2$ with a secondary semiconductor, except for the work function of the semiconductor, also the conduction type (n or p), doping level, and their relative position of the band edges are essential. Favorable junctions can be established, if the energetics aid charge separation (as illustrated in Figure 3c).[42]

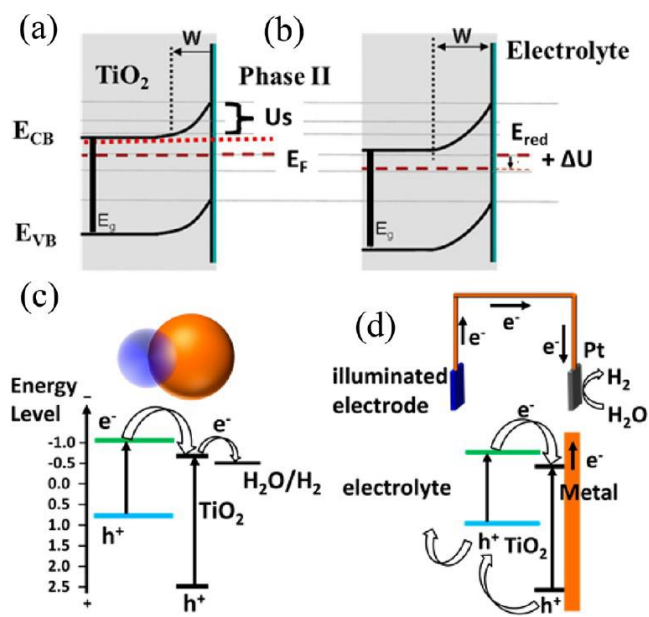

Figure 3. (a,b) Energy diagram of n-type $TiO_2$ semiconductor in contact with second phase for the case (a) $E_{F,sc} > E_{F,redox}$ or $E_{F,sc} > \Phi_M$ leading to Schottky barrier ($U_s$) with width $W$ and (b) after applying anodic bias ($+\Delta U$), further modifying $U_s$ and $W$. (c,d) Illustrations for semi- conductor heterojunctions for (c) open-circuit photocatalytic $H_2$ evolution (e.g., particles in solution) and for (d) photoelectrochemical $H_2$ evolution as photoanode under applied bias.[42] (a,b) Reproduced with permission from ref 7. Copyright 2012 Wiley-VCH Verlag GmbH & Co. KGaA, Weinheim.

In photoelectrochemical reactions, shifts in the Fermi level (and $U_s$ as well as $W$) and thus the thermodynamics and the kinetics can be controlled - to a large extent - by an external applied voltage ($\Delta U$),[55,84,98,101] as illustrated in Figure 3b.[102,32]

A specific form of a junction is a hole or electron transfer cocatalysts that mediate electron or hole transfer not only by electronic junction formation but also by providing additional (chemical) effects. For the electron transfer at the conduction band to the electrolyte, noble metals (M), such as Pt, Pd, or Au, are often used.[29,48,103,93,104] Particularly, in the case of $H_2$ generation from $TiO_2$, Pt nanoparticles also act as a hydrogen recombination center that strongly promotes $H_2$ formation from atomic H.[29,48,103] More recently, economical alternative $H_2$ evolution species such as $MoS_2$,[105–107] MoP/S,[108] as well as intrinsically modified (cocatalyst-free) gray titania[27] have become increasingly investigated. Typical classic $O_2$ evolution catalysts are $IrO_2$ and $RuO_2$,[109] whereas more recent research targets substitutes such as Co-oxides and Co-phosphates.[110,111] Nevertheless, reports on simultaneous successful photocatalytic $O_2$ evolution and $H_2$ evolution from the same particle are scarce.[6] Under OCP conditions, reduction as well as oxidation reactions have to occur at the same photoelectrochemical entity (particle) (Figure 3c). The situation under photoelectrochemical conditions (Figure 3d) is insofar different as the applied bias can be used to extract electrons from the junction to the back contact and into an electrochemical circuit−the relevant reaction on such a photoanode is then the hole transfer to the electrolyte.[42] That is, in this case, $TiO_2$ is mainly decorated with $O_2$ evolution catalysts and/or is used in a hole capture environment.

In general, if charge carriers do not "react away" sufficiently fast, they may accumulate at the semiconductor surface. For hole accumulation in the valence band, this can oxidize the semiconductor, which may lead to dissolution. In contrast to most other semiconductors (such as CdS, CdSe), for $TiO_2$ the electronic nature of the valence band prevents such photo-induced dissolution.[7]

To avoid charge-carrier accumulation, cocatalysts[112–115] can be used to enhance the hole transfer rate. Alternatively, a more efficient redox species (with suitable energy levels and kinetics) may be added to the electrolyte (typical "hole-capture agents" or sacrificial electrolytes are methanol, ethanol, etc. at the valence band or electron-capture agents (e.g., $Ag^+$)) at the conduction band. [95,103,116]

To describe semiconductor junctions (as in Figure 3a,b) quantitatively, the so-called Mott−Schottky approach (eq 1) is most frequently used (this under OCP or under bias). It connects $U_s$ and $W$ as follows:

$$W = \left[\frac{2}{q\varepsilon_0\varepsilon N_d}\left(U_s - \frac{kT}{q}\right)\right]^{1/2} \quad (1)$$

where $W$ is the width of the space charge (depletion) layer induced by the contact with a material of a different work function (metal or electrolyte). $\varepsilon$ denotes the dielectric constant, $\varepsilon_0$ is the vacuum permittivity, $q$ is the charge of the electron, $N_d$ is the donor concentration (for an n-type semiconductor), $k$ is the Boltzmann constant, and $T$ is the absolute temperature.

This concept is the basis to the characterization of junctions in view of Mott−Schottky (MS) plots and photoelectrochemical characterization (photocurrent spectra and photocurrent− voltage behavior).[69]

In MS measurements, impedance techniques are used to evaluate the voltage-dependent space charge capacitance $C(U)$. Typically $C(U)$ is evaluated either from single (high) frequency measurement or from fitting equivalent circuits to impedance measurements. The capacity $C(U)$ is generally represented by an ideal plate condensator with $C(U) = \varepsilon\varepsilon_0 A/W(U)$, $W(U)$ being the width of the space charge layer at a specific applied voltage $U$, $U_s = (U_{fb} - U)$.

In principle from the linear part of a $1/C_{SC}{}^2$ versus $U$ plot the flatband potential (at $1/C^2 = 0$) can be evaluated, and the doping density of the material can be obtained from the slope.[117,118]

However, it has to be pointed out that for titania and a large range of oxide electrodes - although this approach is meanwhile extensively used often - a nonideal behavior is obtained. This, as underlying assumptions in the deduction of the Mott−Schottky treatments, is not met. For example: The MS treatment does not hold (i) if Fermi-level pinning occurs,[119] (ii) for degenerate semiconductors;[120] degeneracy is reached typically if the material has $N_d > 10^{21}$ cm$^{-3}$ (for such high doping levels, the implicit assumption that the space charge capacitance is smaller than the capacitance of the Helmholtz layer ($C_{sc} \ll C_H$) is no longer provided).

In the context of the present paper, it is also important is to note that for nanosize materials such as TiO$_2$ nanotubes[121], $W$ cannot extend infinitely with U (further outlined in section 3.2 for nanotubes).

In photoelectrochemical experiments, the photocurrent ($i_{ph}$) generated in a semiconductor electrode is generally registered as a function of wavelength or as a function of the applied potential $U$. The quantitative description typically follows the Johnson−Gaïtner−Butler approach (eq 2, eq 3, and eq 4).[122−124]

$$i_{ph} \propto \alpha_\lambda \propto \frac{(h\nu - E_g)^n}{h\nu} \quad (2)$$

$$\text{IPCE} = \frac{i_{ph}(\lambda)}{qI(\lambda)} \quad (3)$$

$$i_{ph} \propto (U - U_{fb})^{0.5} \quad (4)$$

where $i_{ph}$ is the generated photocurrent, $\alpha_\lambda$ is absorption coefficient, $h\nu$ is the energy of the incident light, $E_g$ is the band gap energy, $I(\lambda)$ is the incident photon flux, $q$ is the electronic charge, $U$ is the applied potential, and $U_{fb}$ is the flat band potential.

From photocurrent spectra (or IPCE vs $h\nu$)$^{1/n}$ plots and using eq 2, the band gap energy (direct $n = 1/2$ or indirect $n = 2$ of the semiconductor can be evaluated.[123] From the photocurrent− voltage behavior (at a fixed wavelength), the optical flat-band potential of the material (at $i_{ph} = 0$) can be obtained. The analysis of photocurrent transients gives information on carrier kinetics (such as mobility and trapping),[125] and thus, it enables, together with modulated techniques (IMVS, IMPS),[12,83] the extraction of key time scales of semiconductor/electrolyte junctions. In many cases, the presence of a sub-band gap $i_{ph}$ response provides proof of successful band gap modification. In contrast to plain light absorption measurements of the Kubelka−Munk type that evaluate $\alpha_\lambda$ by plain light absorption, $i_{ph}$ measurements show if a light absorbing electronic state is indeed able to couple electronically to the semiconductor - that is, if electrons can be thermally excited to extended bands, and/or show a significant mobility.

Also, please note that eq 2 contains assumptions of eq 1, and thus, nonidealities as discussed for MS plots are also affecting photocurrent data.

Under OCP conditions it should be considered that shining light on a semiconductor mainly enhances the minority carrier density (for n-type this is the $h^+$ concentration); this leads to a photopotential, that is, on $TiO_2$ (n-type semiconductor), shift of the OCP in the negative direction. Thus, the band bending is reduced, and this causes a self-induced change in the reaction rate.[126]

**2.3 Size Effects on Junctions.** Except for the relative positions of the band edges of a secondary semiconductor decoration, the particle sizes compared with hole and electron diffusion length also determine the efficiency of such junctions; that is, electrons and holes should reach the surface and react away before they statistically recombine.[127–129]

A specific case of semiconductive junctions and sensitizers are provided by quantum dots (QDs).[98,130,131] In QDs, the small size of the particle (some nm) favors rapid charge transfer, and additionally, $E_g$ and thus the relative positions of the band edges also depend on the particle size.[131] This is due to quantum confinement that can be described by[46]

$$E_g = \frac{h^2}{8m_0 R^2}\left(\frac{1}{m_e^*} - \frac{1}{m_h^*}\right) - \frac{1.8q^2}{4\pi\varepsilon_0\varepsilon_r R} \qquad (5)$$

where $h$ is Planck's constant, $R$ is the radius of the particle, $q$ is the charge of the electron, $m_0$ is free electron mass, $m_e^*$ is the effective mass of the electron (for $TiO_2$, typically between 5 to 30 $m_0$), $m_h^*$ is the effective mass of the hole (for $TiO_2$ typically between 0.01 to 3.0 $m_0$), $\varepsilon_0$ is the permittivity of vacuum, and $\varepsilon_r$ is the static dielectric constant ($\approx 30-185$).

In general, the gap becomes size-dependent, and thus, for suitable species, a match with the conduction band of $TiO_2$, for example, can be achieved by size adjustment. In semiconductor junctions (including QDs) usually the Fermi-level equilibrates according to the available carriers (doping densities) at a value in between the two levels before contact (if no Fermi level pinning occurs).[117,131,132]

Junction formation with a small metal particle is generally considered in the same terms as macroscopic metal/semi- conductor junctions; that is, using the Schottky approach and implying that the Fermi-level of the semiconductor will equilibrate to the metal Fermi level (this is due to an order of magnitude higher charge-carrier density in the metal compared with a typical semiconductor). Nevertheless, this does not hold for sufficiently small metal nanoparticles (nm size). Namely, if metal nanoparticles are placed on highly doped semiconductors (for example, for various forms of $TiO_2$ of anodic layers, NTs, or reduced titania, doping levels

$N_d$ up to $10^{21}$ cm$^{-3}$ have been reported.[133,134]

A rough estimate for a Pt nanoparticle of 1 nm in diameter on a highly doped TiO$_2$ layer yields a depletion of ca. 1000 nm$^3$ of the semiconductor (assuming all atoms in the Pt to be ionized).[135] This corresponds to a depletion width of 10−20 nm around a Pt particle. (In this case, the number of carriers provided by the doping species of the semiconductor and the number of free carriers in the entire metal particle become comparable). Under these conditions, the Fermi level of the semiconductor will significantly affect the Fermi level of the metal particle, which is commonly not assumed to occur.[135] As a consequence, for such metal particles, a particle-size dependent reactivity (for a few nm) can be expected, which is dependent on the electronic nature of the substrate. Such combined substrate- and size-dependent effects are for example observed for nm sized Au on various substrates.[136−140]

In the context of nanoscale TiO$_2$ (and applications in photocatalysis), it is important to consider size effects on the stability of polymorphs. For bulk TiO$_2$, rutile has the lowest free energy, and hence, given the necessary activation energy, any crystalline form will finally transform into the rutile phase.[118,141,142] However, for nanoscale materials, a large number of experimental and theoretical investigations conclude that for crystallite sizes smaller than approximately 10−30 nm, anatase represents the most stable phase.[142,143] Indeed, for many nanomaterials at moderate annealing temperatures a transition from an amorphous phase to anatase is experimentally reported.[144] For sufficiently large systems (>few 10 nm), anatase to rutile transformation takes place at temperatures of 500−700 °C. The exact conversion temperature depends on several factors, including impurities, primary particle size, texture, and strain in the structure.[142,145,146]

The fact that nanoscale anatase is thermodynamically stable can be used for thermal modifications of nm-size entities without facing a conversion to rutile. This has been exploited for production of efficient anatase-type photocatalysts (e.g., by oxidation of nanoscale powders of nitrides).[28,147,148]

**2.4 Key Examples of Junctions.** In the following, we briefly illustrate some most important types of junctions relevant to TiO$_2$ nanotubes.

*Junctions with a Sensitizer.* Particularly beneficial electronic junctions can be formed by so-called sensitizers. Figure 3c illustrates sensitization of TiO$_2$ under OCP conditions. Light can be absorbed in TiO$_2$ and a decorated smaller gap semiconductor. The smaller gap of the sensitizer allows for electrons to be excited at longer wavelength. Electrons can then be transferred to the conduction band of TiO$_2$, while holes generated at shorter wavelength in TiO$_2$ are transferred to the conduction band of the small gap semiconductor. From the respective bands, electrons and holes are then available to react with the environment.

For TiO$_2$, typical examples of a suitable sensitizer are II−VI semiconductors (CdS, CdSe).[149,150] While such junctions under OCP and photoelectrochemical conditions can provide a quite high photoconversion efficiency, a main practical obstacle remains photocorrosion.

*Junctions by Polymorphs or Facets.* Mixed polymorphs of TiO$_2$ such as anatase/rutile (such as in Degussa P25 commercial catalyst) are of a high importance because they may form internal homojunctions that aid electron−hole separation (leading to a higher photocatalytic activity; see also section 3.2).[94,99]

It should be noted, however, that considerable debate still exists on the relative energetic position of the two most important polymorphs, namely, anatase and rutile (as also illustrated in Figure 2c).[151] Usually band edge positions are determined by vacuum or electrochemical techniques. Overall, there is a tendency that vacuum-type measurements put the rutile band edge energies higher while electrochemical measurements tend to find the opposite.[67−70] This difference between vacuum and electrochemical techniques has only recently been addressed in theory and has been ascribed to the role of OH-termination of TiO$_2$ in aqueous environments that drastically affects the relative position of the Fermi level[152] or the level of band-edge pinning.[84] For a fully hydroxylated surface (i.e., H$^+$ and OH$^-$ adsorption on the undercoordinated surface oxygen and titanium atoms, respectively), $E_F$ is only ∼0.5 eV above the H$^+$/H$_2$ potential in the case of anatase and−depending on the level of reduction− roughly at the same level, or below, for rutile.[152]

In recent years, the formation of junctions between crystal facets has been increasingly studied.[153,154] In general, the photocatalytic activity of titania has been reported to be strongly dependent on the crystal facet exposed to the environment. On anatase, reports particularly point out the high reactivity of {001} facets as opposed to {101} facets.[155,156]

However, most available anatase TiO$_2$ crystals are dominated by the thermodynamically stable {101} facets (more than 94%), rather than {001} facets.[53,141] This holds not only for natural anatase crystals but also for TiO$_2$ nanopowders and crystallites synthesized in Cl$^-$ or SO$_4^{2-}$ solutions.[157,158] However, nano- crystals formed in fluoride solutions show fluorine-terminated surfaces, and the relative stability is reversed: {001} is energetically preferable to {101}.[33,156,159,160] However, Cheng[82] showed that upon a decrease of the fluoride capping, clean {010} facets became the majority and showed the highest activity.

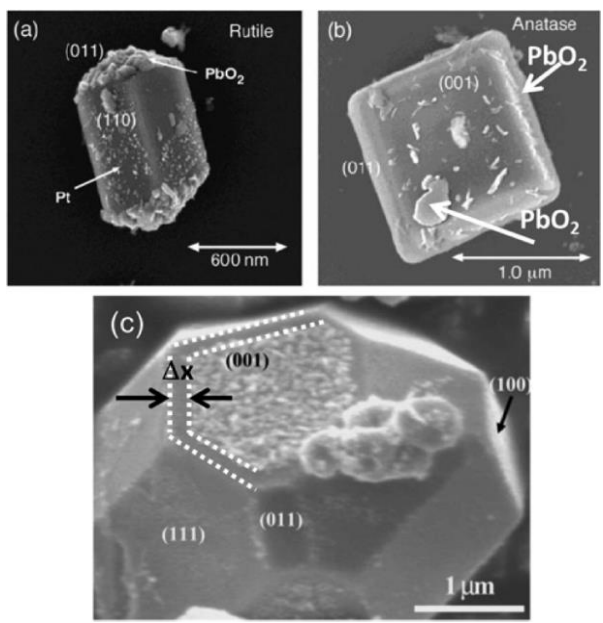

Figure 4. SEM images illustrate facet-induced junctions resulting in selective deposition of Pt (electron capture) and PbO$_2$ (hole capture) on a rutile particle (a) and an anatase particle (b). (c) SEM image of BaTiO$_3$ crystal after photoreaction in aqueous AgNO$_3$ leading to silver deposits. (Please note the empty fringe along the crystal edges.) (a,b) Reproduced with permission from ref 153. Copyright 2002 The Royal Society of Chemistry and the Centre National de la Recherche Scientifique. (c) Reproduced with permission from ref 154. Copyright 2008 Springer Science+Business Media, LLC.

The different reactivity on the different facets of anatase and rutile and their heterojunctions has been elucidated by several reports.[153,154,161−163] Matsumura et al. showed in their remarkable work (Figure 4a,b) that the intrinsic difference in surface energy of crystal faces on titania can lead to charge separation and electron and hole specific reactions on the different facets.[153] The authors investigated 1 μm size particles of rutile and anatase. On these surfaces, preferential electron capture (Pt decoration) could be observed on the (110) plane of rutile. Hole capture (PbO$_2$ deposition) preferentially occurs on the (011) plane of rutile, whereas no selective PbO$_2$ deposition can be observed on anatase.

Regarding this effect of internally generated heterojunctions between two facets of a crystal, it is important to consider the crystal size versus the range of a junction in terms of carrier depleted and accumulated zones (e.g., approximated by a Schottky model such as in eq 1). That is, for particles (or other moieties) of a few nm, the entire particle may be affected by depletion/accumulation of a space charge layer, whereas for larger particles (μm size), it may be only the edges of crystal where the two faces of different energy meet. This effect is illustrated in Figure 4c for photoinduced deposition of Ag on a BaTiO$_3$ crystal.[91,154]

**2.5  Doping.** The electronic structure of TiO$_2$ can be altered by introducing suitable intermediate state(s) in the band gap. The purpose of this may be to change the electrical conductivity of the material (doping) or to narrow the gap itself (band gap engineering).[102,164] In both cases, the states introduced need to lie considerably close to the band edges to act as electric doping species (within the energetic range of thermal activation), otherwise they can only be optically addressed. Various reviews describe general doping strategies in detail.[6,94,102,164,165] Figure 2c summarizes a range of common species considered for doping TiO$_2$. The data given refer to positions relative to the band edges of intrinsic TiO$_2$ (obtained by DOS calculations).

In spite of wide investigations, since the work of Asahi et al.,[166] nitrogen doping remains the most successful approach. Mean- while it is well-established that oxygen-substitutional nitrogen doping N(O, sub.) causes an apparent narrowing of the band gap by introducing N$_{2p}$ states just above the TiO$_2$ valence band,[6,38] as shown in Figure 2b this in contrast to states formed by interstitial nitrogen [N(int.)].[38,167]

Similar to substitutional N, oxygen-substituted carbon, sulfur, and phosphorus form states near the valence band edge.[168−171] Also, substitution of Ti(Ti, subs.) has been widely reported, and various reviews[172−176] deal in depth with advantages (optical response) and disadvantages (recombination centers) of doping states introduced to $TiO_2$.[170,177]

In the context of the present Perspective, a specifically important point is the introduction of $Ti^{3+}$ species into the $TiO_2$ lattice. In classic work on $TiO_2$ reduction these species are reported to form an intermediate state about 0.8 eV below the conduction band of $TiO_2$.[178] However, formation of $Ti^{3+}$ states is in general combined with the formation of oxygen vacancies. $O_v$ have been reported to lie over a wide range of energetic positions near the conduction band of $TiO_2$.

Recently, $Ti^{3+}/O_v$ pairs formed by reduction of $TiO_2$ have generated a considerable amount of revitalized interest in the context of creating "self-doped" titania of a gray to black color with some interesting properties as discussed below.

**Colored Titania: From Yellow to Black.** In general, band gap narrowing treatments produce colored powders. The classic lattice N doping, carbon doping as well as various treatments that result in a surface N-modification (rather than solid state doping) provide a yellow powder.[179,180] More recent approaches lead to red,[20,21] green,[22] black,[23−26] and gray[27,28] material.

Among these, red anatase represents a refined version of N-doping (Figure 5a). The material is produced by a pre-doping of $TiO_2$ with an interstitial boron gradient to improve the solubility of substitutional nitrogen in bulk anatase without introducing nitrogen-related $Ti^{3+}$ as extra electrons from boron can compensate for the charge difference between lattice $O^{2-}$ and substitutional $N^{3-}$.[20] This red $TiO_2$ can absorb the full visible light spectrum and provide an absorption band gap gradient, varying from 1.94 eV on the surface to 3.22 eV in the core. Red $TiO_2$-based photoanodes have been reported to be able to split water under visible light irradiation.[20,21]

While the above selected examples rely on doping or codoping of $TiO_2$ with specific extrinsic donors or acceptors, intrinsic doping the formation of so-called black or gray titania is simply produced by high-temperature treatment of anatase under a reducing atmosphere.[181] The appearance of a black to blue color can be ascribed to $Ti^{3+}$ and $O_v$ species that increase in concentration with an increasing level of reduction (Figure 5a). The origin of color is typically assigned to d−d transitions.[182]

Most convincing proof for $Ti^{3+}/O_v$ formation is electron paramagnetic resonance (EPR). Hydrogen-treated $TiO_2$ usually shows the presence of $O_v$ and $Ti^{3+}$ in the lattice,[183] and depending on the reducing conditions (generally with increasing temperature), the signal intensity of $Ti^{3+}$ increases and changes its signature. Various studies ascribe changes in the signature to different positions (i.e., $Ti^{3+}$ located at regular Ti-lattice sites or interstitial $Ti^{3+}$ sites).[52,184−188]

It is noteworthy that for these classic reduction treatments, generally the $Ti^{3+}$ states are reported to be prone to reoxidation in air or aqueous environments.[189]

Except for high-temperature reduction, synthetic approaches to form "self-doped" $TiO_2$ involve solvothermal treatments,[190] organic lithiation treatments,[191−194] or use of imidazole to react with $O_2$ to form CO and NO as the effective reducing gas.[195]

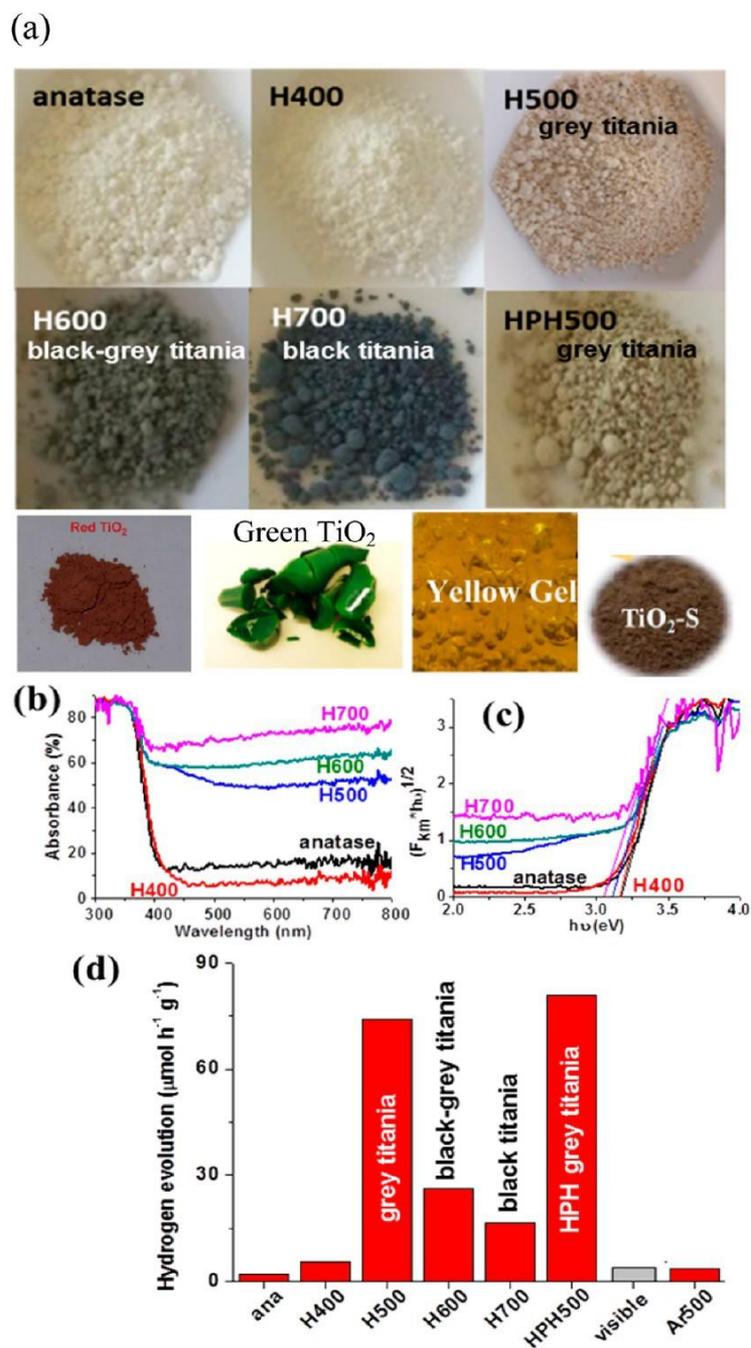

Figure 5. (a) Optical images of anatase nanopowders treated under different hydrogenation conditions (H: flow furnace, HPH: high- pressure hydrogenation) illustrating shades of gray and black coloration. For red, yellow, and green titania, optical images are taken from refs 20,22,24,30. (b) Integrated light reflectance spectra of $TiO_2$ after different hydrogenation treatments. (c) Data of Figure 1b transformed using the Kubelka−Munk function for the extraction of the optical gap energy. (d) Noble-metal-free photocatalytic hydrogen evolution rate under AM 1.5 (100 mW/cm$^2$) illumination for $TiO_2$ nanoparticles after different hydrogenation treatments, a reference treated in argon gas at 500 °C and the most active sample if UV light is blocked (visible). (a−d) Reproduced with permission from ref 27. Copyright 2016 Wiley-VCH Verlag GmbH & Co. KGaA, Weinheim.

(a) Images in the third row from left to right: Reproduced with permission from ref 20. Copyright 2012 The Royal Society of Chemistry; Reproduced with permission from ref 22. Copyright 2017 Wiley-VCH Verlag GmbH & Co. KGaA, Weinheim. Reproduced with permission from ref 24. Copyright 2013 American Chemical Society; Reproduced with permission from ref 30. Copyright 2014 The Royal Society of Chemistry.

Particularly high attention has been triggered by an approach pioneered by Chen and Mao[23] that used mesoporous $TiO_2$ particles that were exposed to a high-pressure/high-temperature treatment in pure $H_2$. After several days, the powder turned black, and when decorated with Pt nanoparticles, a very high water splitting activity was obtained (under open-circuit conditions). The authors ascribed the high activity to a narrowing of the band gap from 3.2 to 1.8 eV, given the formation of an amorphous H-containing shell around a crystalline core of the anatase nanoparticle.[23] This finding triggered a large amount of follow-up work, and a considerable number of beneficial effects of "blackening" $TiO_2$ were reported, including photoelectrochemical $H_2$ production, photocatalytic pollution degradation, and effective supercapacitor electrodes.[44,133,134,196]

Another remarkable property of high-temperature hydro- genated $TiO_2$ nanoparticles was reported by Liu et al.,[25,197,198] that is, in the absence of any noble metal cocatalyst, the treated material is able to photocatalytically generate $H_2$. This intriguing feature has been reported for nanotubes,[197] powders and nanoparticles of $TiO_2$,[25] as well as for $Ti/TiO_2$ core−shell structures.[199] In other words, a suitable hydrogenation treatment not only leads to a visible light absorption as reported by Chen and Mao,[23] but also creates an intrinsic cocatalytic active center in $TiO_2$,[25] similar to the cocatalytic effect obtained by noble metal decoration. This center has been ascribed to stabilized $Ti^{3+}$ states that are energetically close to the conduction band of $Ti^{3+}$,[25,26,147] or alternatively to the beneficial effect of surface hydroxides.[27,152] More recently, it was shown that this cocatalytic activation requires the presence of defects (high index places), which does not occur on single crystal low index planes of anatase.[200] In view of noble-metal-free $H_2$ evolution, it is noteworthy that the activity of powder shows a maximum efficiency at intermediate reduction treatments (Figure 5b−d); that is, gray titania is significantly more active than black titania. In other words, the enhanced visible light absorption of black titania seems not mechanistically connected to the effect causing noble-metal-free hydrogen evolution.[27,28]

It is however important that in any case, various changes of properties occur when $TiO_2$ nanostructures are "blackened". Namely, the formation of reduced states also drastically increases the electrical conductivity of the material, which is likely the primary cause of property improvements in various reports.[134,181,201−204]

## 3 TIO₂ NANOTUBES AND WHY NANOTUBE ARRAYS?

Over the past decade, highly defined 1D $TiO_2$ morphologies (such as nanotubes or -wires) have become widely explored for their photocatalytic performance and were found in many cases to be superior to nanoparticles.[44−46,133,205] Various synthesis approaches are available to produce 1D assemblies in solution.[206]

Nevertheless, a main drawback of any powder form (e.g., loose assemblies of nanowires, nanotubes, etc.) is that either photocatalytic processes have to be conducted in a suspension (which requires postreaction separation of the loose material from the solvent) or the photocatalyst has to be immobilized on a carrier by compacting or sintering, thus providing a random orientation on the substrate. In contrast, $TiO_2$ nanotubes that are perpendicularly aligned and directly back contacted on a conductive surface can be grown by self-organizing anodization (SOA) of a metallic Ti substrate as illustrated in Figure 1a.

As grown, the tubes are amorphous but can be annealed (crystallized) to anatase or rutile. Anodic tube layers can be grown to defined geometries[207] where the tube geometry is to a large extent defined by the anodization parameters (voltage, time, electrolyte).[54,102,208] These tube layers have meanwhile been exploited widely for OCP as well as for photo- electrochemical photocatalysis.

The wide interest in such nanotube layers is due to various advantages that are inherent to these structures:

*i).* For anodic self-organized tubes, a key feature is the fact that they are fabricated from the metal; that is, no immobilization process is needed, and the tubes can be directly used as back contacted photoelectrodes.

*ii).* Directionality for charge separation (i.e., as described in Figure 6a) and orthogonal separation of charge carriers can be exploited.

*iii).* Easy control of photocatalytically relevant parameters (diameter, length, wall thickness) is provided.

*iv).* Controlled doping via substrate can be achieved.

*v).* Defined chemical or electronic gradients or junctions can be fabricated.

*vi).* Metallic substrates (even with complex geometries) can be conformally coated (as illustrated in Figure 6b; showing a wire mesh that is fully and uniformly coated with a perpendicular tube layer after anodizing in a simple two- electrode arrangement).

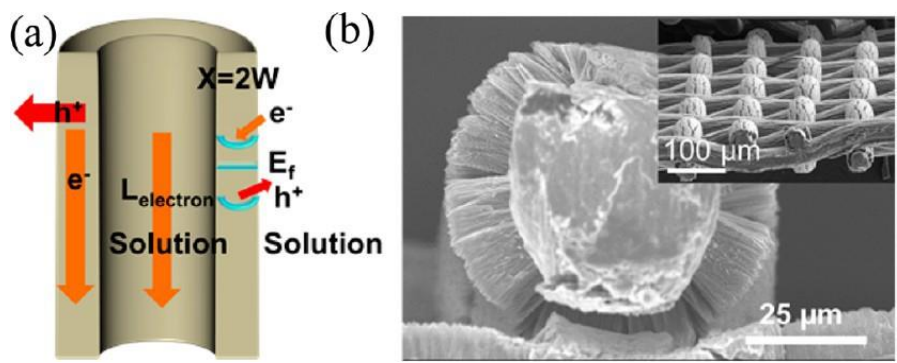

Figure 6. (a) Illustration of orthogonal electron−hole separation in a nanotube wall. (b) Illustration of conformal coating of a complex substrate morphology with $TiO_2$ nanotubes (NTs) using a metal wire mesh as an example (inset: lower magnification of the coated Ti-wire mesh).

As a result, the anodic structures can be used directly not only in static reactors but also, more importantly, as photoanodes in electrochemically assisted photocatalytic processes.

Among the above aspects, a most important point for photocatalysis is that in these structures an orthogonal carrier separation is facilitated (Figure 6a), i.e. electrons and holes are spatially separated within the tube wall (that is, e⁻ are collected in the wall center and transported to the back contact while h⁺ are driven to the wall/electrolyte interface).[7,209] This is particularly beneficial to overcome limitations due to the short diffusion length of holes in $TiO_2$ (~10 nm) while exploiting the comparably long electron diffusion length (~20 μm in $TiO_2$ nanotubes).[83]

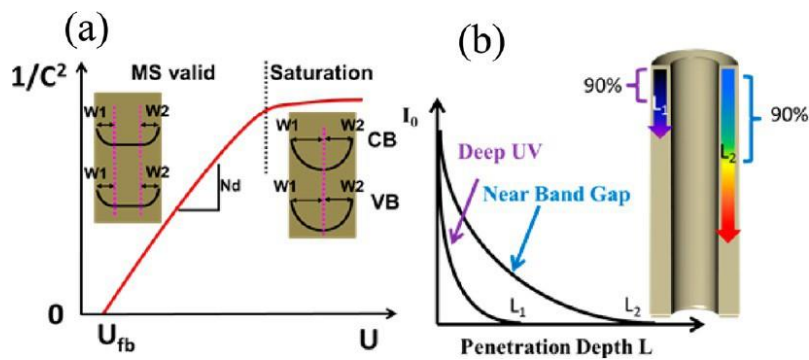

Figure 7. (a) Depletion of tube walls with increasing anodic voltage. At low voltage, Mott−Schottky behavior holds; at high voltage, the space charge layers overlap leading to a complete carrier depletion of the tube walls. (b) Schematic of light penetration depth $L$ into a tube layer for different wavelengths (note that UV light is absorbed in the outermost part; longer wavelengths penetrate deeper).

In terms of photoelectrochemical and capacitance properties it is noteworthy that for a typical $TiO_2$ nanotube wall thicknesses (say 20 nm), already at a relatively moderate bias (~0.3 V anodic to the flat−band potential), the depletion layers overlap as shown in Figure 7a.[83]

In this case, charge-carrier depletion occurs over the entire nanostructure (and a saturation in $W$, the photocurrent, and capacitance, respectively, is observed).[46,210]

In other words, the space−charge layer follows the wall contours only up to this threshold voltage (and may most effectively aid charge separation only in this voltage range).

Considering transport of carriers orthogonal to the tube-wall direction migration toward the surface of the semiconductor has a transient time ($\tau$) that can be expressed by eq 6:[211]

$$\tau = L^2/\pi D \tag{6}$$

where $L$ is the relevant travel distance of charge (e.g., the wall thickness) and $D$ is the diffusion coefficient of the excited charge carriers; for example, for $L \approx 20$ nm and $D = 5 \times 10^{-3}$ cm² s⁻¹, the transient time is around a hundred picoseconds (if no trapping/ detrapping with bulk states occurs).

That is, for carriers driven to the tube surface, trapping and reaction (both have a time scale in the microsecond range) are much more reaction-rate-determining than hole transport.

In photoelectrochemical or photocatalytic experiments with tube layers, another important point is to consider the penetration depth of light (as illustrated in Figure 7b).

In many cases, the light absorption characteristics (profile) for different wavelengths is crucial, as the light absorption coefficient, α, is much higher for deep UV than for lower light energies.[83] In TiO$_2$ nanotubes, deep UV light is absorbed in the outermost part (i.e., hundreds of nanometers) of a tube layer, while near-band-gap light penetrates on the order of micro- meters.[83] As a result, UV-generated electrons have to travel much farther to the back contact than electrons generated by longer wavelengths (holes usually get surface trapped and react with the electrolyte or "wait" to recombine with a passing electron;[83] see also Figure 6a).

Additionally, it should be considered what light source is used for excitation (e.g., a broad spectral UV/vis lamp (such as a solar simulator) or pure UV (e.g., a laser)). This is of special importance because a solar simulator spectrum possesses a strong intensity in the range of 3.0 to 3.2 eV. In other words, the small difference in band gap between rutile and anatase considerably influences the results. This is not the case if a single-wavelength-deep UV source is used.

However, it has to be pointed out that the exact tube geometry can affect light absorption and reflection greatly; not only can interference effects by partial modifications be intentionally[212] or randomly[213] created in tubes,[100] but also, the overall tube shape can be tuned to maximize light absorption.[214,215]

TiO$_2$ represents a relatively good electron conduction material (long electron lifetime in anatase TiO$_2$)[145,146] compared with other oxide semiconductors. However, in TiO$_2$ nanotubes, the carrier lifetime (diffusion length) depends to a large extent on the quality of the layer used for photoelectrochemical investigations.[214,216,217]

Charge-carrier transport in TiO$_2$ nanotubes can be signifi- cantly different from comparable nanoparticulate systems, as bulk states present in anodic nanotube structures are found to strongly affect the trapping/detrapping transport of majority carrier transport.[83,218] The electron diffusion length under UV illumination for nanotubes is much higher than for comparable nanoparticle layers because of lower surface recombination.[83] However, for both tubes and particles, the overall photocurrent is to a large extent determined by surface recombination effects.

Regarding the limiting factor for electron conduction in TiO$_2$ nanotubes, somewhat conflicting reports exist. Investigations by Richter et al. indicate that TiO$_2$ nanotubes and nanoparticle films have equally low electron transport rates.[182] This is in the case of TiO$_2$ nanotubes ascribed to the presence of exciton-like trap states.

Other reports[219] investigated the electric transport properties of single TiO$_2$ nanotubes separated from an anodic titania nanotube array. The temperature dependence of the resistance measured with a four point method along a single tube show a Mott-variable-range-hopping behavior. Impedance spectroscopy in the frequency range of 40 Hz to 1 MHz carried out at room temperature indicates that the electronic transport of these polycrystalline tubes is dominated by the grain cores (i.e., intrinsic defects in the TiO$_2$ matrix). These authors conclude that the low mobility in TiO$_2$ nanotubes is not due to scattering from grain boundaries but is ascribed to native defects such as Ti$^{3+}$ states.[182]

Very interesting is the comparison of these conductivity measurements where 4 point contacts are placed along single tubes, to measurements carried out between a top contact and the Ti metal back contact. In the latter case, an order of magnitude higher resistivity in the tube layers is observed than measurements along a single (separated) tube. This was found to originate in many cases from a rutile layer formed during thermal processing. This rutile layer at the metal/tube interface strongly reduces charge transport to the back contact[220] (a main strategy to overcome this effect is to anneal lifted-off-oxide membranes, i.e., without a metal layer underneath[144]). In spite of this back contact issue, reports by Jennings et al.[216] and Lynch et al.[83] show strongly beneficial effects in nanotubes compared with particles due to reduced surface recombination rates and an electron diffusion length of 25−100 μm.

With respect to photoelectrochemical approaches (as opposed to OCP), the electron conductivity is highly important as the photogenerated electrons have to travel along the tubes to the back contact. As mentioned, for different forms of TiO$_2$ electrodes, an extremely wide large range of electron mobilities has been reported,[214] which to a large extent must be ascribed not only to a different crystallinity of the different structures but also to variations in preparation techniques of the photo- electrodes. Conductivities are found to depend strongly on the type of tubes[221,222] (and in particular on applied reduction treatments, forming Ti$^{3+}$ states,[27] as well as the presence of tube contamination[223]).

**3.1 Growth and Morphology of Nanotubes (NTs).** Anodic NT layer are grown by self-organizing anodization (SOA) that can be carried out in a simple anode/cathode arrangement, where metallic Ti (e.g., a foil) serves as anode. The oxide growth process is based on the oxidation of M (M → M$^{z+}$ + ze$^−$) and its conversion to metal oxide (MO$_{z/2}$) under an applied voltage, the source of oxygen ions being typically H$_2$O in the electrolyte.[208] SOA has been reported to be successful on a wide range of metals and alloys, and a number of reviews are available that focus on the growth of these arrays.[61,62,208,225,226]

Self-organized TiO$_2$ nanotubes can be grown under a wide range of electrochemical anodization conditions,[61,62,209,227−230] generally in aqueous or organic solvents (ethylene glycol, glycerol, triethylene glycol) with some fluoride addition, to lengths ranging from a few tens of nanometeres to several hundred micrometers, with tube diameters between ≈10 nm and 800 nm, and wall thicknesses of 10−100 nm. They can grow with a hexagonally close packed or a spaced configuration (Figure 1). Not only membranes such as in Figure 8a,b but also a number of "rippled", branched, or spaced morphologies can be pro- duced.[7,54,60,64] Under many conditions, the tubes consist of a two-shell structure with an inner shell containing species from the electrolyte, such as carbon and fluorides (Figure 9a left).[221]

It was found that the inner shell in "double-walled" tubes significantly affects the properties of the tubes. Particularly, the inner shell is detrimental for electron conductivity (Figure 9d).[231] The double-walled morphology of anodic tubes becomes even more apparent after annealing the as-grown tubes; the tube walls then typically consist of small grains of 5−20 nm (Figure 9b). This inner shell can be avoided in certain electrolytes. The most frequently used electrolytes for SOA are based on ethylene glycol (EG) and result in a "double-walled" structure, whereas tubes formed in dimethyl sulfoxide (DMSO)-based electrolytes show a single-walled tube morphology.[222,231]

More recently, a decoring process was introduced that can selectively remove the inner shell from double-walled tubes and thus provides a single wall morphology as shown in Figure 9a (right).[221,222,231] Figure 9b,c shows TEM images for single- and double-walled tubes after the same annealing procedure. Clearly much larger (>100 nm) and thinner grains are provided in single- walled tubes, and this leads to a drastically improved (2 orders of magnitude) electron conductivity (Figure 9d) and a significantly enhanced electron transport (Figure 9e) in the single-walled tubes.[231]

### 3.2 Key Factors That Influence the Photocatalytic Activity.

*General.* Important factors that affect the photo- catalytic performance of $TiO_2$ nanotubes under open-circuit conditions or as a photoelectrode are the crystallinity of the tubes, a rational optimization of the geometry,[7,236] and any sort of potential gradient that aids electron−hole separation (junctions).[7] A number of key factors that influence the photocatalytic activity are compiled in Figure 10a−d: this in terms of open-circuit dye degradation (Figure 10a,b), open- circuit $H_2$ generation (Figure 10c), and photoelectrochemical $H_2$ generation (Figure 10d). Early work showed that nanotube layers can have a higher dye degradation efficiency than comparable compacted nanoparticle layers (Figure 10a).[209] With respect to nanotube layers (generally for nanoparticle suspensions), for low reactant concentrations, a Langmuir−Hinshelwood kinetics is observed.[238] For particles, as expected from a point of zero charge of $TiO_2$ of approximately 6−7, for acidic pH values typically a better adsorption of COO – containing dyes (for example AO7) takes place, and typically an increased photocatalytic kinetics is observed.[227]

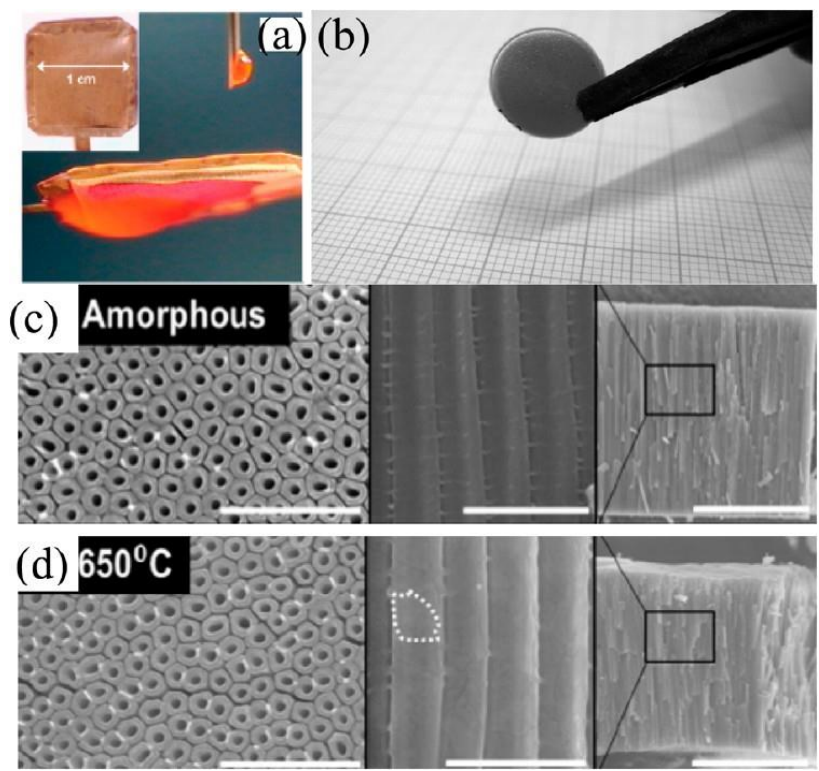

Figure 8. (a) Photograph showing a completely anodized metal anode leading to a both-side-open membrane. A colored water droplet demonstrates permeation of $H_2O$. (b) Optical image of a free-standing membrane. (c,d) Top (left) and cross-sectional (right) view SEM images of free-standing $TiO_2$ nanotube membranes annealed at different temperatures: (c) amorphous and (d) anatase at 650 °C. Scale bars are 1 μm, 500 nm, and 10 μm from left to right column, respectively. (a) Reproduced with permission from ref 224. Copyright 2010 American Chemical Society. (b) Reproduced with permission from ref 230. Copyright 2008 Wiley-VCH Verlag GmbH & Co. KGaA, Weinheim. (c,d) Reproduced with permission from ref 144. Copyright 2016 Elsevier B.V.

Regarding dye (pollution) degradation using $TiO_2$ powders as well as nanotubes, several reviews exist that deal with photocatalytic activity for different lengths, diameters, and types of titania nanostructures (crystallinity, doping, decoration with cocatalysts, etc.).[7,79,239,240]

*Geometry and Conductivity.* In general, a strong increase in the degradation kinetics of organics can be observed with increasing tube length up to a certain limit[7,241] (Figure 10b) (as the open-circuit decomposition to a large extent relates with the amount of absorbed superband-gap light). As mentioned, the penetration depth of light ($hv > E_g$) in typical tubes is several micrometers, and therefore, open-circuit efficiencies for photocatalysis usually saturate at ∼7−10 μm (see also Figure 7b).

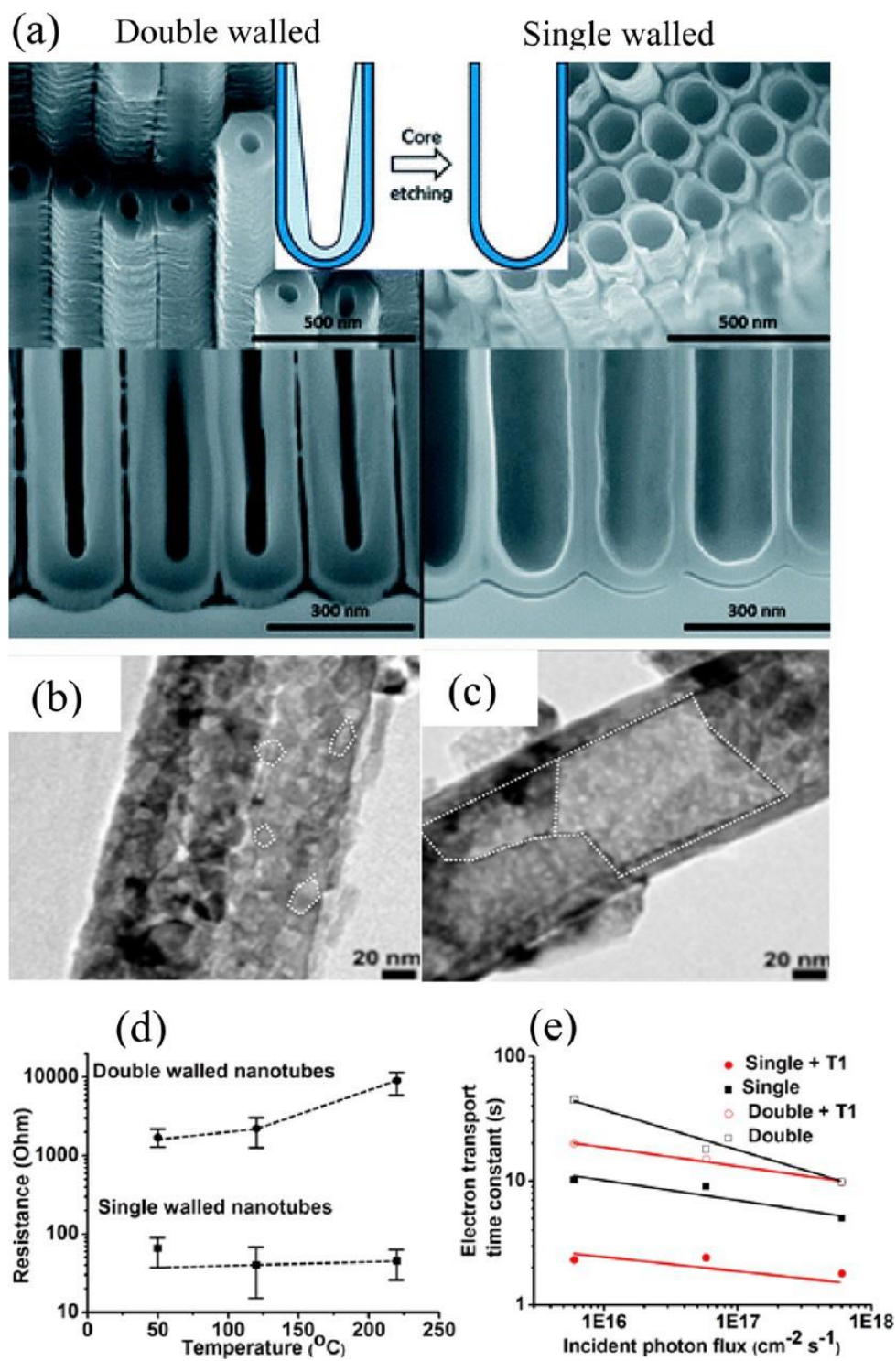

Figure 9. (a) SEM images of tubes showing typical double-walled morphology (left) and single-walled morphology (right). The single- walled tubes were obtained by a core removal process.[221] (b,c) TEM images of annealed tubes (450 °C, air) showing strongly different anatase crystallite sizes for double-walled (b) and single-walled (c) nanotubes. (d) Comparison of electrical resistance of the single-walled and double-walled nanotube layers of 15 nm thickness at different temperatures. (e) Electron transfer time ($\tau_c$) constants from IMPS measurements under UV light (350 nm) for single-walled and double- walled $TiO_2$ nanotubes with and without T($n$) layers of TiO$_2$ nanoparticle decoration (TiCl$_4$ treatment) ($n$: number of repeated TiCl$_4$ treatments). (a,e) Reproduced with permission from ref 221. Copyright 2015 the Royal Society of Chemistry. (b−d) Reproduced with permission from ref 222. Copyright 2013 the Royal Society of Chemistry.

A similar influence of length of nanotubes can also be observed for photocatalytic H$_2$ evolution under OCP (Figure 10c) and photoelectrochemical conditions (Figure 10d). Under OCP, significant amounts of H$_2$ can only be obtained, if cocatalysts, for example, Pt-nanoparticles, are decorated on the tubes (Figure 10c), or if the tubes have been exposed to activating treatments (H$_2$-annealing,[25] ion-implantation,[26,147] etc.)

In photoelectrochemical applications, the tube conductivity becomes crucial, and as a result, doping with Nb, Ta, Ru (Figure 10d)[232,233] or exposure to reductive treatments affects the efficiencies; whether single or double-walled tubes are used also determines the efficiency (Figure 9e). Single-walled tubes, particularly after an additional passivating treatment in TiCl$_4$, show a much faster electron transport than double-walled tubes (Figure 9e).[221,242] Namely, a controlled layer-by-layer TiCl$_4$ treatment leads to improved electron transport characteristics in TiO$_2$ nanotubes.[221]

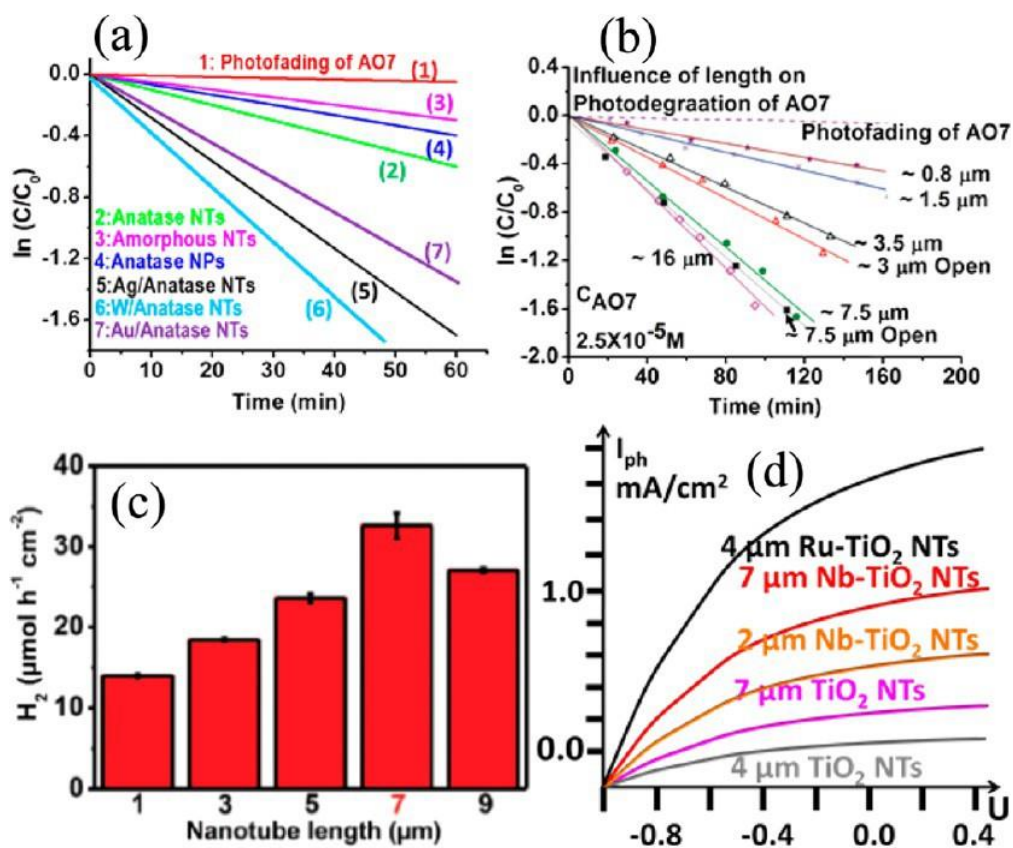

Figure 10. (a) Comparative compilation of photocatalytic decom- position rates of (AO7) azo-dye for different TiO$_2$ photocatalytic layers. (b) Influence of TiO$_2$ nanotube layer thickness on decay of AO7. (c) Photocatalytic H$_2$ evolution measured for TiO$_2$ NTs with different length after decoration with TiO$_2$ nanoparticles and Pt

cocatalyst. (d) Photoelectrochemical "water splitting" in KOH: compiled photocurrent data from different reports on TiO$_2$ nanotube layers.[232,233] (a) Reproduced with permission from ref 234. Copyright 2007 Elsevier B.V. Reproduced with permission from ref 235. Copyright 2012 Springer-Verlag. (b) Reproduced with permission 236. Copyright 2010 Wiley-VCH Verlag GmbH & Co. KGaA, Weinheim. (c) Reproduced with permission from ref 237. Copyright 2016 The Royal Society of Chemistry.

Other geometric factors that have been found to influence the intrinsic photocatalytic activity of TiO$_2$ NTs are the top-geometry of the tubes, their diameter, side wall corruga- tion,[243,244] and other factors that not only affect the electronic properties of the tubes but also influence the reflectivity of a nanotube layer surface.[7]

In extreme cases, morphologies can be tailored to show a photonic crystal type of properties, as shown by Song et al.[213] and later by Zhang et al.,[245,246] or tubes can be grown with conical profiles that allow for an optimized light absorption.[215]

*Polymorphs.* As-formed (amorphous) TiO$_2$ nanotubes show a significantly lower photocatalytic activity than tubes annealed to anatase or rutile.[118,216,247–249] Generally, for samples annealed in air,[235] the photocatalytic activity increases with increasing annealing temperatures (above 300 °C), first due to anatase formation at 300 °C and second due to a higher crystallinity. Above ≈500 °C, the rutile phase starts forming, and a highest photocatalytic activity is observed for tubes annealed at ≈650 °C (Figure 8c,d, i.e., when a mixed anatase/rutile structure is present). This holds not only for dye degradation and OCP H$_2$ evolution under solar light but also for UV laser illumination. Therefore, anatase/rutile junctions, because of band offsets, are a more plausible explanation for the benefit of mixed polymorphs than an enhanced light absorption in rutile.[36,250]

It should, however, be mentioned again that if tubes are annealed on their substrate in O$_2$-containing atmosphere, except for a conversion from an amorphous to anatase structure, the formation of a thermal rutile layer underneath the tubes (due to a direct oxidation of Ti metal) can be observed.[118,228,230,247,248] In order to eliminate this substrate effect, high-temperature annealing experiments can be carried out with lifted-off membranes (Figure 8).[144] The fabrication of defect-free crystalline TiO$_2$ nanotube (NT) membranes that maintain a full anatase phase composition has been reported up to an annealing temperature of 950 °C (Figure 8c,d).[144]

In view of the electronic properties, annealing of the amorphous tubes to a crystalline structure mainly changes the conductivity and lifetime of charge carriers. It is noteworthy that also so-called "water annealing" was reported to convert amorphous TiO$_2$ nanotubes to crystalline material,[251] and similarly, some other low-temperature approaches are provided in the literature.[238] However, in these approaches, conversion to anatase is only partial, and the efficiency in photocatalysis or solar cell applications remains normally far below thermal annealing.[238]

*Facets.* As outlined in section 2.4, the exposure of different crystal planes of anatase and rutile may influence the reactivity of TiO$_2$ crystallites for photocatalysis; this occurs not only by the intrinsically different energetic nature of the different planes but also by microjunction formation.[153,154] Efforts to exploit these features in TiO$_2$ nanotubes are scarce, but some recent work indicates that anatase TiO$_2$ nanotube arrays with predominantly exposed highly energetic {001} facets can be prepared by using NH$_4$F as a capping agent during anodization.[33,252] For such tube arrays, the photodegradation of rhodamine B, has been found to increase in accordance with an increase in the degree of exposed {001} facets.[252] The {001} facets of nanotube arrays have been considered to provide sites that are particularly efficient to produce active oxygen species such as OH$^•$, $^•$O$_2^-$, and H$_2$O$_2$ upon UV irradiation.[252]

*Doping.* Many approaches to dope TiO$_2$ nanotubes are essentially adopted from nanoparticle strategies. Powders and accordingly tubes can be doped by carbon,[180,253] nitrogen,[254] boron,[255] sulfur or fluorine by a thermal treatment (annealing) in a respective environment, i.e. for carbon doping annealing in CO or acetylene or ashing of organic compounds is used, for nitrogen heat treatments in NH$_3$ or for S in H$_2$S are used.[166,180,253,256,257]

Ion implantation is the most effective method to dope nitrogen into the TiO$_2$ lattice at lower to medium doping levels[254,258] (max of about 10$^{18}$ ions/cm$^2$), which leads to substitutional doping (with an according XPS N 1s peak at 396 eV). However, the technique suffers from limitations considering ion penetration depth - at maximum, on the order of micrometers - at MeV acceleration energies, and often an inhomogeneous dopant distribution is obtained.[259,260] Nevertheless, implantation profiles can be exploited to create well-defined buried junctions into TiO$_2$ nanotube walls (see section 2.5).[258]

Most unique to anodic TiO$_2$ nanotubes is that they can be doped by anodization of a homogeneous TiX alloy (i.e., titanium (Ti) alloy with the dopant (X)). For example, W-, Mo-, Nb-, Ru-, or Ta-doped TiO$_2$ nanotubes; noble-metal-containing tubes;[261−263] or even N-doped TiO$_2$ nanotubes (from Ti−N alloys) have been obtained by growth from the respective alloys.[179,264] Typically the substrate is prepared by arc-melting of pure Ti and the dopant metal,[265] or the alloy is produced on a substrate by cosputtering Ti and the dopant metal. Such doped tubes may show strongly enhanced photocatalytic properties; for example, for intrinsic W- and Mo-doped tubes, a strong increase of the photocatalytic activity under OCP was found.[236,263] This beneficial effect for W and Mo could not be explained by a better charge transport in the tubes but has been ascribed to modification of the band- or surface-state distribution of the doped nanotubes.[266−268] Nb and Ta doping are particularly effective to increase the conductivity of tubes, and this has been shown to improve photoelectrochemical photocatalytic reactions such as H$_2$ generation from methanol−water electrolytes.[233,269] Similarly, Ru can be integrated into tube walls. Ru in TiO$_2$ can either act as a dopant[232] (see also section 2.5) or be present as RuO$_2$ to act as a cocatalyst for O$_2$ evolution.[270] In photo- electrochemical experiments, indeed for intrinsically Ru-doped tubes, very high light to H$_2$ conversion efficiencies can be observed (Figure 10d).[232]

Additionally, doping of TiO$_2$ is reported to take place by ion incorporation from the anodization electrolyte (e.g., for phosphorus by anodization in a phosphate-containing electrolyte or N-compounds from N-containing electrolytes).[271] However, such attempts targeting nitrogen doping[272] mostly lead to XPS peaks at 400 eV (corresponding to adsorbed species, see e.g. ref54) and/or do not show convincingly electronic coupling of the doping species. For nanotubes prepared in organic electrolytes, carbon-contamination in the inner shell can take place due to the decomposition of the organic electrolyte under the applied voltage,[184,230] and commonly an enhanced visible absorption is observed. Additionally, a number of reports show doping of tubes with Cr,[273] C,[180] and V[274] with more or less beneficial effects to the photocatalytic properties.

Fe-doped TiO$_2$ nanotubes showed increased activity for the photodegradation of methyl orange.[275,276] Pt- and N-doped nanotubes[277] were reported to provide a higher activity for H$_2$ evolution than nanoparticles. Gadolinium and nitrogen-codoped TiO$_2$ nanotubes[278] have been shown to possess higher catalytic activity in the Rhodamine B degradation reaction (the presence of Gd$^{3+}$ has been reported to yield a higher crystallinity,[279] to sensitize the surface of the nanotube,[278,280] and to enhance the photocatalytic activity of TiO$_2$ in the visible light region). An increase in the photocatalytic activity was also observed for C,N,S-tridoped TiO$_2$ nanotubes.[281] Silica coated nanotubes[282] annealed at 650 °C showed higher photoactivity than nanoparticles. Titania nanotubes modified with 4 wt % WO [283,284] and annealed at 380 °C also enhanced the photocatalytic activity, compared with nondoped materials.

*Self-Doping and Gray/Black TiO$_2$ NTs.* As for nanopowders (section 2.5, various reduction treatments of TiO$_2$ such as heating in vacuum or reduction with hydrogen at elevated temperatures lead to self-doping of TiO by the formation of lattice defects.[202,285,286]

For TiO$_2$ nanotubes, such reduced tubes show an increased activity in dye degradation measurements.[202] The effect of Ti$^{3+}$ formation has been attributed to a higher conductivity (better charge separation) or the formation of surface states that facilitate charge transfer.[238]

Moreover, also electrochemically reduced TiO$_2$ nanotubes were reported to show a remarkably improved photoelectro-chemical water-splitting performance.[133,287] It should be noted, however, that under photoelectrochemical conditions, simply the improved conductivity of the tubes may be responsible for observed improved performance. This seems plausible in particular for various annealing treatments used to reduce TiO$_2$ nanowires and nanotubes, which reported a higher photoelectrochemical water splitting performance.[133,288] Open- circuit H$_2$ evolution measurements (as for powders) almost exclusively use decoration with Pt as a cocatalyst on the tubes.[23]

Only the most recent work by Liu et al.[25,197,289] shows a remarkable activation of reduced (black) TiO$_2$ nanotubes for noble-metal-free photocatalytic H$_2$ generation (Figure 11).

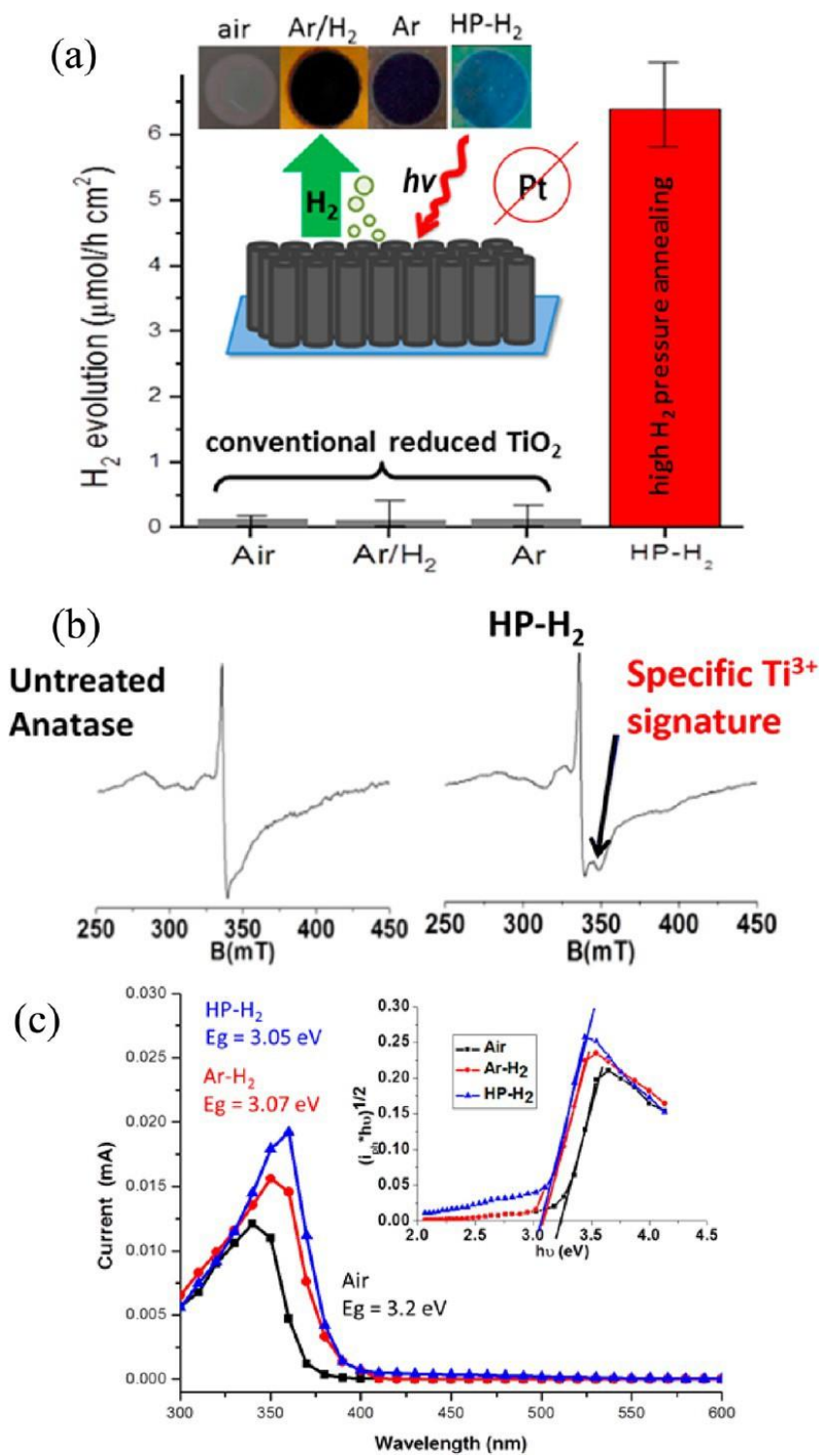

Figure 11. (a) Noble-metal-free photocatalytic $H_2$ production under open-circuit conditions in methanol/water (50/50 vol %) using $TiO_2$ nanotubes heat-treated in different atmospheres (air at 450 °C; Ar at 500 °C; Ar/$H_2$ at 500 °C; HP-$H_2$, heat treatment in $H_2$ at 20 bar at 500 °C. (Inset: optical images for the differently treated samples). (b) EPR spectra for anatase and hydrogenated anatase. (c) Photocurrent spectra of $TiO_2$ nanotube layers before and after treatment under different annealing conditions. (a) Reproduced with permission from ref 197. Copyright 2014 American Chemical Society. (b) Reproduced with permission from ref 25. Copyright

In this work, it was reported that with respect to TiO$_2$ NTs exposed to a high-pressure hydrogen treatment[197] or by an appropriate treatment at elevated temperature in H$_2$,[27] a modification of TiO$_2$ takes place that yields intrinsic "co-catalytic centers" for H$_2$ evolution. In spite of color changes observed for the modified tube layers (Figure 11a), this center in anatase is only active for illumination with $h\nu > 3.2$ eV. Extensive investigations using electron paramagnetic resonance (EPR), photoluminescence (PL), among others, ascribe the effect to Ti$^{3+}$ states located energetically close to the conduction band (Figure 11b).

In general, these tubes show a unique EPR feature (Figure 11b) that is characteristic for the titania intrinsically activated for H$_2$ evolution catalysis. While the tubes show strong visible light absorbance up to $\lambda > 800$ nm,[197] a significant photocurrent (and H$_2$ evolution) can generally only be measured above the anatase band gap (Figure 11c). Nevertheless, for reduced tubes, a slight shift to lower energies in the photocurrent onset can be observed (in line with the presence of active states close to the conduction band).[197]

The cocatalytic effect can be enhanced by using electrochemi- cally reduced tubes.[286] It was shown that by using anatase powder[25] (Figure 5) and later H-ion and N- implanted tubes[26,100] this cocatalytic effect can be created.

Usually surface Ti$^{3+}$ states are not considered to be stable and readily oxidize in air atmosphere. However, on one hand, it has been reported that Ti$^{3+}$ states in anatase have a tendency to be buried in a subsurface configuration,[53] possibly in combination with vacancy condensation which may stabilize these configurations.[27] On the other hand, Ti$^{3+}$ states can be stabilized by N states in lattice by charge-transfer resonance. Hoang et al.[44] reported on a synergistic effect using a hydrogenation and nitration cotreatment of a TiO$_2$ nanowire (NW) array that improved the Pt-catalyzed water photooxidation performance and the stability of Ti$^{3+}$ states. Later Zhou et al. showed that appropriately oxidized TiN as well as N-implanted TiO$_2$ nanotubes[100,147] also show a noble-metal-free activation for H$_2$ evolution.

**Conversion and Formation of Core/Shell Structures.** Core−shell structures of tubes can be formed by decoration or conversion of the tube walls over their entire length. TiO$_2$ nanotubes can be comparably easily converted to a perovskite oxide by hydrothermal treatments or by heat treatments to oxy- carbides or nitrides.[290−292] Particularly, conversion to other semiconductive materials such as SrTiO$_3$, in the context of photocatalysis and specifically for H$_2$ generation, is highly interesting.[291,293]

By a partial conversion of the TiO$_2$ nanotube wall, core−shell structures as illustrated in Figure 1g can be formed,[74,294] where a heterojunction between SrTiO$_3$ and TiO$_2$ is created (such as in Figure 12a,b). SrTiO$_3$ has a slightly higher conduction and valence band position than anatase. Thus, photogenerated charge carriers are separated in the field of the junction driving electrons to the TiO$_2$ core and h$^+$ to the SrTiO$_3$ shell (Figure 12c, in analogy to Figure 6a). For such structures, a significant enhancement of the photoelectrochemical properties has been found (Figure 12d).[74,295] More recently, there are reports that this core−shell structure is particularly efficient when doped. For these structures, photoelectrochemical water splitting perform- ance is reported to be much higher compared with samples without Nb doping (Figure 12e).[232] Other work shows Cr-doped $SrTiO_3$/$TiO_2$ NT core−shell structures to be active for visible light photocatalysis (for example for the degradation of RhB). Here Cr-doping is proposed to shift the valence band edge in $SrTiO_3$ to more positive values and thus activate visible light absorption in the $SrTiO_3$ shell.

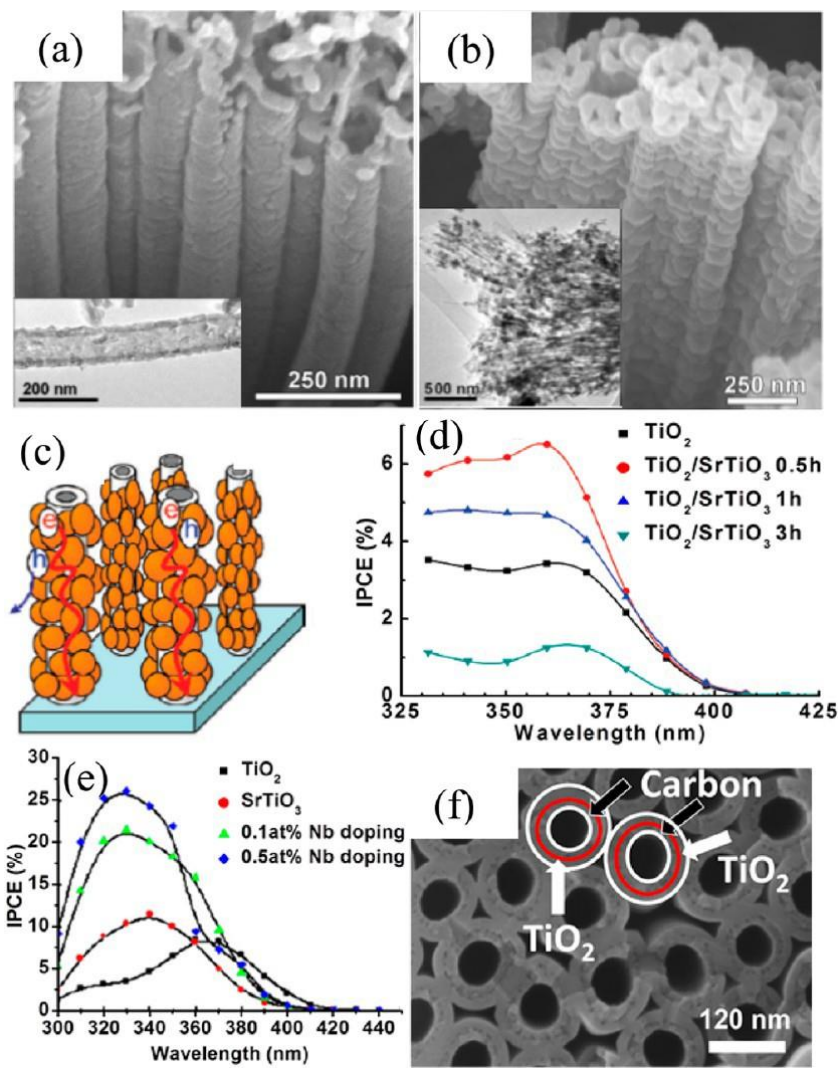

Figure 12. (a) SEM images of $TiO_2$ nanotube array after annealing at 450 °C and TEM image of a single nanotube (inset). (b) $TiO_2$/$SrTiO_3$ core−shell nanocomposite obtained after 40 h hydrothermal treatment and TEM image of the nanotubes after 1 h hydrothermal treatment (inset); (c) Depiction of $TiO_2$/$SrTiO_3$ heterostructure after hydro- thermal treatment. (d) IPCE spectra of $TiO_2$ nanotube and $TiO_2$/ $SrTiO_3$ nanocomposite electrodes and (e) spectra of Nb-doped $TiO_2$/$SrTiO_3$ heterostructure nanotubes. (f) SEM images of $TiO_2$ nanotube core structure with graphitized walls. (a−d) Reproduced with permission from ref

74. Copyright 2010 American Chemical Society. (e) Reproduced with permission from ref 295. Copyright 2012 Elsevier B.V. (f) Reproduced with permission from ref 297. Copyright 2015 The Royal Society of Chemistry.

An elegant and simple way to create core−shell tubes is to leave remnants of an organic electrolyte, in which the tubes were formed, and thermally graphitize the organics.[296] Graphitization can be achieved by annealing the tubes in $O_2$-free environments (for example, the carbon in the inner shell present in tubes formed in EG electrolytes, Figure 12f). Such core−shell structures have tube walls cladded with a thin layer of graphite.[290] The graphite layer on the one hand can specifically enhance the tube conductivity. On the other hand, it allows the modification of the tube walls with attachment principles based on established carbon surface chemistry.[223,297]

A particularly useful scaffold is provided by spaced nanotubes (Figure 13a,c).[295,298] In such templates, the interspace between the tubes can be adjusted, thus facilitating filling selectively using various deposition techniques. In Figure 13d, the tubes are decorated with a multilayer of $TiO_2$ nanoparticles (using a $TiCl_4$ treatment) in order to create a hierarchical $TiO_2$ nanoparticle on nanotube structure.[237] Such tube morphologies combine a high surface area with faster electron transport properties and were shown to provide an enhanced photocatalytic performance (Figure 13e).

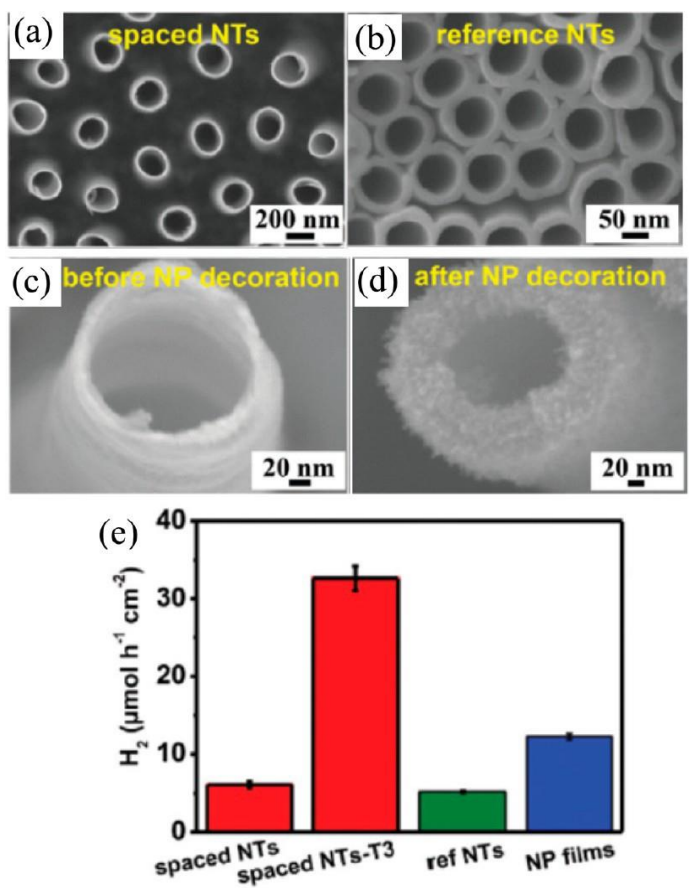

Figure 13. SEM images of (a) spaced $TiO_2$ NTs and (b) reference $TiO_2$ NTs; (c) spaced $TiO_2$ NTs before and (d) after $TiO_2$ nanoparticle decoration. (e) Photocatalytic $H_2$ evolution measured for spaced $TiO_2$ NTs

with/without TiO$_2$ nanoparticle decoration and reference samples on FTO (films); all samples were decorated with 1 nm thick Pt cocatalyst. Reproduced with permission from ref 237. Copyright 2016 The Royal Society of Chemistry.

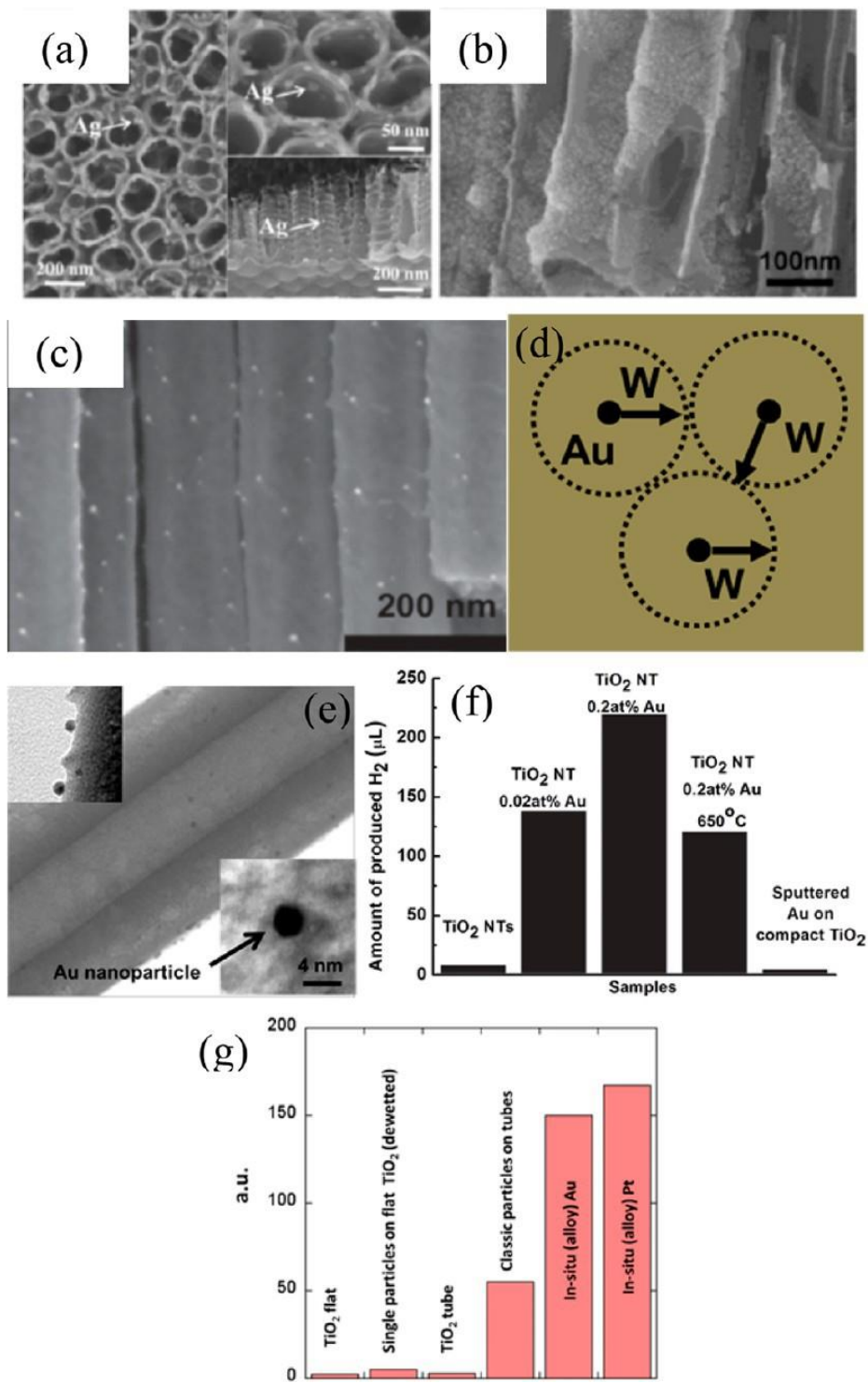

Figure 14. SEM images of (a) Ag nanoparticles and (b) Pt nanoparticles photodecorated on TiO$_2$ nanotube layers. (c) SEM cross-sectional image of self-decorated nanotubes formed by anodizing a Ti−Au alloy (0.2 at-

% Au): (d) Schematic drawing of the space charge layer (W) around an Au particle on a $TiO_2$ NT surface. This suggests an ideal noble metal decoration on $TiO_2$ to be achievable for 2W ≈ 30−60 nm for typical anodic $TiO_2$ parameters.[310] (e) TEM image of sample shown in (c). (f) Photocatalytic $H_2$ evolution activities for different Ti−Au alloys and reference samples. (g) Comparison of open-circuit $H_2$ production for intrinsically (from alloys) Pt or Au decorated tubes under AM1.5 illumination (100 mW/cm$^2$) in comparison with reference data. (a) Reproduced with permission from ref 304. Copyright 2007 Elsevier B.V. (b) Reproduced with permission from ref 308. Copyright 2011 Elsevier B.V. (c−f) Reproduced with permission from ref 210. Copyright 2013 Wiley-VCH Verlag GmbH & Co. KGaA, Weinheim.

*Decoration with Nanoparticles.* Decoration of $TiO_2$ nano-tubes with nanoparticles (metals, semiconductors, polymers) is frequently used to achieve property improvements[46] due to (i) heterojunction formation and sensitization as described in section 2.4, (ii) suitable surface mediators attached for an enhanced charge transfer with the surrounding (cocatalytic effects), and (iii) surface plasmon effects that lead to field enhancement in the vicinity of metal particles and thus allow for example for a more efficient charge harvesting.

Methods for decorating $TiO_2$ nanotubes involve dip-coating,[299,300] spin-coating,[301] physical vapor deposition (PVD),[302] and electrodeposition reactions.[303] Partial decoration of $TiO_2$ nanotubes by noble metal nanoparticles (such as, Au, Ag, Pt, Pd, AuPd) is very frequently carried out in order to achieve cocatalytic effects.[234,304−307] Ag or Pt nanoparticles can be deposited on the tube wall by exploiting the photocatalytic effect to reduce Ag or Pt compounds on a $TiO_2$ surface by UV illumination (Figure 14a,b).[304,308] Other metal nanoparticles are preferably deposited by UHV evaporation or chemical reduction techniques.[234,304,309] Noble-metal-decorated Ag/$TiO_2$, Au/$TiO_2$, or Pt/$TiO_2$ nanotubes show a significantly higher photocatalytic activity compared with plain nanotubes not only for OCP $H_2$ generation (Figure 14f,g) but also for pollution degradation (usually ascribed to junction formation and mediation of electron transfer to $O_2$),[234] as illustrated in Figure 10a.

Oxide nanoparticle decoration of $TiO_2$ nanotubes by, for example, $WO_3$,[284] or tungstates,[112] $Cu_2O$,[311−313] $Fe_2O_3$,[41,275] $CuInS_2$,[314] $ZnO$,[315,316] $Bi_2O_3$,[317] $ZnTe$,[318] or $TiO_2$[299,314] has been obtained by slow hydrolysis of precursors, electrochemically, or by CVD/PVD deposition, and for these junctions higher photocatalytic rates have been reported. One of the most followed up schemes to establish useful p−n heterojunctions for solid-state solar energy devices is deposition of $Cu_2O$.[319] Due to the high conduction band level of $Cu_2O$, this p-type material is expected to be able (and has accordingly been reported) to promote the photocatalytic conversion of $CO_2$.[319] Other common narrow band gap semiconductors that are decorated on $TiO_2$ nanotubes are CdS, CdSe, PbS, and their quantum dots.[98,130,131,150,320] These species are commonly deposited on the nanotube wall electrochemically or by sequential chemical bath deposition methods. CdS and CdSe have band gap values of 2−2.4 eV[321] (i.e., they absorb visible light) and have a conduction band position that allows the injection of excited electrons into the $TiO_2$ conduction band (i.e., act as efficient sensitizer).

Similar to the hydrolysis of $TiCl_4$ shown in Figure 13d,[299] $WO_3$ nanoparticles can be deposited, leading to an additional junction formation between $TiO_2$ and the misaligned bands of $WO_3$ that can be beneficially exploited.[284]

In more recent work, tubes have been decorated using C60,[324] graphene,[325] Ag/AgCl or AgBr,[36,326] or BiOI[327] that beneficially affect the photocatalytic activity. In these works, Ag halogenides can essentially be photoreduced directly but are then reoxidized (refreshed) by the environment. C60 or graphene represents not only classic electron transfer mediators but also may effectively transport charge away from the surface. Decoration with nickel oxide nanoparticles has recently been shown to exhibit significant photoelectrochemical activity under visible light (possibly by charge injection from NiO states to the conduction band of $TiO_2$).[300] A simple but very successful approach for particle decoration is magnetic filling of the $TiO_2$ nanotubes with a suspension of $Fe_3O_4$ nanoparticles (Figure 15a).[19] This provides tubes with magnetic guidance features (see also Figure 15b,c).[328]

A most unique decoration approach for anodic nanotubes with noble metal particles is SOA of low concentration Ti-X (X = Au, Pt) alloys[210,329] that can provide very uniform and defined particle diameters, as well as a controllable distribution of particles over $TiO_2$ walls.

Figure 14c,e show the in situ (during anodization) formation of Au nanoclusters on $TiO_2$ nanotubes grown from Au- containing titanium alloys.[210] These clusters are regularly spread and have a typical particle size of ≈5−7 nm. The decoration density (i.e., the cluster spacing) can be controlled by the amount of Au in the alloy and the anodization time. This permits to tune the cluster interspacing for an optimum activity for photo-catalytic open-circuit $H_2$ production. The results are in line with the concept shown in Figure 14d (i.e., an ideally decorated wall that has a particle spacing of twice the space charge layer).[210] It is also remarkable that such ideally noble-metal-decorated tubes show a higher $H_2$ production than tubes that are decorated with traditional methods (Figure 14f,g).[210] The "alloying" principle for intrinsic noble metal decoration has been demonstrated also for Pt (but not for Au) and is likely transferrable to other noble metals and alloys.

$TiO_2$ surfaces can be modified with a wide range of organic monolayers. Particularly, carboxylates, silanes phosphonates, or hydroxamic acid can be well-anchored on $TiO_2$ surfaces.[330–332] While carboxylates are usually the preferred anchor group in dye- sensitized solar cells, because of their charge-transfer mediating nature, usually silanes or particularly phosphonates adhere considerably better to titania. If phosphonate- or silane-decorated layers are exposed to UV light, scission of the organic chain usually occurs after their linker group, leaving the inorganics attached to the $TiO_2$.[333] In the context of photo- catalysis, this chain scission reactions can be used to photo- catalytically release payloads as illustrated in Figure 15.[19]

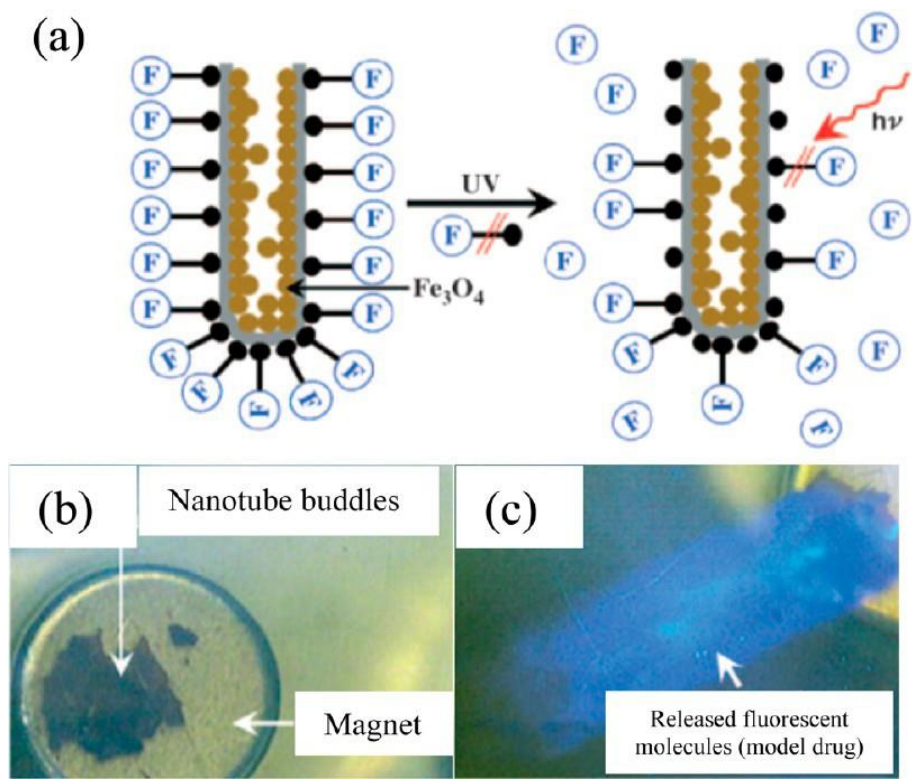

Figure 15. (a) Diagram showing the release principle of active molecules (monolayer with model drug) from functionalized magnetic TiO$_2$ nanotubes upon irradiation with UV light. A fluorescent dye (active molecule) was attached to the TiO$_2$ nanotubes with a siloxane linker. (b,c) Sequence of images showing the release of the fluorescent dye into the surrounding after switching on UV light. The movement of the tube layers in medium was guided by a permanent magnet underneath the Petri dish. Reproduced with permission from ref 19. Copyright 2009 Wiley-VCH Verlag GmbH & Co. KGaA, Weinheim.

3.3 **Site-Selective Junctions.** Nanotube layers provide unique possibilities to deposit or embed specific functionality with a high degree of site selectivity either at the tube mouth (crown position), at the tube bottom, or at specific locations in the tube wall.[129,322,323] This includes the placement of cocatalysts, cocatalytic sites, or heterojunctions. In the following, we give some examples where the localized placement of a cocatalytic site leads to efficient light-harvesting/reactive-site combinations, where the charge carriers from the light harvesting site are directed to the reactive site.

*Co-Catalyst at Tube Mouth.* Particularly, using metal sputtering at a shallow angle allows a large range of metals to be deposited only at the mouth of the tubes.[107] Over the past years, a range of metals have been explored as photocatalytic cocatalysts decorated in this crown position (e.g, see Figure 16a,c).[129,322,323] In view of minimizing the amount of cocatalyst used for photocatalytic efficiency, crown position decoration has been found to be superior to a deposition deeper into the tubes (the latter can be achieved by changing the sputtering angle). A key consideration is that noble metal decoration at the top induces a gradient in the semiconductor

Fermi level ($E_F$) (Figure 16f). Here $E_F$ is pinned by Pt at a lower level than in the rest of the tube which directs excited electrons produced in the underlying TiO$_2$ tube toward the Pt decorated end of the tube.[323]

As mentioned, the absorption depth (into TiO$_2$ nanotubes) of light with an energy in the band gap region of anatase is a few micrometers,[83,334] and anatase tubes provide an electron diffusion length in the range of several tens of micrometers.[216] As a result, electrons can be harvested from several-micron-long tubes and thus can significantly contribute to an overall H$_2$ evolution production.[100,322,335−337]

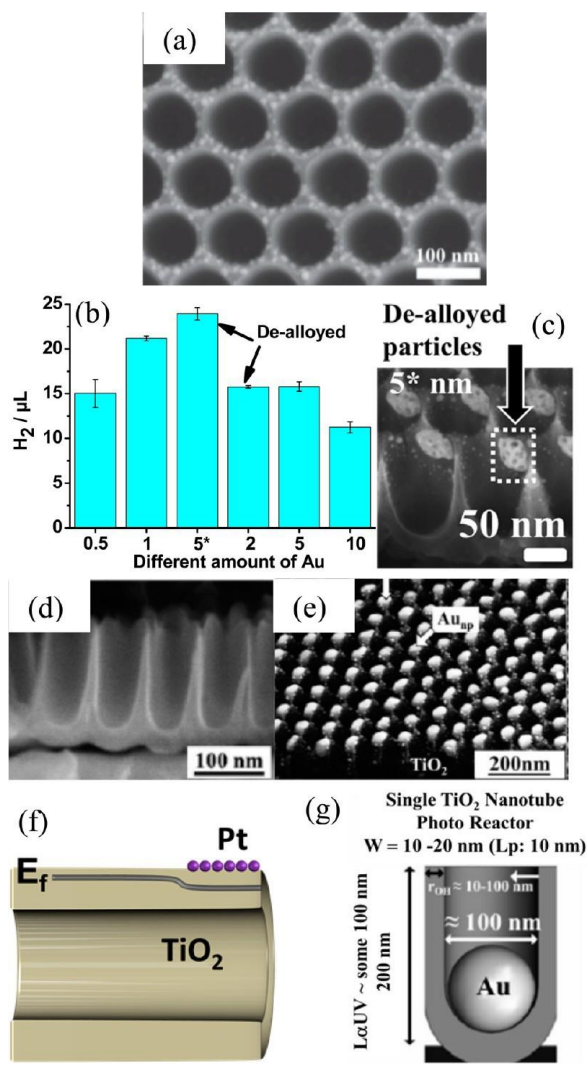

Figure 16. (a) Selective tube top decoration (crown position) of Au particles by small angle sputtering. (b) Comparison of photocatalytic H$_2$ evolution measured for Ag−Au alloys with constant Ag/Au ratio of 2:1; all samples were dewetted or additionally dealloyed in HNO$_3$ for 2 h at −15 °C. Please note the dealloyed morphology in (c). (d) SEM images of a highly ordered TiO$_2$ cavity array after dewetting a 50 nm thick Au layer (e) filling with exactly one Au catalyst particle per cavity is achieved. (f) Scheme of $E_F$ after Pt metal decoration of TiO$_2$ nanotube top. (g) Critical dimensions in a one particle photocatalytic reactor in view of UV absorption ($L_\alpha$ vs tube length), hole diffusion length ($L_p$ vs wall thickness), and range of photocatalytically generated radicals (rOH* vs tube inner diameter). (a−c) Reproduced with permission from ref 322. Copyright

2015 Wiley-VCH Verlag GmbH & Co. KGaA, Weinheim. (d,e,g) Reproduced with permission from ref 129. Copyright 2013 Wiley-VCH Verlag GmbH & Co. KGaA, Weinheim.

Nevertheless, if top illumination (onto the tube layers) is used, then shading effects of the top-deposited noble metal have to be considered.[107] That is, with an increasing amount of cocatalyst, activity loss due to an increasingly less transparent tube top can be observed (Figure 17a,b). In-depth investigations using various configurations of cocatalyst (top-bottom) as well as front and back-side illumination[338] indicate the most efficient use of noble metal loading to be decoration of an optimal amount of cocatalyst at the top combined with illumination from the top. This may be ascribed to a most efficient electron−hole separation within the field of the space charge layer of the metal/$TiO_2$ junction.[84]

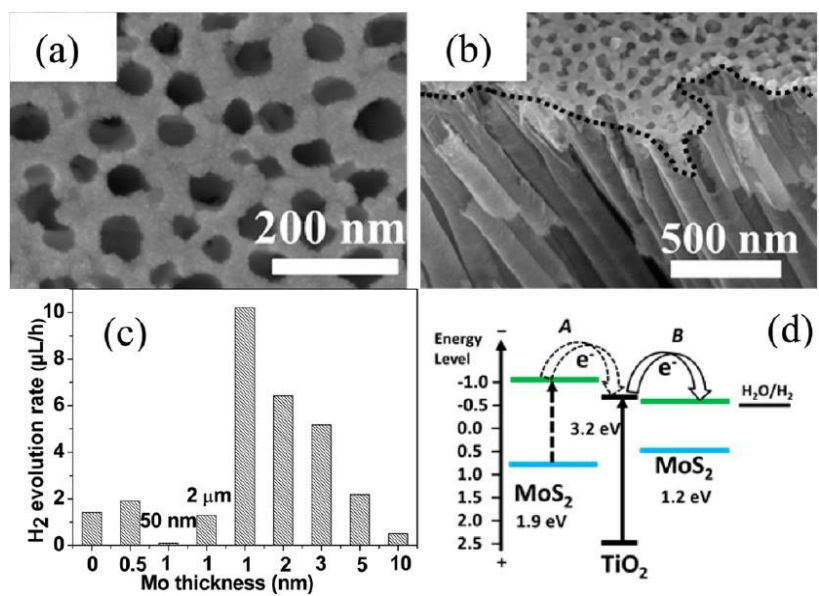

Figure 17. SEM images of (a) the top and (b) cross-sectional for $TiO_2$ NT layer with 10 nm $MoS_2$ top decoration. (c) $H_2$ evolution activity for different $MoS_2$ thickness and different $TiO_2$ nanotube length with 1 nm $MoS_2$ top layer under AM1.5 conditions. (d) Band gap alignments of $TiO_2$ and $MoS_2$ for different thicknesses of the $MoS_2$ layer. Reproduced with permission from ref 107. Copyright 2016 Elsevier B.V.

However, the significance of a harvesting length is evident from the observation that anatase tubes with a high electron diffusion length of >7 μm length (with only a few nanometers of top noble metal decoration) lead to the highest open-circuit photocatalytic activities.[338] As a low-cost alternative to noble metals, other cocatalysts such as $MoS_2$ have been explored on $TiO_2$ nanotubes.[45,339,340] This catalyst can be created selectively at the tube tops, using a Mo-sputter coating and conversion to $MoS_2$.[107] Figure 17a,b show a nominally 10 nm thick Mo layer, deposited by shallow angle sputtering, which has then been converted to $MoS_2$ by thermal sulfurization. Thin $MoS_2$ layers in combination with anatase nanotubes show an interesting dual behavior as sensitizer and electron transfer mediator. This is due to a comparably easily achievable quantum confinement effect that leads to band gap widening for $MoS_2$ layers at thicknesses of <2 nm.[107] Accordingly, $MoS_2$ acts as a sensitizer when the band gap is widened by quantum size features (see scheme in Figure 17d left), or as an electron-transfer mediator (Figure 17d right) when the layers are thick enough to show a bulk $E_g$ value of 1.2 eV. Layers in the range of 1–3 nm thickness were found to act both as sensitizer and mediator (corresponding to thickness variations of the layer on the NT tops, Figure 17c). Most efficient is the deposition a 1–3 nm-thick layer on a $TiO_2$ tube layer of ≈6 μm thickness, as again (as in the case of Pt described above) an optimized light absorber (electron harvester)/cocatalyst geometry is provided.[107]

*Self-Ordered Dewetting.* Another possibility to arrange metal particles on tubes is temperature-induced dewetting of metal films.[129,322,323] It can be used to change nanoparticle arrange- ments that are deposited by sputtering on the tube top; in this case, typically very thin layers of 0.5 to 5 nm are used (Figure 16a,c).[129,322] Moreover, it can be used to produce unique noble metal configurations when using thicker layers (several 10 nm) on ordered substrates.

To exploit metal dewetting for maximized self-ordering, the deposited-layer-thickness needs to be in the range of the self- ordering length scale of the substrate. In Figure 16e, this has been used to form arrays of 50 nm-sized Au NPs arrays. One single cocatalytic NP is present per each photocatalytic $TiO_2$ nanotube in a highly ordered TiNT substrate (Figure 16d). The fabrication process is highly reliable, and the arrays are filled with virtually 100% success rate over large surface areas (several $cm^2$).[129]

This approach can provide Au NPs on the $TiO_2$ surface with a tunable decoration density (typically much higher than obtained on smooth $TiO_2$). Such Au/$TiO_2$ structures are promising not only as an efficient photocatalyst but also potentially as functional electrodes, high-density memory device, or as a plasmonic platform.[341]

This approach can also further be exploited using alloys that first are dewetted and then are dealloyed.[322] In the example of Figure 16c, first an Au/Ag alloy was deposited that then was dealloyed by selective dissolution of Ag using $HNO_3$. For such dealloyed and crown positioned Au-networks, the highest photocatalytic activities for $H_2$ evolution have been measured if deposited on >7 μm anatase tubes.[322]

*Buried Junctions by Ion Implantation.* A very versatile tool to modify tube-walls of their tops is ion implantation with accelerated ions in the KeV–MeV range.[259,260] It allows us to embed chemical (doping) or defect (vacancy/interstitial) profiles into the tube walls (such as in Figure 18a), which are mainly adjustable in dose (ion flux) and depth (ion energy).[26]

Recent work using such an approach showed that hydrogen implantation can be used to create noble-metal-free $H_2$ evolution activity in $TiO_2$ nanotubes. In the context of earlier work, on high-pressure hydrogenated nanotubes, this means that H ion implantation represents an alternative method to create cocatalytic activity for $H_2$ evolution by controlled H-introduction into tubes while at the same time creating defects (see section 2.5).[26,100]

In particular, low-dose ion implantation (Figure 18a) was found to induce an ion and damage profile in $TiO_2$ nanotubes that leads to this "co-catalytic" activity.

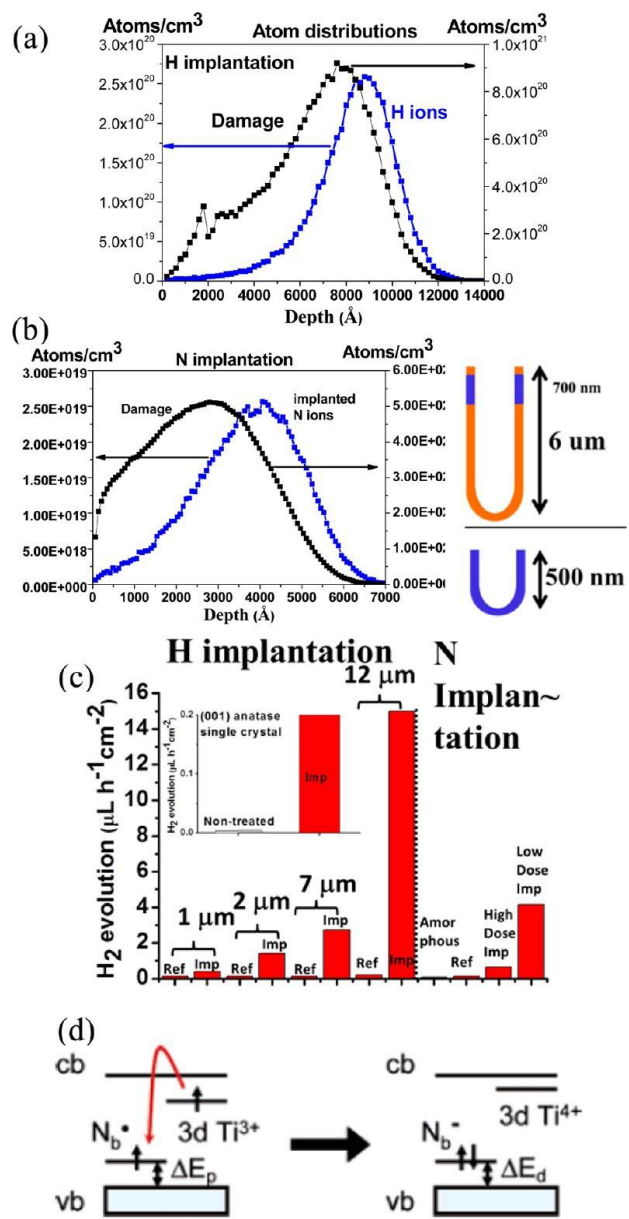

Figure 18. Implanted active zones at the tube tops created by ion implantation. (a) SRIM simulation of depth-distribution of H at 30 keV and (b) N ions implanted at 60 keV and accompanied damage (vacancies) created in a $TiO_2$ nanotube target. Corresponding illustrations in panel b show that for short tube length, the entire tube can be implanted (at the used energies), whereas for longer tubes, only the top ∼1000 nm is affected. (c) Photocatalytic $H_2$ production under open-circuit conditions in methanol/water (50/50 vol %) of different

$TiO_2$ nanotube layers before and after H-implantation (measured under AM 1.5, 100 mW/cm$^2$) (inset: photocatalytic $H_2$ production of (001) single crystal anatase before and after H-implantation), and from N-implanted $TiO_2$ NTs for two doses of $8 \times 10^{14}$ ions/cm$^2$ and $1 \times 10^{16}$ ions/cm$^2$, nonimplanted reference (annealed) and amorphous (non-annealed) $TiO_2$ nanotube layers. (d) Model of stabilization of $Ti^{3+}$ states with nitrogen states N˙ (N˙ or N˙) by charge-transfer resonance.[185] (a−c) Reproduced with permission from ref 100. Copyright 2016 Wiley-VCH Verlag GmbH & Co. KGaA, Weinheim. Reproduced with permission from ref 26. Copyright 2015 American Chemical Society. (d) Reproduced with permission from ref 185. Copyright 2006 American Chemical Society.

The proton-implanted region shows the same type of specific defect characteristics and modifications of the electronic properties as described for high-pressure hydrogenation.[26] The activation for photocatalytic $H_2$ generation also in this case has been ascribed to the formation of $Ti^{3+}$ states (in a surface near region) that are energetically close to the conduction band of anatase. As described above for top decorated tubes, also in these top implanted tubes a strong synergetic effect between implanted region (catalyst) and implant-free tube segment (absorber) was observed with a maximum efficiency for >12 μm tubes (the optimized absorber/cocatalyst geometry is shifted to higher tube length than in the top decoration case likely due to the significantly deeper penetration of the top catalyst layer (Figure 18c).[26]

Additional support for this finding is provided by experiments using low-dose nitrogen-ion implantation.[100] Figure 18b shows the ion distribution and corresponding damage profile for N ion implantation at 60 keV at a dose of $8 \times 10^{14}$ ions/cm$^2$. These tubes were also shown to provide noble-metal-free photocatalytic $H_2$ evolution (Figure 18c) and to provide a similar defect signature as optimally hydrogenated, or hydrogen implanted tubes. In contrast to plain hydrogen implantation, nitrogen not only leads to an active zone only at the top part of the tubes but also acts as a doping species and thus induces a beneficial Fermi-level gradient (junction) at the tube top (Figure 18b).[100] An additional consideration is that N-species are reported to beneficially interact with $Ti^{3+}$ states via electron transfer resonance,[185] which contributes to the stabilization of $Ti^{3+}$ species as illustrated in Figure 18d.

The coupling of this top layer and the underlying non-implanted part of the nanotubes strongly contributes to an efficient carrier separation and thus to a significantly enhanced $H_2$ generation (Figure 18c). Short tubes (1 μm) that were uniformly implanted showed only a minute amount of $H_2$ generation, while the implanted zone at the top of 7 μm long tubes yielded a maximum efficiency.[100]

*Buried Junctions by Anodizing Metal Multilayers.* A most efficient way to create embedded tube wall modifications is to exploit perpendicular growth of anodic tubes into a metallic substrate.[342] Figure 19a shows a highly defined oxide-stack-nanotube array grown by simple but optimized SOA of a commercially available metal multilayer of Ti and Ta.[342] In the example, the sputtered Ti/Ta multilayer contains alternating Ti/ Ta metallic layers (the metallic layer thicknesses are 7.5 nm for Ti and 2 nm for Ta). After SOA of the multilayer, nanotube layers consist of $TiO_2/Ta_2O_5$ heterojunctions in the tube walls. The oxide segments in the wall have expanded to 9 nm-thickness for $Ta_2O_5$ and 11 nm-thickness for $TiO_2$ (Figure 19b,c). Such nanotubular structures with modulated tube walls can strongly alter the optical, electrical, electronic, and surface chemistry properties of nanometer-sized materials[342] The material in the example was shown to provide enhanced photocatalytic properties in the deep UV.[342] The strategy, however, is not limited to this example, but particularly, it can be used to fabricate a wide range of transition metal oxide nanotube arrays with superlattice or heterojunction features. Metal deposition with PVD techniques currently can reach a precision in the nm range[206,302,343] Therefore, these desired metallic substrates allow for designing an extremely large variety and high control of wall- engineered tubes.

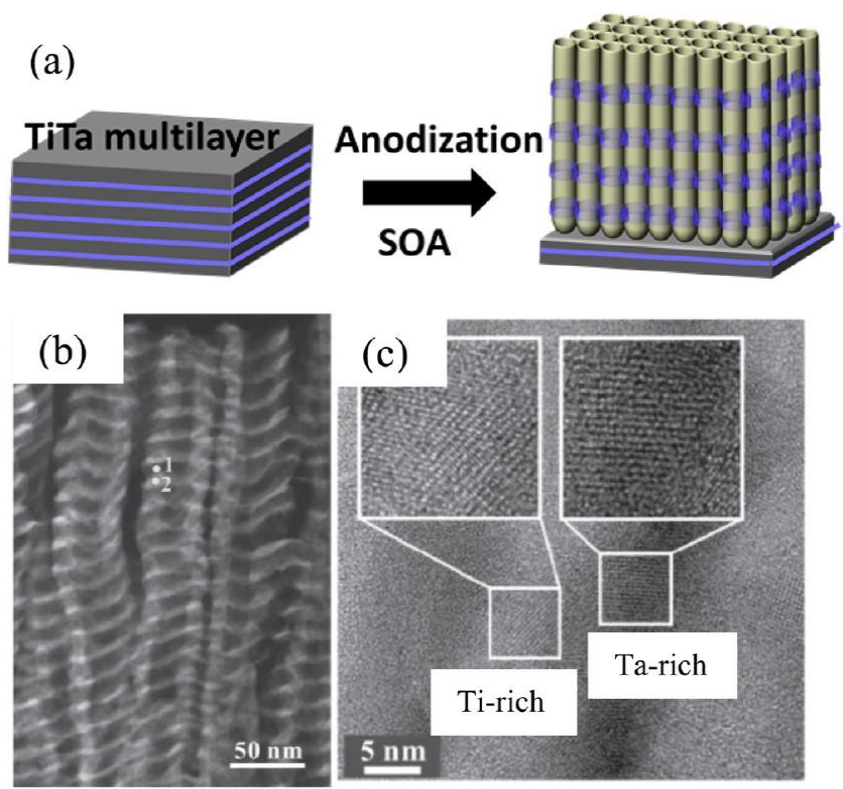

Figure 19. (a) Schematic drawing of anodization of a metal multilayer (Ti−Ta) to create stacked oxide junction tubes. (b) HAADF-STEM image of the formed $TiO_2/Ta_2O_5$ "superlattice" nanotube array. (c) HRTEM images for sample annealed at 630 °C showing fringes for $TiO_2$ anatase and $Ta_2O_5$. Reproduced with permission from ref 342. Copyright 2010 Wiley-VCH Verlag GmbH & Co. KGaA, Weinheim.

3.4    **Selectivity and Selective Capturing of Photo- catalytic Reactants.** A fundamental obstacle for an even wider use of photocatalytic reactions is their inherent lack of selectivity.[5,18] This is not only due to the nonspecificity of generated reaction species (e.g., OH• radicals) but also due to a lack of control of the interaction time in nanoparticle suspensions.[81] Regarding the latter aspect, the defined geometries of nanotubes present promising features.[7] Particularly, the use of flow-through membranes that combine size-selective features with a defined reaction time may pave the way to high- selectivity photoreactors.[144] Currently such membranes have been shown to possess self-cleaning features[223] (e.g., allow for example the photocatalytic reopening of clogged tube-based protein-filters or create refreshenable harvesting systems).[223]

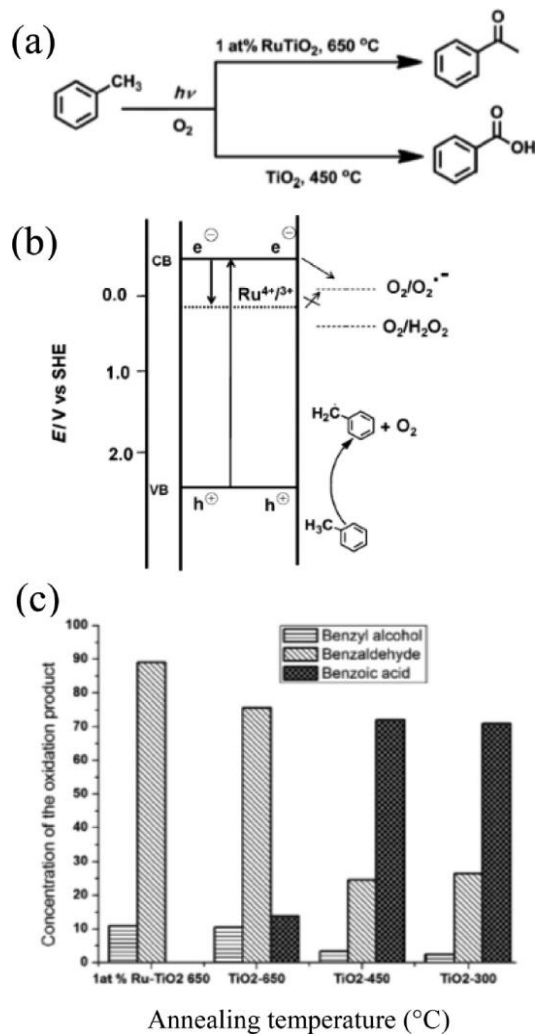

Figure 20. Tuning reaction selectivity: (a) illustration of different photocatalytic pathways on neat $TiO_2$ nanotube samples, and mixed Ru- $TiO_2$ tubes. The two different catalyst layers lead to selective benzaldehyde or benzoic acid formation, respectively. (b) Illustration of the band position during toluene oxidation on $TiO_2$ and ruthenium- doped $TiO_2$ nanotubes under UV irradiation. CB = conduction band, VB = valence band. The presence of the Ru-state prevents peroxide radical formation and thus changes selectivity. (c) Distribution of the products from toluene oxidation after 4 h UV exposure of different $TiO_2$ nanotubes or Ru-$TiO_2$ nanotubes. Reproduced with permission from ref 18. Copyright 2014 Wiley-VCH Verlag GmbH & Co. KGaA, Weinheim.

Possibly an even higher potential lies in the use of photogenerated electron−hole pairs for redox- or radical-based organic synthetic reactions. Up to now, only a small number of successful attempts have been

reported.[344−346] To a large extent, this is ascribed to the fact that a multitude of reaction pathways become accessible when a photoinduced electron or hole transfer from $TiO_2$ to an organic species occurs.[347,348] For example, a wide range of radical species can be initiated at the $TiO_2$ valence or conduction band, which generally leads to a high degree of nonselectivity and thus a wide product distribution. Conventional experiments use commercial $TiO_2$ nanoparticles and the main approaches to enhance the reaction selectivity is based on optimizing solution solvents to steer lifetime and speciation of radicals.[347,348] A conceptually entirely different approach is to alter the photocatalysts' electronic properties and its nanoscopic geometry (thus steering carrier energetics and lifetimes).[347,348]

Figure 20 shows Ru-doped $TiO_2$ nanotubes that were used to achieve a drastic change in the selectivity of a photocatalytic reaction.[18] For the photocatalytic oxidation of toluene, depend- ing on the electronic properties of $TiO_2$ (anatase, rutile, or Ru- doped tubes), a strong change in the main reaction product (namely, benzoic acid versus benzaldehyde) can be achieved, and certain undesired reaction pathways can be completely shut down.[18] This is mainly based on changed electronic properties (as illustrated in Figure 20a,b) for the Ru-modified tubes. The introduced Ru level prevents the formation of intermediate superoxide radicals and thus the reaction of toluene to benzoic acid (Figure 20c).[18] This example may illustrate how further tailoring the electronic levels may be beneficially used but the combination with the exploitation of geometric features of $TiO_2$ nanotubes (such as site specific junctions) and reactor design provides a large potential for further improvements.[18]

Another particularly challenging problem in selective photo- catalysis is the application of the method to pollutants that are present in the environment only at low concentrations. In this case, selective accumulation of the pollutant in/at the photo- catalyst is desired, in order to then apply an efficient (selective) destruction step.[357]

Various works have aimed to improve the photoefficiency by using pollutant harvesters such as silica, alumina, zeolites, and activated carbon in combination with titania.[282,349−351] These adsorbents are usually attached to titania to accumulate the target species within the photocatalytic reaction range. Zeolites are a most attractive candidate due to the uniform pore and channel size (3−8 Å), their high adsorption capacity, and their hydrophobic and hydrophilic properties.[357] These features allow for selective exclusion of undesired molecules or ions. Furthermore, zeolites provide a high thermal stability and more importantly a photostable inorganic framework.[282,351−357]

Figure 21a shows top of $TiO_2$ nanotubes after filling with a zeolite (ZSM5). In the example of Figure 21b, upon illumination, very efficient destruction of a small concentration of acetophenone can be achieved either in ethanol or water-based solutions.[357] Here acetophone is present only in parts-per- million concentrations. The used zeolite provides cavities that enable capturing the acetophone selectively versus the back- ground smaller solution molecules (MeOH, EtOH). The modified tubes show an overall 10 times faster destruction rate of the target molecule compared with unmodified tubes (Figure 21b).[357]

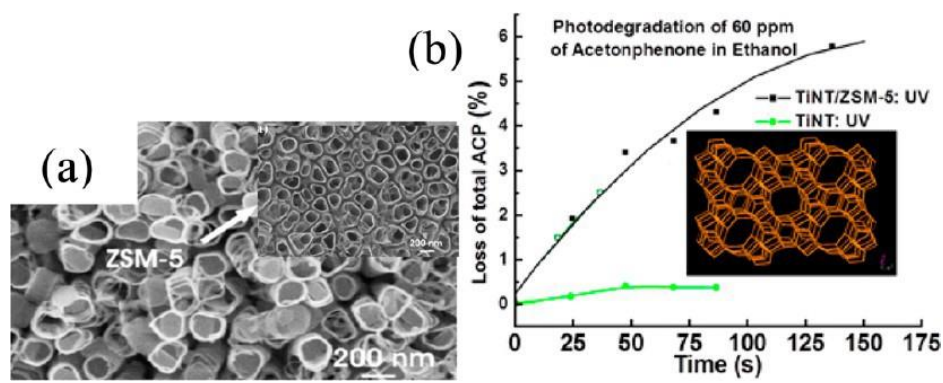

Figure 21. Capturing target molecules: (a) SEM images of top of $TiO_2$ nanotubular structure after loading with ZSM-5 nanocrystals (inset: prior to ZSM-5 loading). (b) Comparison of photodegradation of 60 ppm of acetophenone (aromatic organic pollutant) in ethanol for $TiO_2$ nanotubes with and without loaded with ZSM-5 zeolite: TiNT/ZSM-5; $TiO_2$ nanotube: TiNT) (inset: model of ZSM-5 molecular structure). Reproduced with permission from ref 357. Copyright 2009 IOP Publishing Ltd.

This example illustrates an effective harvesting system. Such systems are promising not only for selective pollutant degradation on demand (e.g., considering harvesting during the night, destruction during the day) but also for an improved selectivity in photocatalytic synthesis.[357]

## 4. SUMMARY AND OUTLOOK

Photocatalytic reactions on $TiO_2$ have been studied for almost 50 years. Over this time, significant progress has been made toward the understanding of key principles and critical factors controlling photocatalytic effects and the overall activity of titania-based photocatalysts. Photocatalysis is generally inves- tigated in particle-suspensions or particle-layers assembled to photoelectrodes (photoelectrochemistry). A large body of fundamental work has been carried out on single crystals (rutile) and aided to extract key mechanistic aspects of the involved photoinduced reaction steps. At the same time, an enormous amount of investigations has been performed on $TiO_2$ nanopowders and their modifications in view of an improved efficiency toward applications such as pollution degradation, hydrogen generation, adjusting surface hydroxylation or wetting properties, and accordingly, numerous excellent literature reviews are available.

In the past 10 years, increasingly, research interest has shifted from powders to novel type of nanostructures, namely, 1D nanotubes, that provide a highly defined catalyst geometry and allow for an unprecedented control over light harvesting and charge-carrier-separation management.

The present Perspective first gives a brief overview of some fundamental features of photocatalysis and mechanistic motives to improve the performance of photocatalysts with the intent to focus on points relevant to $TiO_2$ nanotube structures. While we aim at giving here a relatively comprehensive overview, we highlight (in the sense of a feature article) aspects that seem to us to be most interesting, surprising, or unexploited ground with considerable open potential. We emphasize the important role of semiconductor electrochemistry and junction formation to describe and engineer open-circuit potential (OCP) and photoelectrochemical photocatalysis. We then focus on the most investigated $TiO_2$ morphology over the past 10 years, that is, anodic $TiO_2$ nanotube arrays. We give the key reasons (and expectations) for the high interest in these structures and highlight their unique features in photocatalysis. We describe how geometric factors, crystalline structure, and composition of the tubes can be controlled and affect reactivity and give an outline of concepts that target a further improvement of the nanotube layers. While many principles to modify and steer activity and selectivity were adopted from findings on powders (doping, junction formation, cocatalysts), other features are exclusive to these anodic tubes and allow the design of unique geometries of photocatalytically active sites.

We emphasize the role of site-specific modifications and junctions embedded in the tube wall that can be assembled using metals, other suitable semiconductors, intrinsic doping, or faceting. These junctions may play a key role for further improvement as site-selective modifications that can be used not only toward an enhanced photocatalytic efficiency but also to achieve an improved selectivity.

Particularly in flow-through reactors (both-side-open membranes), this leaves a high potential for designing, with nanometer precision, reaction profiles by constructing sequences of catalytic centers, which, namely, in organic synthesis, may provide unprecedented control of the specificity of reaction product(s).

In synthesis and design of anodic $TiO_2$ nanotubes, a main open point is to address the origin for the still comparably low electron mobility in the nanotube walls (orders of magnitude lower than in single crystalline material). A key progress in this direction likely lies in optimized annealing strategies that "heal" the intrinsically high defect density in anodic nanotubes. Advanced annealing approaches also may allow for defined tube walls with a higher degree of control over crystallization (ideally to a single crystalline anatase 1D-structure).

Similarly challenging is the growth of sufficiently thin and defined tube walls that provide electronic quantum size effects this not only in view of a further improvement of electronic transport properties but also as a tool for band gap engineering that would additionally allow to establish intrinsic beneficial $TiO_2$ junctions.

Except for advancements that are specific to anodic nanotubes, the integration of findings from other $TiO_2$ research fields (morphologies, nanostructures, and polymorphs) are crucial steps toward a continuous improvement of anodic $TiO_2$ nanotubes in a broad range of photocatalytic applications. Colorful titania modifications and their unexpected reactivity may represent one of these perspectives that can provide a pathway to highly beneficial defect and band gap engineering.

Overall, we hope to have provided an overview of the current state of the use of $TiO_2$ nanotubes and to give a perspective for further advancing the structures toward tailoring photocatalytic reactivity and specificity in view of most important photocatalytic applications. Furthermore, we tried to describe the most relevant underlying concepts that strongly stimulate current research efforts. These efforts are not only in classic fields such as hydrogen generation or pollution degradation but also in reusable microstructures, biomedical devices, or high-selectivity microreactors.

## Notes

The authors declare no competing financial interest.

## ACKNOWLEDGMENTS

The authors would like to thank the DFG, ERC, and EAM cluster for financial support.

## REFERENCES

1. Hoffmann, M. R.; Martin, S. T.; Choi, W. Y.; Bahnemann, D. W. *Chem. Rev.* 1995, *95*, 69−96.
2. Hagfeldt, A.; Grätzel, M. *Chem. Rev.* 1995, *95*, 49−68.
3. Fujishima, A.; Honda, K. *Nature* 1972, *238*, 37−38.
4. Ni, M.; Leung, M. K. H.; Leung, D. Y. C.; Sumathy, K. *Renewable Sustainable Energy Rev.* 2007, *11*, 401−425.
5. Linsebigler, A. L.; Lu, G. Q.; Yates, J. T. *Chem. Rev.* 1995, *95*, 735−758.
6. Fujishima, A.; Zhang, X.; Tryk, D. A. *Surf. Sci. Rep.* 2008, *63*, 515− 582.
7. Paramasivam, I.; Jha, H.; Liu, N.; Schmuki, P. *Small* 2012, *8*, 3073− 3103.
8. Habisreutinger, S. N.; Schmidt-Mende, L.; Stolarczyk, J. K. *Angew. Chem., Int. Ed.* 2013, *52*, 7372−7408.
9. Song, Y. Y.; Gao, Z.; Lee, K.; Schmuki, P. *Electrochem. Commun.* 2011, *13*, 1217−1220.
10. White, J. L.; Baruch, M. F.; Pander, J. E.; Hu, Y.; Fortmeyer, I. C.; Park, J. E.; Zhang, T.; Liao, K.; Gu, J.; Yan, Y.; Shaw, T. W.; Abelev, E.; Bocarsly, A. B. *Chem. Rev.* 2015, *115*, 12888−12935.
11. Ferrin, P.; Nilekar, A. U.; Greeley, J.; Mavrikakis, M.; Rossmeisl, J. *Surf. Sci.* 2008, *602*, 3424−3431.
12. Grätzel, M. *Nature* 2001, *414*, 338−344.
13. Rajeshwar, K.; Osugi, M. E.; Chanmanee, W.; Chenthamarakshan, C. R.; Zanoni, M. V. B.; Kajitvichyanukul, P.; Krishnan-Ayer, R. *J. Photochem. Photobiol., C* 2008, *9*, 171−192.
14. Abe, R. *J. Photochem. Photobiol., C* 2010, *11*, 179−209.
15. Bauer, S.; Schmuki, P.; von der Mark, K.; Park, J. *Prog. Mater. Sci.* 2013, *58*, 261−326.
16. Park, J.; Bauer, S.; Von Der Mark, K.; Schmuki, P. *Nano Lett.* 2007, *7*, 1686−1691.
17. Oh, S.; Brammer, K. S.; Li, Y. S.; Teng, D.; Engler, A. J.; Chien, S.; Jin, S. *Proc. Natl. Acad. Sci. U. S. A.* 2009, *106*, 2130−2135.
18. Tripathy, J.; Lee, K.; Schmuki, P. *Angew. Chem., Int. Ed.* 2014, *53*, 12605−12608.
19. Shrestha, N. K.; Macak, J. M.; Schmidt-Stein, F.; Hahn, R.; Mierke, C. T.; Fabry, B.; Schmuki, P. *Angew. Chem., Int. Ed.* 2009, *48*, 969−972.
20. Liu, G.; Yin, L.-C.; Wang, J.; Niu, P.; Zhen, C.; Xie, Y.; Cheng, H.- M. *Energy Environ. Sci.* 2012, *5*, 9603−9610.
21. Zhao, J.; Zhang, L.; Xing, W.; Lu, K. *J. Phys. Chem. C* 2015, *119*, 7732−7737.
22. Tian, L.; Xu, J.; Alnafisah, A.; Wang, R.; Tan, X.; Oyler, N. A.; Liu, L.; Chen, X. *Chem. - Eur. J.* 2017, DOI: 10.1002/chem.201606027. (23) Chen, X.; Liu, L.; Yu, P. Y.; Mao, S. S. *Science* 2011, *331*, 746−750.
23. Grabstanowicz, L. R.; Gao, S.; Li, T.; Rickard, R. M.; Rajh, T.; Liu, D.; Xu, T. *Inorg. Chem.* 2013, *52*, 3884−3890.
24. Liu, N.; Schneider, C.; Freitag, D.; Venkatesan, U.; Marthala, V. R. R.; Hartmann, M.; Winter, B.; Spiecker,

E.; Osvet, A.; Zolnhofer, E. M.; Meyer, K.; Nakajima, T.; Zhou, X.; Schmuki, P. *Angew. Chem., Int. Ed.* 2014, *53*, 14201−14205.
25. Liu, N.; Haeublein, V.; Zhou, X.; Venkatesan, U.; Hartmann, M.; MacKovic, M.; Nakajima, T.; Spiecker, E.; Osvet, A.; Frey, L.; Schmuki, P. *Nano Lett.* 2015, *15*, 6815−6820.
26. Liu, N.; Zhou, X.; Nguyen, N. T.; Peters, K.; Zoller, F.; Hwang, I.; Schneider, C.; Miehlich, M. E.; Freitag, D.; Meyer, K.; Fattakhova- Rohlfing, D.; Schmuki, P. *ChemSusChem* 2017, *10*, 62−67.
27. Zhou, X.; Liu, N.; Schmidt, J.; Kahnt, A.; Osvet, A.; Romeis, S.; Zolnhofer, E. M.; Marthala, V. R. R.; Guldi, D. M.; Peukert, W.; Hartmann, M.; Meyer, K.; Schmuki, P. *Adv. Mater.* 2017, *29*, 1604747.
28. Liu, X.; Zhu, G.; Wang, X.; Yuan, X.; Lin, T.; Huang, F. *Adv. Energy Mater.* 2016, *6*, 1600452.
29. Lin, T.; Yang, C.; Wang, Z.; Yin, H.; Lü, X.; Huang, F.; Lin, J.; Xie, X.; Jiang, M. *Energy Environ. Sci.* 2014, *7*, 967−972.
30. Zheng, Z.; Huang, B.; Meng, X.; Wang, J.; Wang, S.; Lou, Z.; Wang, Z.; Qin, X.; Zhang, X.; Dai, Y. *Chem. Commun.* 2013, *49*, 868− 870.
31. Chen, X.; Shen, S.; Guo, L.; Mao, S. S. *Chem. Rev.* 2010, *110*, 6503−6570.
32. Yang, H. G.; Sun, C. H.; Qiao, S. Z.; Zou, J.; Liu, G.; Smith, S. C.; Cheng, H. M.; Lu, G. Q. *Nature* 2008, *453*, 638−641.
33. Zuo, F.; Bozhilov, K.; Dillon, R. J.; Wang, L.; Smith, P.; Zhao, X.; Bardeen, C.; Feng, P. *Angew. Chem.* 2012, *124*, 6327−6330.
34. Dette, C.; Pérez-Osorio, M. A.; Kley, C. S.; Punke, P.; Patrick, C.
35. E.; Jacobson, P.; Giustino, F.; Jung, S. J.; Kern, K. *Nano Lett.* 2014, *14*, 6533−6538.
36. Yu, J.; Dai, G.; Huang, B. *J. Phys. Chem. C* 2009, *113*, 16394−16401.
37. Cheng, X.; Liu, H.; Chen, Q.; Li, J.; Wang, P. *Electrochim. Acta* 2013, *103*, 134−142.
38. Asahi, R.; Morikawa, T.; Irie, H.; Ohwaki, T. *Chem. Rev.* 2014, *114*, 9824−9852.
39. Kisch, H.; Sakthivel, S.; Janczarek, M.; Mitoraj, D. *J. Phys. Chem. C* 2007, *111*, 11445−11449.
40. Zhu, Y.; Chen, Z.; Gao, T.; Huang, Q.; Niu, F.; Qin, L.; Tang, P.; Huang, Y.; Sha, Z.; Wang, Y. *Appl. Catal., B* 2015, *163*, 16−22.
41. Han, H.; Riboni, F.; Karlicky,́ F.; Kment, S.; Goswami, A.; Sudhagar, P.; Yoo, J.; Wang, L.; Tomanec, O.; Petr, M.; Haderka, O.; Terashima, C.; Fujishima, A.; Schmuki, P.; Zboril, R. *Nanoscale* 2017, *9*, 134−142.
42. Choudhary, S.; Upadhyay, S.; Kumar, P.; Singh, N.; Satsangi, V. R.; Shrivastav, R.; Dass, S. *Int. J. Hydrogen Energy* 2012, *37*, 18713− 18730.
43. Henderson, M. A. *Surf. Sci. Rep.* 2011, *66*, 185−297.
44. Hoang, S.; Berglund, S. P.; Hahn, N. T.; Bard, A. J.; Mullins, C. B. *J. Am. Chem. Soc.* 2012, *134*, 3659−3662.
45. Zhou, W.; Yin, Z.; Du, Y.; Huang, X.; Zeng, Z.; Fan, Z.; Liu, H.; Wang, J.; Zhang, H. *Small* 2013, *9*, 140−147.
46. Lee, K.; Mazare, A.; Schmuki, P. *Chem. Rev.* 2014, *114*, 9385−9454.
47. Schneider, J.; Matsuoka, M.; Takeuchi, M.; Zhang, J.; Horiuchi, Y.; Anpo, M.; Bahnemann, D. W. *Chem. Rev.* 2014, *114*, 9919−9986.
48. Ma, Y.; Wang, X. L.; Jia, Y. S.; Chen, X. B.; Han, H. X.; Li, C. *Chem. Rev.* 2014, *114*, 9987−10043.
49. Chen, X.; Mao, S. S. *Chem. Rev.* 2007, *107*, 2891−2959.
50. Wang, X.; Li, Z.; Shi, J.; Yu, Y. Chem. Rev. 2014, 114, 9346−9384.
51. Kapilashrami, M.; Zhang, Y.; Liu, Y.-S.; Hagfeldt, A.; Guo, J. *Chem. Rev.* 2014, *114*, 9662−9707.
52. Sang, L.; Zhao, Y.; Burda, C. *Chem. Rev.* 2014, *114*, 9283−9318.
53. Diebold, U. *Surf. Sci. Rep.* 2003, *48*, 53−229.
54. Roy, P.; Berger, S.; Schmuki, P. *Angew. Chem., Int. Ed.* 2011, *50*, 2904−2939.
55. Rajeshwar, K. *Encyclopedia of Electrochemistry*; Wiley-VCH Verlag GmbH & Co. KGaA: Weinheim, Germany, 2007; pp 1−53.
56. Sasahara, A.; Onishi, H. *Solid State Phenom.* 2010, *162*, 115−133.
57. Hans-Joachim, L.; Laurie, P. *Photoelectrochemical Water Splitting: Materials, Processes and Architectures*; The Royal Society of Chemistry: London, U.K., 2014.
58. Li, X.; Yu, J.; Low, J.; Fang, Y.; Xiao, J.; Chen, X. *J. Mater. Chem. A* 2015, *3*, 2485−2534.

59. Moniz, S. J. A.; Shevlin, S. A.; Martin, D. J.; Guo, Z.-X.; Tang, J. *Energy Environ. Sci.* 2015, *8*, 731−759.
60. Regonini, D.; Bowen, C. R.; Jaroenworaluck, A.; Stevens, R. *Mater. Sci. Eng., R* 2013, *74*, 377−406.
61. Zhou, X.; Nguyen, N. T.; Ozkan, S.; Schmuki, P. *Electrochem. Commun.* 2014, *46*, 157−162.
62. Ghicov, A.; Schmuki, P. *Chem. Commun.* 2009, 2791−2808.
63. Mor, G. K.; Varghese, O. K.; Paulose, M.; Shankar, K.; Grimes, C. A. *Sol. Energy Mater. Sol. Cells* 2006, *90*, 2011−2075.
64. Su, Z.; Zhou, W. *J. Mater. Chem.* 2011, *21*, 8955−8970.
65. Lim, J. H.; Choi, J. *Small* 2007, *3*, 1504−1507.
66. Huo, K.; Gao, B.; Fu, J.; Zhao, L.; Chu, P. K. *RSC Adv.* 2014, *4*, 17300−17324.
67. Hurum, D. C.; Agrios, A. G.; Gray, K. A.; Rajh, T.; Thurnauer, M. C. *J. Phys. Chem. B* 2003, *107*, 4545−4549.
68. Buckeridge, J.; Butler, K. T.; Catlow, C. R. A.; Logsdail, A. J.; Scanlon, D. O.; Shevlin, S. A.; Woodley, S. M.; Sokol, A. A.; Walsh, A. *Chem. Mater.* 2015, *27*, 3844−3851.
69. Kavan, L.; Grätzel, M.; Gilbert, S. E.; Klemenz, C.; Scheel, H. J. *J. Am. Chem. Soc.* 1996, *118*, 6716−6723.
70. Xiong, G.; Shao, R.; Droubay, T. C.; Joly, A. G.; Beck, K. M.; Chambers, S. A.; Hess, W. P. *Adv. Funct. Mater.* 2007, *17*, 2133−2138.
71. Di Paola, A.; Bellardita, M.; Ceccato, R.; Palmisano, L.; Parrino, F. *J. Phys. Chem. C* 2009, *113*, 15166−15174.
72. Gai, L.; Duan, X.; Jiang, H.; Mei, Q.; Zhou, G.; Tian, Y.; Liu, H. *CrystEngComm* 2012, *14*, 7662−7671.
73. Kim, C. W.; Suh, S. P.; Choi, M. J.; Kang, Y. S. Y. S.; Kang, Y. S. Y. S. *J. Mater. Chem. A* 2013, *1*, 11820−11827.
74. Zhang, J.; Bang, J. H.; Tang, C.; Kamat, P. V. *ACS Nano* 2010, *4*, 387−395.
75. Ha, M. N.; Zhu, F.; Liu, Z.; Wang, L.; Liu, L.; Lu, G.; Zhao, Z. *RSC Adv.* 2016, *6*, 21111−21118.
76. Reihl, B.; Bednorz, J. G.; Müller, K. A.; Jugnet, Y.; Landgren, G.; Morar, J. F. *Phys. Rev. B: Condens. Matter Mater. Phys.* 1984, *30*, 803− 806.
77. Li, J.; Wu, N. *Catal. Sci. Technol.* 2015, *5*, 1360−1384.
78. Ishibashi, K.; Fujishima, A.; Watanabe, T.; Hashimoto, K. *J. Photochem. Photobiol., A* 2000, *134*, 139−142.
79. Fujishima, A.; Rao, T. N.; Tryk, D. A. *J. Photochem. Photobiol., C* 2000, *1*, 1−21.
80. Gaya, U. I.; Abdullah, A. H. *J. Photochem. Photobiol., C* 2008, *9*, 1−12.
81. Hirakawa, T.; Nosaka, Y. *Langmuir* 2002, *18*, 3247−3254.
82. Pan, J.; Liu, G.; Lu, G. Q.; Cheng, H. M. *Angew. Chem., Int. Ed.* 2011, *50*, 2133−2137.
83. Lynch, R. P.; Ghicov, A.; Schmuki, P. *J. Electrochem. Soc.* 2010, *157*, G76−G84.
84. Morrison, S. R. *Electrochemistry at Semiconductor and Oxidized Metal Electrodes*; Plenum Press: New York, 1980.
85. *Photoelectrochemical Water Splitting: Materials, Processes and Architectures*; Lewerenz, H.-J., Peter, L., Eds.; The Royal Society of Chemistry: London, U.K., 2013; pp 289332
86. Tu, W.; Zhou, Y.; Zou, Z. *Adv. Mater.* 2014, *26*, 4607−4626.
87. Kumar, B.; Llorente, M.; Froehlich, J.; Dang, T.; Sathrum, A.; Kubiak, C. P. *Annu. Rev. Phys. Chem.* 2012, *63*, 541−569.
88. Marchand, R.; Brohan, L.; Tournoux, M. *Mater. Res. Bull.* 1980, *15*, 1129−1133.
89. Muscat, J.; Swamy, V.; Harrison, N. M. *Phys. Rev. B: Condens. Matter Mater. Phys.* 2002, *65*, 224112.
90. Dubrovinsky, L. S.; Dubrovinskaia, N. A.; Swamy, V.; Muscat, J.; Harrison, N. M.; Ahuja, R.; Holm, B.; Johansson, B. *Nature* 2001, *410*, 653−654.
91. Li, L.; Salvador, P. a; Rohrer, G. S. *Nanoscale* 2014, *6*, 24−42. (92) Zhang, Z.; Yates, J. T. *Chem. Rev.* 2012, *112*, 5520−5551.
92. Kudo, A.; Miseki, Y. *Chem. Soc. Rev.* 2009, *38*, 253−278.
93. Kumar, S. G.; Devi, L. G. *J. Phys. Chem. A* 2011, *115*, 13211−13241.
94. Maeda, K. *ACS Catal.* 2013, *3*, 1486−1503.
95. Zhou, P.; Yu, J.; Jaroniec, M. *Adv. Mater.* 2014, *26*, 4920−4935.
96. Tran, P. D.; Wong, L. H.; Barber, J.; Loo, J. S. C. *Energy Environ. Sci.* 2012, *5*, 5902−5918.

97. Sun, W.-T.; Yu, Y.; Pan, H.-Y.; Gao, X.-F.; Chen, Q.; Peng, L.-M. *J. Am. Chem. Soc.* 2008, *130*, 1124−1125.
98. Kawahara, T.; Konishi, Y.; Tada, H.; Tohge, N.; Nishii, J.; Ito, S. *Angew. Chem., Int. Ed.* 2002, *41*, 2811−2813.
99. Zhou, X.; Häublein, V.; Liu, N.; Nguyen, N. T.; Zolnhofer, E. M.; Tsuchiya, H.; Killian, M. S.; Meyer, K.; Frey, L.; Schmuki, P. *Angew. Chem., Int. Ed.* 2016, *55*, 3763−3767.
100. Yun, J.-H.; Ng, Y. H.; Huang, S.; Conibeer, G.; Amal, R. *Chem. Commun.* 2011, *47*, 11288−11290.
101. Nah, Y. C.; Paramasivam, I.; Schmuki, P. *ChemPhysChem* 2010, *11*, 2698−2713.
102. Connelly, K.; Wahab, a. K.; Idriss, H. *Mater. Renew. Sustain. Energy* 2012, *1*, 3.
103. Galinska, A.; Walendziewski, J. *Energy Fuels* 2005, *19*, 1143− 1147.
104. Shen, M.; Yan, Z.; Yang, L.; Du, P.; Zhang, J.; Xiang, B. *Chem. Commun.* 2014, *50*, 15447−15449.
105. Ambrosi, A.; Sofer, Z.; Pumera, M. *Chem. Commun.* 2015, *51*, 8450−8453.
106. Zhou, X.; Licklederer, M.; Schmuki, P. *Electrochem. Commun.* 2016, *73*, 33−37.
107. Kibsgaard, J.; Jaramillo, T. F. *Angew. Chem., Int. Ed.* 2014, *53*, 14433−14437.
108. Burrows, I. R.; Denton, D. A.; Harrison, J. A. *Electrochim. Acta* 1978, *23*, 493−500.
109. Cook, T. R.; Dogutan, D. K.; Reece, S. Y.; Surendranath, Y.; Teets, T. S.; Nocera, D. G. *Chem. Rev.* 2010, *110*, 6474−6502.
110. Kanan, M. W.; Nocera, D. G. *Science* 2008, *321*, 1072−1075.
111. Xie, Y. *Adv. Funct. Mater.* 2006, *17*, 3340−3346.
112. Xiang, Q.; Yu, J.; Jaroniec, M. *J. Am. Chem. Soc.* 2012, *134*, 6575−6578.
113. Gao, Z.-D.; Qu, Y.-F.; Zhou, X.; Wang, L.; Song, Y.-Y.; Schmuki, P. *ChemistryOpen* 2016, *5*, 197−200.
114. Maeda, K.; Domen, K. *J. Phys. Chem. Lett.* 2010, *1*, 2655−2661.
115. Bumajdad, A.; Madkour, M. *Phys. Chem. Chem. Phys.* 2014, *16*, 7146−7158.
116. Tsuchiya, H.; Macak, J. M.; Ghicov, A.; Rader, A. S.; Taveira, L.; Schmuki, P. *Corros. Sci.* 2007, *49*, 203−210.
117. Beranek, R.; Tsuchiya, H.; Sugishima, T.; Macak, J. M.; Taveira, L.; Fujimoto, S.; Kisch, H.; Schmuki, P. *Appl. Phys. Lett.* 2005, *87*, 243114.
118. Bard, A. J.; Bocarsly, A. B.; Fan, F. R. F.; Walton, E. G.; Wrighton, M. S. J. Am. Chem. Soc. 1980, 102, 3671−3677.
119. Fabregat-Santiago, F.; Garcia-Belmonte, G.; Bisquert, J.; Bogdanoff, P.; Zaban, A. J. Electrochem. Soc. 2003, 150, E293−E298.
120. Schottky, W. Naturwissenschaften 1938, 26, 843−843.
121. Gärtner, W. W. *Phys. Rev.* 1959, *116*, 84−87.
122. Spanier, J. E.; Fridkin, V. M.; Rappe, A. M.; Akbashev, A. R.;
123. Polemi, A.; Qi, Y.; Young, S. M.; Gu, Z.; Hawley, C. J.; Imbrenda, D.; Xiao, G.; Bennett-Jackson, A. L.; Johnson, C. L. Nat. Photonics 2016, 10, 611−616.
124. Butler, M. A. J. Appl. Phys. 1977, 48, 1914−1920.
125. Beranek, R.; Macak, J. M.; Gärtner, M.; Meyer, K.; Schmuki, P. Electrochim. Acta 2009, 54, 2640−2646.
126. van de Lagemaat, J.; Park, N.-G.; Frank, A. J. J. Phys. Chem. B 2000, 104, 2044−2052.
127. Murdoch, M.; Waterhouse, G. I. N.; Nadeem, M. A.; Metson, J. B.; Keane, M. A.; Howe, R. F.; Llorca, J.; Idriss, H. Nat. Chem. 2011, 3, 489−492.
128. Jovic, V.; Chen, W. T.; Sun-Waterhouse, D.; Blackford, M. G.; Idriss, H.; Waterhouse, G. I. N. J. Catal. 2013, 305, 307−317.
129. Yoo, J. E.; Lee, K.; Altomare, M.; Selli, E.; Schmuki, P. Angew. Chem., Int. Ed. 2013, 52, 7514−7517.
130. Plass, R.; Pelet, S.; Krueger, J.; Grätzel, M.; Bach, U. J. Phys. Chem. B 2002, 106, 7578−7580.
131. Kongkanand, A.; Tvrdy, K.; Takechi, K.; Kuno, M.; Kamat, P. V. J. Am. Chem. Soc. 2008, 130, 4007−4015.
132. Barquinha, P.; Pereira, L.; Águas, H.; Fortunato, E.; Martins, R. Mater. Sci. Semicond. Process. 2004, 7, 243−247.
133. Wang, G.; Wang, H.; Ling, Y.; Tang, Y.; Yang, X.; Fitzmorris, R. C.; Wang, C.; Zhang, J. Z.; Li, Y. Nano Lett. 2011, 11, 3026−3033.

134. Lu, X.; Wang, G.; Zhai, T.; Yu, M.; Gan, J.; Tong, Y.; Li, Y. Nano Lett. 2012, 12, 1690−1696.
135. Sze, S. M.; Ng, K. K. Physics of Semiconductor Devices, 2nd ed.; John Wiley & Sons: Hoboken, NJ, 1981.
136. Schubert, M. M.; Hackenberg, S.; van Veen, A. C.; Muhler, M.; Plzak, V.; Behm, R. J. J. Catal. 2001, 197, 113−122.
137. Andreeva, D.; Idakiev, V.; Tabakova, T.; Andreev, A.; Giovanoli, R. Appl. Catal., A 1996, 134, 275−283.
138. Okumura, M.; Nakamura, S.; Tsubota, S.; Nakamura, T.; Azuma, M.; Haruta, M. Catal. Lett. 1998, 51, 53−58.
139. Hvolbæk, B.; Janssens, T. V. W.; Clausen, B. S.; Falsig, H.; Christensen, C. H.; Nørskov, J. K. Nano Today 2007, 2, 14−18.
140. Schimpf, S.; Lucas, M.; Mohr, C.; Rodemerck, U.; Brückner, A.; Radnik, J.; Hofmeister, H.; Claus, P. Catal. Today 2002, 72, 63−78.
141. Lazzeri, M.; Vittadini, A.; Selloni, A. Phys. Rev. B: Condens. Matter Mater. Phys. 2001, 63, 155409.
142. Bauer, S.; Pittrof, A.; Tsuchiya, H.; Schmuki, P. Electrochem. Commun. 2011, 13, 538−541.
143. Ranade, M. R.; Navrotsky, A.; Zhang, H. Z.; Banfield, J. F.; Elder, S. H.; Zaban, A.; Borse, P. H.; Kulkarni, S. K.; Doran, G. S.; Whitfield, H. J. Proc. Natl. Acad. Sci. U. S. A. 2002, 99, 6476−6481.
144. So, S.; Hwang, I.; Riboni, F.; Yoo, J. E.; Schmuki, P. Electrochem. Commun. 2016, 71, 73−78.
145. Ghicov, A.; Albu, S. P.; Hahn, R.; Kim, D.; Stergiopoulos, T.;
146. Kunze, J.; Schiller, C. A.; Falaras, P.; Schmuki, P. Chem. - Asian J. 2009, 4, 520−525.
147. Tang, H.; Prasad, K.; Sanjinès, R.; Schmid, P. E.; Lévy, F. J. Appl. Phys. 1994, 75, 2042−2047.
148. Zhou, X.; Zolnhofer, E. M.; Nguyen, N. T.; Liu, N.; Meyer, K.; Schmuki, P. Angew. Chem., Int. Ed. 2015, 54, 13385−13389.
149. Al-Thabaiti, S. A.; Hahn, R.; Liu, N.; Kirchgeorg, R.; So, S.;
150. Schmuki, P.; Basahel, S. N.; Bawaked, S. M. Chem. Commun. 2014, 50, 7960−7963.
151. Sant, P. A.; Kamat, P. V. Phys. Chem. Chem. Phys. 2002, 4, 198− 203.
152. Yang, L.; Luo, S.; Liu, R.; Cai, Q.; Xiao, Y.; Liu, S.; Su, F.; Wen, L. J. Phys. Chem. C 2010, 114, 4783−4789.
153. Scanlon, D. O.; Dunnill, C. W.; Buckeridge, J.; Shevlin, S. A.; Logsdail, A. J.; Woodley, S. M.; Catlow, C. R. A.; Powell, M. J.; Palgrave, R. G.; Parkin, I. P.; Watson, G. W.; Keal, T. W.; Sherwood, P.; Walsh, A.; Sokol, A. A. Nat. Mater. 2013, 12, 798−801.
154. Deák, P.; Kullgren, J.; Aradi, B.; Frauenheim, T.; Kavan, L. Electrochim. Acta 2016, 199, 27−34.
155. Ohno, T.; Sarukawa, K.; Matsumura, M. New J. Chem. 2002, 26, 1167−1170.
156. Giocondi, J. L.; Rohrer, G. S. Top. Catal. 2008, 49, 18−23.
157. Liu, S.; Yu, J.; Jaroniec, M. J. Am. Chem. Soc. 2010, 132, 11914−11916.
158. Yang, H. G.; Liu, G.; Qiao, S. Z.; Sun, C. H.; Jin, Y. G.; Smith, S. C.; Zou, J.; Cheng, H. M.; Lu, G. Q. J. Am. Chem. Soc. 2009, 131, 4078− 4083.
159. Shibata, H.; Ogura, T.; Mukai, T.; Ohkubo, T.; Sakai, H.; Abe, M. J. Am. Chem. Soc. 2005, 127, 16396−16397.
160. Dai, S.; Wu, Y.; Sakai, T.; Du, Z.; Sakai, H.; Abe, M. Nanoscale Res. Lett. 2010, 5, 1829−1835.
161. Selloni, A. Nat. Mater. 2008, 7, 613−615.
162. Han, X.; Kuang, Q.; Jin, M.; Xie, Z.; Zheng, L. J. Am. Chem. Soc. 2009, 131, 3152−3153.
163. Giocondi, J. L.; Salvador, P. a.; Rohrer, G. S. Top. Catal. 2007, 44, 529−533.
164. Li, R.; Han, H.; Zhang, F.; Wang, D.; Li, C. Energy Environ. Sci. 2014, 7, 1369−1376.
165. Munprom, R.; Salvador, P. A.; Rohrer, G. S. J. Mater. Chem. A 2015, 3, 2370−2377.
166. Di Valentin, C.; Finazzi, E.; Pacchioni, G.; Selloni, A.; Livraghi, S.; Paganini, M. C.; Giamello, E. Chem. Phys. 2007, 339, 44−56.
167. Roose, B.; Pathak, S.; Steiner, U. Chem. Soc. Rev. 2015, 44, 8326−8349.
168. Asahi, R.; Morikawa, T.; Ohwaki, T.; Aoki, K.; Taga, Y. Science 2001, 293, 269−271.
169. Zhao, Z.; Liu, Q. J. Phys. D: Appl. Phys. 2008, 41, 025105.
170. Wang, H.; Lewis, J. P. J. Phys.: Condens. Matter 2006, 18, 421− 434.
171. Yang, K.; Dai, Y.; Huang, B.; Whangbo, M. H. J. Phys. Chem. C 2009, 113, 2624−2629.

172. Yang, K.; Dai, Y.; Huang, B. J. Phys. Chem. C 2007, 111, 18985−18994.
173. Finazzi, E.; Di Valentin, C.; Pacchioni, G. J. Phys. Chem. C 2009, 113, 220−228.
174. Umebayashi, T.; Yamaki, T.; Itoh, H.; Asai, K. J. Phys. Chem. Solids 2002, 63, 1909−1920.
175. Osorio-Guillen, J.; Lany, S.; Zunger, A. Phys. Rev. Lett. 2008, 100, 036601.
176. Wang, Y.; Doren, D. J. Solid State Commun. 2005, 136, 142−146.
177. Valentin, C. Di; Pacchioni, G.; Onishi, H.; Kudo, A. Chem. Phys. Lett. 2009, 469, 166−171.
178. Shao, G. J. Phys. Chem. C 2009, 113, 6800−6808.
179. Yang, K.; Dai, Y.; Huang, B. Phys. Rev. B: Condens. Matter Mater. Phys. 2007, 76, 195201.
180. Czoska, A. M.; Livraghi, S.; Chiesa, M.; Giamello, E.; Agnoli, S.; Granozzi, G.; Finazzi, E.; Di Valentin, C.; Pacchioni, G. J. Phys. Chem. C 2008, 112, 8951−8956.
181. Kim, D.; Tsuchiya, H.; Fujimoto, S.; Schmidt-Stein, F.; Schmuki, P. J. Solid State Electrochem. 2012, 16, 89−92.
182. Hahn, R.; Ghicov, A.; Salonen, J.; Lehto, V.-P.; Schmuki, P. Nanotechnology 2007, 18, 105604.
183. Chen, X.; Liu, L.; Huang, F. Chem. Soc. Rev. 2015, 44, 1861− 1885.
184. Richter, C.; Schmuttenmaer, C. A. Nat. Nanotechnol. 2010, 5, 769−772.
185. Bavykin, D.; Walsh, F. Titanate and Titania Nanotubes Synthesis, Properties and Applications; School of Engineering Sciences, University of Southampton: Southampton, U.K., 2010.
186. Khomenko, V. M.; Langer, K.; Rager, H.; Fett, A. Phys. Chem. Miner. 1998, 25, 338−346.
187. Livraghi, S.; Paganini, M. C.; Giamello, E.; Selloni, A.; Di Valentin, C.; Pacchioni, G. J. Am. Chem. Soc. 2006, 128, 15666−15671.
188. Liu, H.; Ma, H. T.; Li, X. Z.; Li, W. Z.; Wu, M.; Bao, X. H. Chemosphere 2003, 50, 39−46.
189. Pan, X.; Yang, M.-Q.; Fu, X.; Zhang, N.; Xu, Y.-J. Nanoscale 2013, 5, 3601−3614.
190. Brückner, A. Chem. Soc. Rev. 2010, 39, 4673−4684.
191. Zhou, X.; Liu, N.; Schmuki, P. Electrochem. Commun. 2014, 49, 60−64.
192. Sasikala, R.; Shirole, A.; Sudarsan, V.; Sakuntala, T.; Sudakar, C.; Naik, R.; Bharadwaj, S. R. Int. J. Hydrogen Energy 2009, 34, 3621−3630.
193. Zhang, K.; Wang, L.; Kim, J. K.; Ma, M.; Veerappan, G.; Lee, C.- L.; Kong, K.; Lee, H.; Park, J. H. Energy Environ. Sci. 2016, 9, 499−503.
194. Zhang, K.; Ravishankar, S.; Ma, M.; Veerappan, G.; Bisquert, J.; Fabregat-Santiago, F.; Park, J. H. Adv. Energy Mater. 2017, 7, 1600923.
195. Zhang, K.; Park, J. H. J. Phys. Chem. Lett. 2017, 8, 199−207.
196. Cho, I. S.; Choi, J.; Zhang, K.; Kim, S. J.; Jeong, M. J.; Cai, L.; Park, T.; Zheng, X.; Park, J. H. Nano Lett. 2015, 15, 5709−5715.
197. Zuo, F.; Wang, L.; Wu, T.; Zhang, Z.; Borchardt, D.; Feng, P. J. Am. Chem. Soc. 2010, 132, 11856−11857.
198. Naldoni, A.; Allieta, M.; Santangelo, S.; Marelli, M.; Fabbri, F.; Cappelli, S.; Bianchi, C. L.; Psaro, R.; Dal Santo, V. J. Am. Chem. Soc. 2012, 134, 7600−7603.
199. Liu, N.; Schneider, C.; Freitag, D.; Hartmann, M.; Venkatesan, U.; Müller, J.; Spiecker, E.; Schmuki, P. Nano Lett. 2014, 14, 3309−3313.
200. Liu, N.; Schneider, C.; Freitag, D.; Hartmann, M.; Venkatesan, U.; Müller, J.; Spiecker, E.; Schmuki, P. Nano Lett. 2014, 14, 3309−3313.
201. Nikitenko, S. I.; Chave, T.; Cau, C.; Brau, H.-P.; Flaud, V. ACS Catal. 2015, 5, 4790−4795.
202. Liu, N.; Steinrück, H.-G.; Osvet, A.; Yang, Y.; Schmuki, P. Appl. Phys. Lett. 2017, 110, 072102.
203. Li, H.; Chen, Z.; Tsang, C. K.; Li, Z.; Ran, X.; Lee, C.; Nie, B.; Zheng, L.; Hung, T.; Lu, J.; Pan, B.; Li, Y. Y. J. Mater. Chem. A 2014, 2, 229−236.
204. Yang, Y.; Hoffmann, M. R. Environ. Sci. Technol. 2016, 50, 11888−11894.
205. Dong, J.; Han, J.; Liu, Y.; Nakajima, A.; Matsushita, S.; Wei, S.; Gao, W. ACS Appl. Mater. Interfaces 2014, 6, 1385−1388.
206. Cui, H.; Zhao, W.; Yang, C.; Yin, H.; Lin, T.; Shan, Y.; Xie, Y.; Gu, H.; Huang, F. J. Mater. Chem. A 2014, 2, 8612−8616.
207. Ge, M.; Cao, C.; Huang, J.; Li, S.; Chen, Z.; Zhang, K.-Q.; Al- deyab, S. S.; Lai, Y. J. Mater. Chem. A

2016, 4, 6772−6801.
208. Tian, J.; Zhao, Z.; Kumar, A.; Boughton, R. I.; Liu, H. Chem. Soc. Rev. 2014, 43, 6920−6937.
209. Kowalski, D.; Kim, D.; Schmuki, P. Nano Today 2013, 8, 235−264.
210. Riboni, F.; Nguyen, N. T.; So, S.; Schmuki, P. Nanoscale Horiz. 2016, 1, 445−466.
211. Macak, J. M.; Zlamal, M.; Krysa, J.; Schmuki, P. Small 2007, 3, 300−304.
212. Lee, K.; Hahn, R.; Altomare, M.; Selli, E.; Schmuki, P. Adv. Mater. 2013, 25, 6133−6137.
213. Grätzel, M.; Frank, A. J. J. Phys. Chem. 1982, 86, 2964−2967.
214. Kim, D.; Ghicov, A.; Albu, S. P.; Schmuki, P. J. Am. Chem. Soc. 2008, 130, 16454−16455.
215. Song, Y. Y.; Schmuki, P. Electrochem. Commun. 2010, 12, 579−582.
216. So, S.; Schmuki, P. Angew. Chem., Int. Ed. 2013, 52, 7933−7935.
217. So, S.; Kriesch, A.; Peschel, U.; Schmuki, P. J. Mater. Chem. A 2015, 3, 12603−12608.
218. Jennings, J. R.; Ghicov, A.; Peter, L. M.; Schmuki, P.; Walker, A. B. J. Am. Chem. Soc. 2008, 130, 13364−13372.
219. Tighineanu, A.; Albu, S. P.; Schmuki, P. Phys. Status Solidi RRL 2014, 8, 158−162.
220. Kelly, J. J.; Vanmaekelbergh, D. Electrochim. Acta 1998, 43, 2773−2780.
221. Stiller, M.; Barzola-Quiquia, J.; Lorite, I.; Esquinazi, P.; Kirchgeorg, R.; Albu, S. P.; Schmuki, P. Appl. Phys. Lett. 2013, 103, 173108.
222. Ozkan, S.; Mazare, A.; Schmuki, P. Electrochim. Acta 2015, 176, 819−826.
223. So, S.; Hwang, I.; Schmuki, P. Energy Environ. Sci. 2015, 8, 849−854.
224. Mirabolghasemi, H.; Liu, N.; Lee, K.; Schmuki, P. Chem. Commun. 2013, 49, 2067−2069.
225. Xu, J.; Yang, L.; Han, Y.; Wang, Y.; Zhou, X.; Gao, Z.; Song, Y. Y.; Schmuki, P. ACS Appl. Mater. Interfaces 2016, 8, 21997−22004.
226. Roy, P.; Dey, T.; Lee, K.; Kim, D.; Fabry, B.; Schmuki, P. J. Am. Chem. Soc. 2010, 132, 7893−7895.
227. Masuda, H.; Fukuda, K. Science 1995, 268, 1466−1468. (226) Bummel, T. Eur. Phys. J. A 1936, 99, 518−551.
228. Tsuchiya, H.; Macak, J. M.; Ghicov, A.; Schmuki, P. Small 2006, 2, 888−891.
229. MacAk, J. M.; Sirotna, K.; Schmuki, P. Electrochim. Acta 2005, 50, 3679−3684.
230. Albu, S. P.; Ghicov, A.; Macak, J. M.; Schmuki, P. Phys. Status Solidi RRL 2007, 1, R65−R67.
231. Albu, S. P.; Ghicov, A.; Aldabergenova, S.; Drechsel, P.; LeClere, D.; Thompson, G. E.; Macak, J. M.; Schmuki, P. Adv. Mater. 2008, 20, 4135−4139.
232. Liu, N.; Mirabolghasemi, H.; Lee, K.; Albu, S. P.; Tighineanu, A.; Altomare, M.; Schmuki, P. Faraday Discuss. 2013, 164, 107−116.
233. Roy, P.; Das, C.; Lee, K.; Hahn, R.; Ruff, T.; Moll, M.; Schmuki, P. J. Am. Chem. Soc. 2011, 133, 5629−5631.
234. Das, C.; Roy, P.; Yang, M.; Jha, H.; Schmuki, P. Nanoscale 2011, 3, 3094−3096.
235. Paramasivam, I.; Macak, J. M.; Schmuki, P. Electrochem. Commun. 2008, 10, 71−75.
236. Liu, N.; Paramasivam, I.; Yang, M.; Schmuki, P. J. Solid State Electrochem. 2012, 16, 3499−3504.
237. Agarwal, P.; Paramasivam, I.; Shrestha, N. K.; Schmuki, P. Chem.- Asian J. 2010, 5, 66−69.
238. Nguyen, N. T.; Ozkan, S.; Hwang, I.; Mazare, A.; Schmuki, P. Nanoscale 2016, 8, 16868−16873.
239. Meskin, P. E.; Gavrilov, A. I.; Maksimov, V. D.; Ivanov, V. K.; Churagulov, B. P. Russ. J. Inorg. Chem. 2007, 52, 1648−1656.
240. Pelaez, M.; Nolan, N. T.; Pillai, S. C.; Seery, M. K.; Falaras, P.; Kontos, A. G.; Dunlop, P. S. M.; Hamilton, J. W. J.; Byrne, J. A.; O'Shea, K.; Entezari, M. H.; Dionysiou, D. D. Appl. Catal., B 2012, 125, 331− 349.
241. Wang, M.; Ioccozia, J.; Sun, L.; Lin, C.; Lin, Z. Energy Environ. Sci. 2014, 7, 2182−2202.
242. Wang, J.; Lin, Z. Chem. Mater. 2010, 22, 579−584.
243. Hwang, I.; So, S.; Mokhtar, M.; Alshehri, A.; Al-Thabaiti, S. A.; Mazare, A.; Schmuki, P. Chem. - Eur. J. 2015, 21, 9204−9208.
244. Zhang, G.; Huang, H.; Zhang, Y.; Chan, H. L. W.; Zhou, L. Electrochem. Commun. 2007, 9, 2854−2858.
245. Zhang, G.; Huang, H.; Liu, Y.; Zhou, L. Appl. Catal., B 2009, 90, 262−267.
246. Zhang, Z.; Wang, P. Energy Environ. Sci. 2012, 5, 6506−6512.

247. Zhang, Z.; Zhang, L.; Hedhili, M. N.; Zhang, H.; Wang, P. Nano Lett. 2013, 13, 14−20.
248. Tighineanu, A.; Ruff, T.; Albu, S.; Hahn, R.; Schmuki, P. Chem. Phys. Lett. 2010, 494, 260−263.
249. Ghicov, A.; Tsuchiya, H.; Macak, J. M.; Schmuki, P. Phys. Status Solidi A 2006, 203, R28−R30.
250. Yu, J.; Dai, G.; Cheng, B. J. Phys. Chem. C 2010, 114, 19378−19385.
251. Gesenhues, U. J. Phys. Chem. Solids 2007, 68, 224−235.
252. Vaenas, N.; Bidikoudi, M.; Stergiopoulos, T.; Likodimos, V.; Kontos, A. G.; Falaras, P. Chem. Eng. J. 2013, 224, 121−127.
253. Lee, H.; Park, T.-H.; Jang, D.-J. New J. Chem. 2016, 40, 8737−8744.
254. Park, J. H.; Kim, S.; Bard, A. J. Nano Lett. 2006, 6, 24−28.
255. Ghicov, A.; Macak, J. M.; Tsuchiya, H.; Kunze, J.; Haeublein, V.; Frey, L.; Schmuki, P. Nano Lett. 2006, 6, 1080−1082.
256. Siuzdak, K.; Szkoda, M.; Lisowska-Oleksiak, A.; Grochowska, K.; Karczewski, J.; Ryl, J. Appl. Surf. Sci. 2015, 357, 942−950.
257. Barborini, E.; Conti, A. M.; Kholmanov, I.; Piseri, P.; Podestà, A.; Milani, P.; Cepek, C.; Sakho, O.; Macovez, R.; Sancrotti, M. Adv. Mater. 2005, 17, 1842−1846.
258. Morikawa, T.; Asahi, R.; Ohwaki, T.; Aoki, K.; Taga, Y. Jpn. J. Appl. Phys. 2001, 40, L561−L563.
259. Ghicov, A.; Macak, J. M.; Tsuchiya, H.; Kunze, J.; Haeublein, V.; Kleber, S.; Schmuki, P. Chem. Phys. Lett. 2006, 419, 426−429.
260. Kobayashi, N.; Tanoue, H. Nucl. Instrum. Methods Phys. Res., Sect. B 1989, 39, 746−749.
261. Tsujide, T.; Nojiri, M.; Kitagawa, H. J. Appl. Phys. 1980, 51, 1605−1610.
262. Ghicov, A.; Yamamoto, M.; Schmuki, P. Angew. Chem., Int. Ed. 2008, 47, 7934−7937.
263. Nah, Y.; Ghicov, A.; Kim, D.; Berger, S.; Schmuki, P. J. Am. Chem. Soc. 2008, 130, 16154−16155.
264. Das, C.; Paramasivam, I.; Liu, N.; Schmuki, P. Electrochim. Acta 2011, 56, 10557−10561.
265. Aruna, S. T.; Tirosh, S.; Zaban, A. J. Mater. Chem. 2000, 10, 2388−2391.
266. Kim, D.; Fujimoto, S.; Schmuki, P.; Tsuchiya, H. Electrochem. Commun. 2008, 10, 910−913.
267. Zaban, A.; Aruna, S. T.; Tirosh, S.; Gregg, B. A.; Mastai, Y. J. Phys. Chem. B 2000, 104, 4130−4133.
268. Sabin, F.; Türk, T.; Vogler, A. J. Photochem. Photobiol., A 1992, 63, 99−106.
269. Muneer, M.; Philip, R.; Das, S. Res. Chem. Intermed. 1997, 23, 233−246.
270. Altomare, M.; Lee, K.; Killian, M. S.; Selli, E.; Schmuki, P. Chem. - Eur. J. 2013, 19, 5841−5844.
271. Yoo, H.; Choi, Y. W.; Choi, J. ChemCatChem 2015, 7, 643−647.
272. Nakata, K.; Liu, B.; Ishikawa, Y.; Sakai, M.; Saito, H.; Ochiai, T.; Sakai, H.; Murakami, T.; Abe, M.; Takagi, K.; Fujishima, A. Chem. Lett. 2011, 40, 1107−1109.
273. Su, Y.; Zhang, X.; Zhou, M.; Han, S.; Lei, L. J. Photochem. Photobiol., A 2008, 194, 152−160.
274. Mishra, T.; Wang, L.; Hahn, R.; Schmuki, P. Electrochim. Acta 2014, 132, 410−415.
275. Yang, Y.; Kim, D.; Schmuki, P. Electrochem. Commun. 2011, 13, 1021−1025.
276. Kontos, A. I.; Likodimos, V.; Stergiopoulos, T.; Tsoukleris, D. S.; Falaras, P.; Rabias, I.; Papavassiliou, G.; Kim, D.; Kunze, J.; Schmuki, P. Chem. Mater. 2009, 21, 662−672.
277. Deng, L.; Wang, S.; Liu, D.; Zhu, B.; Huang, W.; Wu, S.; Zhang, S. Catal. Lett. 2009, 129, 513−518.
278. Slamet; Tristantini, D.; Valentina; Ibadurrohman, M. Int. J. Energy Res. 2013, 37, 1372−1381.
279. Liu, H.; Liu, G.; Xie, G.; Zhang, M.; Hou, Z.; He, Z. Appl. Surf. Sci. 2011, 257, 3728−3732.
280. Xu, J.; Ao, Y.; Fu, D.; Yuan, C. Colloids Surf., A 2009, 334, 107−111.
281. Singhal, B.; Porwal, A.; Sharma, A.; Ameta, R.; Ameta, S. C. J. Photochem. Photobiol., A 1997, 108, 85−88.
282. Xiao, Q.; Ouyang, L. J. Phys. Chem. Solids 2011, 72, 39−44.
283. Wang, S.; Wang, T.; Ding, Y.; Xu, Y.; Su, Q.; Gao, Y.; Jiang, G.; Chen, W. J. Nanomater. 2012, 2012, 909473.
284. Grandcolas, M.; Cottineau, T.; Louvet, A.; Keller, N.; Keller, V. Appl. Catal., B 2013, 138−139, 128−140.
285. Benoit, A.; Paramasivam, I.; Nah, Y. C.; Roy, P.; Schmuki, P. Electrochem. Commun. 2009, 11, 728−732.
286. Di Valentin, C.; Pacchioni, G.; Selloni, A. J. Phys. Chem. C 2009, 113, 20543−20552.

287. Zhang, Z.; Hedhili, M. N.; Zhu, H.; Wang, P. Phys. Chem. Chem. Phys. 2013, 15, 15637−15644.
288. Weber, M. F.; Schumache, L. C.; Dignam, M. J. J. Electrochem. Soc. 1982, 129, 2022−2028.
289. Chen, X.; Liu, L.; Liu, Z.; Marcus, M. a; Wang, W.-C.; Oyler, N. a; Grass, M. E.; Mao, B.; Glans, P.-A.; Yu, P. Y.; Guo, J.; Mao, S. S. Sci. Rep. 2013, 3, 1510.
290. Howe, R. F.; Grätzel, M. J. Phys. Chem. 1985, 89, 4495−4499.
291. Hahn, R.; Schmidt-Stein, F.; Salonen, J.; Thiemann, S.; Song, Y. Y.; Kunze, J.; Lehto, V.-P.; Schmuki, P. Angew. Chem., Int. Ed. 2009, 48, 7236−7239.
292. Wei, X.; Vasiliev, A. L.; Padture, N. P. J. Mater. Res. 2005, 20, 2140−2147.
293. Zhang, X.; Gao, B.; Hu, L.; Li, L.; Jin, W.; Huo, K.; Chu, P. K. CrystEngComm 2014, 16, 10280−10285.
294. Zhang, X.; Huo, K.; Hu, L.; Wu, Z.; Chu, P. K. J. Am. Ceram. Soc. 2010, 93, 2771−2778.
295. Jiao, Z.; Chen, T.; Xiong, J.; Wang, T.; Lu, G.; Ye, J.; Bi, Y. Sci. Rep. 2013, 3, 2720.
296. Yang, Y.; Lee, K.; Kado, Y.; Schmuki, P. Electrochem. Commun. 2012, 17, 56−59.
297. Yang, L.; Luo, S.; Liu, S.; Cai, Q. J. Phys. Chem. C 2008, 112, 8939−8943.
298. Gao, Z.-D.; Zhu, X.; Li, Y.-H.; Zhou, X.; Song, Y.-Y.; Schmuki, P. Chem. Commun. 2015, 51, 7614−7617.
299. Ozkan, S.; Nguyen, N. T.; Mazare, A.; Cerri, I.; Schmuki, P. Electrochem. Commun. 2016, 69, 76−79.
300. Roy, P.; Kim, D.; Paramasivam, I.; Schmuki, P. Electrochem. Commun. 2009, 11, 1001−1004.
301. Shrestha, N. K.; Yang, M.; Nah, Y. C.; Paramasivam, I.; Schmuki, P. Electrochem. Commun. 2010, 12, 254−257.
302. Salazar, R.; Altomare, M.; Lee, K.; Tripathy, J.; Kirchgeorg, R.; Nguyen, N. T.; Mokhtar, M.; Alshehri, A.; Al-Thabaiti, S. a.; Schmuki, P. ChemElectroChem 2015, 2, 824−828.
303. Honciuc, A.; Laurin, M.; Albu, S.; Sobota, M.; Schmuki, P.; Libuda, J. Langmuir 2010, 26, 14014−14023.
304. Macak, J. M.; Gong, B. G.; Hueppe, M.; Schmuki, P. Adv. Mater. 2007, 19, 3027−3031.
305. Paramasivam, I.; Macak, J. M.; Ghicov, A.; Schmuki, P. Chem. Phys. Lett. 2007, 445, 233−237.
306. Macak, J. M.; Barczuk, P. J.; Tsuchiya, H.; Nowakowska, M. Z.; Ghicov, A.; Chojak, M.; Bauer, S.; Virtanen, S.; Kulesza, P. J.; Schmuki, P. Electrochem. Commun. 2005, 7, 1417−1422.
307. Mohapatra, S. K.; Kondamudi, N.; Banerjee, S.; Misra, M. Langmuir 2008, 24, 11276−11281.
308. Yu, H.; Wang, X.; Sun, H.; Huo, M. J. Hazard. Mater. 2010, 184, 753−758.
309. Song, Y. Y.; Gao, Z.; Schmuki, P. Electrochem. Commun. 2011, 13, 290−293.
310. Macak, J. M.; Schmidt-Stein, F.; Schmuki, P. Electrochem. Commun. 2007, 9, 1783−1787.
311. Lde, D. R. CRC Handbook of Chemistry and Physics, 87th ed.; CRC Press: Boca Raton, FL, 2006.
312. Mazare, A.; Liu, N.; Lee, K.; Killian, M. S.; Schmuki, P. ChemistryOpen 2013, 2, 21−24.
313. Liu, N.; Jha, H.; Hahn, R.; Schmuki, P. ECS Electrochem. Lett. 2012, 1, H29−H31.
314. Hou, Y.; Li, X.; Zou, X.; Quan, X.; Chen, G. Environ. Sci. Technol. 2009, 43, 858−863.
315. Chang, H.-Y.; Tzeng, W.-J.; Cheng, S.-Y. Solid State Ionics 2009, 180, 817−821.
316. Yang, H. Y.; Yu, S. F.; Lau, S. P.; Zhang, X.; Sun, D. D.; Jun, G. Small 2009, 5, 2260−2264.
317. Lei, Y.; Zhao, G.; Liu, M.; Zhang, Z.; Tong, X.; Cao, T. J. Phys. Chem. C 2009, 113, 19067−19076.
318. Pan, H. L.; Yang, T.; Yao, B.; Deng, R.; Sui, R. Y.; Gao, L. L.; Shen, D. Z. Appl. Surf. Sci. 2010, 256, 4621−4625.
319. Liu, Y.; Zhang, X.; Liu, R.; Yang, R.; Liu, C.; Cai, Q. J. Solid State Chem. 2011, 184, 684−689.
320. Wu, L.; Tsui, L. K.; Swami, N.; Zangari, G. J. Phys. Chem. C 2010, 114, 11551−11556.
321. Gerischer, H.; Lübke, M. J. Electroanal. Chem. Interfacial Electrochem. 1986, 204, 225−227.
322. Zhang, Y.; Zhu, J.; Yu, X.; Wei, J.; Hu, L.; Dai, S. Sol. Energy 2012, 86, 964−971.
323. Nguyen, N. T.; Altomare, M.; Yoo, J.; Schmuki, P. Adv. Mater. 2015, 27, 3208−3215.
324. Altomare, M.; Nguyen, N. T.; Schmuki, P. Chem. Sci. 2016, 7, 6865−6886.
325. Lin, J.; Zong, R.; Zhou, M.; Zhu, Y. Appl. Catal., B 2009, 89, 425−431.
326. Liu, C.; Teng, Y.; Liu, R.; Luo, S.; Tang, Y.; Chen, L.; Cai, Q. Carbon 2011, 49, 5312−5320.
327. Shrestha, N. K.; Yang, M.; Paramasivam, I.; Schmuki, P. Semicond. Sci. Technol. 2011, 26, 092002.
328. Liu, J.; Ruan, L.; Adeloju, S. B.; Wu, Y. Dalton trans. 2014, 43, 1706−1715.
329. Jeon, T. H.; Choi, W.; Park, H. J. Phys. Chem. C 2011, 115, 7134−7142.

330. Basahel, S. N.; Lee, K.; Hahn, R.; Schmuki, P.; Bawaked, S. M.; Al-Thabaiti, S. A. Chem. Commun. 2014, 50, 6123−6125.
331. Song, Y. Y.; Schmidt-Stein, F.; Berger, S.; Schmuki, P. Small 2010, 6, 1180−1184.
332. Song, Y. Y.; Roy, P.; Paramasivam, I.; Schmuki, P. Angew. Chem., Int. Ed. 2010, 49, 351−354.
333. Brennan, B. J.; Koenigsmann, C.; Materna, K. L.; Kim, P. M.; Koepf, M.; Crabtree, R. H.; Schmuttenmaer, C. A.; Brudvig, G. W. J. Phys. Chem. C 2016, 120, 12495−12502.
334. Song, Y.; Schmidt-stein, F.; Bauer, S.; Schmuki, P. J. Am. Chem. Soc. 2009, 131, 4230−4232.
335. Cha, G.; Schmuki, P.; Altomare, M. Chem. - Asian J. 2016, 11, 789−797.
336. Nguyen, N. T.; Yoo, J.; Altomare, M.; Schmuki, P. Chem. Commun. 2014, 50, 9653−9656.
337. Nguyen, N. T.; Altomare, M.; Yoo, J. E.; Taccardi, N.; Schmuki, P. Adv. Energy Mater. 2016, 6, 1501926.
338. Yoo, J. E.; Altomare, M.; Mokhtar, M.; Alshehri, A.; Al-Thabaiti, S. A.; Mazare, A.; Schmuki, P. Phys. Status Solidi A 2016, 213, 2733− 2740.
339. Cha, G.; Lee, K.; Yoo, J.; Killian, M. S.; Schmuki, P. Electrochim. Acta 2015, 179, 423−430.
340. Merki, D.; Hu, X. Energy Environ. Sci. 2011, 4, 3878−3888.
341. Kibsgaard, J.; Jaramillo, T. F.; Besenbacher, F. Nat. Chem. 2014, 6, 248−253.
342. Yoo, J. E.; Lee, K.; Schmuki, P. Electrochem. Commun. 2013, 34, 351−355.
343. Wei, W.; Jha, H.; Yang, G.; Hahn, R.; Paramasivam, I.; Berger, S.; Spiecker, E.; Schmuki, P. Adv. Mater. 2010, 22, 4770−4774.
344. Rafieerad, A. R.; Bushroa, A. R.; Nasiri-Tabrizi, B.; Vadivelu, J.; Baradaran, S.; Zalnezhad, E.; Amiri, A. RSC Adv. 2016, 6, 10527−10540.
345. Jiao, W.; Wang, L.; Liu, G.; Lu, G. Q.; Cheng, H. M. ACS Catal. 2012, 2, 1854−1859.
346. Liu, G.; Pan, J.; Yin, L.; Irvine, J. T.; Li, F.; Tan, J.; Wormald, P.; Cheng, H. M. Adv. Funct. Mater. 2012, 22, 3233−3238.
347. Zhang, M.; Wang, Q.; Chen, C.; Zang, L.; Ma, W.; Zhao, J. Angew. Chem., Int. Ed. 2009, 48, 6081−6084.
348. Palmisano, G.; Augugliaro, V.; Pagliaro, M.; Palmisano, L. Chem. Commun. 2007, 33, 3425−3437.
349. Ravelli, D.; Dondi, D.; Fagnoni, M.; Albini, A. Chem. Soc. Rev. 2009, 38, 1999−2011.
350. Takeda, N.; Torimoto, T.; Sampath, S.; Kuwabata, S.; Yoneyama, H. J. Phys. Chem. 1995, 99, 9986−9991.
351. Anderson, C.; Bard, A. J. J. Phys. Chem. B 1997, 101, 2611−2616.
352. Matos, J.; Laine, J.; Herrmann, J. M. Appl. Catal., B 1998, 18, 281−291.
353. Anandan, S.; Yoon, M. J. Photochem. Photobiol., C 2003, 4, 5−18. (353) Kim, Y.; Yoon, M. J. Mol. Catal. A: Chem. 2001, 168, 257−263.
354. Zampieri, A.; Dubbe, A.; Schwieger, W.; Avhale, A.; Moos, R. Microporous Mesoporous Mater. 2008, 111, 530−535.
355. Ha, K.; Lee, Y.-J.; Lee, H. J.; Yoon, K. B. Adv. Mater. 2000, 12, 1114−1117.
356. Mabande, G. T. P.; Ghosh, S.; Lai, Z.; Schwieger, W.; Tsapatsis, M. Ind. Eng. Chem. Res. 2005, 44, 9086−9095.
357. Paramasivam, I.; Avhale, A.; Inayat, A.; Bösmann, A.; Schmuki, P.; Schwieger, W. Nanotechnology 2009, 20, 225607.